\documentclass[aps,rmp,reprint,amsmath,amssymb,graphicx,longbibliography]{revtex4-1}
\pdfoutput=1
\usepackage{bm}
\usepackage{graphicx}
\usepackage{xcolor}
\usepackage{mathtools}
\usepackage{chronology}

\newcommand{\btheta}{\mbox{\boldmath $\theta$}}
\renewcommand{\d}{{\bf d}}

\newcommand{\p}{\mathbb{P}}
\renewcommand{\d}{{\bf d}}
\newcommand{\e}{{\rm e}}
\newcommand{\n}{{\bf n}}
\newcommand{\h}{{\bf h}}
\renewcommand{\S}{{\bf S}}
\newcommand{\I}{{\bf I}}

\begin{document}

\title{Parameter Estimation with Gravitational Waves}

\author{Nelson Christensen}
\affiliation{Artemis, Universit\'e C\^ote d'Azur, Observatoire de la C\^ote d'Azur, Nice 06300, France}
\author{Renate Meyer} 
\affiliation{Department of Statistics, University of Auckland, Auckland 1142, New Zealand}

\date{\today{}}

\begin{abstract}
The new era of gravitational wave astronomy  truly began on September 14, 2015 with the detection of GW150914, the sensational first direct observation of gravitational waves from the inspiral and merger of two black holes  by the two Advanced LIGO detectors.  In the subsequent  first three observing runs of the LIGO/Virgo network, gravitational waves from $\sim 50$ compact binary mergers have been announced, with more results to come. The events have mostly been produced by binary black holes, but two binary neutron star mergers have so far been observed, as well as the mergers of two neutron star - black hole systems. Furthermore, gravitational waves emitted by core-collapse supernovae, pulsars and the stochastic gravitational wave background are within the LIGO/Virgo/KAGRA sensitivity band and are likely to be observed in future observation runs. Beyond signal detection, a major challenge has been the development of statistical and computational methodology for  estimating the physical waveform parameters and quantifying their uncertainties in order to accurately  characterise the emitting system. These methods depend on the sources of the gravitational waves and the gravitational waveform model that is used. This article reviews the main waveform models and parameter estimation methods used to extract physical parameters from gravitational wave signals detected to date by LIGO and Virgo and from those expected to be observed in the future, which will include KAGRA, and how these methods interface with various aspects of LIGO/Virgo/KAGRA science. 
Also presented are the statistical methods used by LIGO and Virgo to estimate detector noise, test general relativity, and draw conclusions about the rates of compact binary mergers in the universe. Furthermore, a summary of major publicly available gravitational wave parameter estimation software packages is  given.
\end{abstract}



\maketitle

\tableofcontents{}

\section{Introduction}
\label{intro}
While we can see the universe with electromagnetic radiations, we can now also listen to the universe with gravitational waves. After decades of work, Advanced LIGO~\cite{TheLIGOScientific:2014jea} and Advanced Virgo~\cite{TheVirgo:2014hva} have made direct detections~\cite{LIGOScientific:2018mvr,Abbott:2020niy,LIGOScientific:2021qlt}. The detections in and of themselves were the confirmation of the prediction made by Albert Einstein a century before~\cite{Einstein:1916b,Einstein:1918btx} as a consequence of general relativity~\cite{Einstein:1916a}. The direct detection of gravitational waves was a fundamental physics result of tremendous significance. It should be noted that the existence of gravitational waves was already established by the observation of the decay of the orbit of a binary neutron star system, exactly at the rate predicted by general relativity~\cite{1982ApJ...253..908T,taylor:1989,2010ApJ...722.1030W,0004-637X-829-1-55}. Gravitational waves are a new means to observe the universe, and are providing important astrophysical and cosmological information, with much more to come with future observations.

Advanced LIGO and Advanced Virgo have now completed three observational runs.
The first observing run, O1, occurred from September 12, 2015 until January 19, 2016, and involved only Advanced LIGO.  Advanced LIGO's second observing run, O2, started on November 30, 2016, then went until August 25, 2017. On August 1, 2017, Advanced Virgo formally joined O2, although it was acquiring some engineering data leading up to that date. Advanced Virgo also observed in O2 up until August 25, 2017, thereby providing the first time three detectors were used to search for gravitational waves. The third observing run, O3, commenced on April 1, 2019, and ended on March 27, 2020.

\begin{figure}
\includegraphics[width=0.5\textwidth]{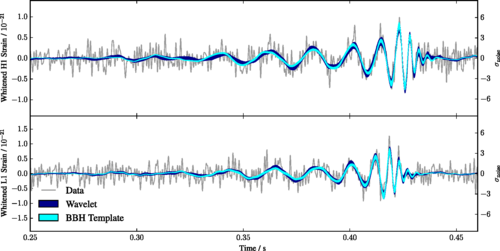}
\caption{The measured detector strain time-series of the first detected gravitational wave signal by Advanced LIGO, GW150914~\cite{Abbott:2016blz}, as observed in H1 (top) and L1 (bottom). The times displayed are with respect to September 14, 2015, 09:50:45 UTC.
The shaded regions are the 90\% credible regions for the reconstructed waveforms.  The dark blue comes from a model that does not assume a particular waveform morphology (template agnostic) and employs sine-Gaussian wavelets, namely BayesWave, to be described below and in~\citet{Cornish:2014kda}. The cyan corresponds to the modeled analyses (templated analysis) using IMRPhenom~\cite{Hannam:2013oca,Schmidt:2014iyl,Husa:2015iqa,Khan:2015jqa,Bohe:PPv2,Ajith:2007kx,Pan:2007nw,Ajith_2007} and EOBNR~\cite{Taracchini:2013rva,Pan:2013rra} template waveforms (to be described below). Grey traces represent the data.
From~\citet{TheLIGOScientific:2016wfe}; see there for more details.}
\label{fig:GW150914time}
\end{figure}

\begin{figure}
\includegraphics[width=0.5\textwidth]{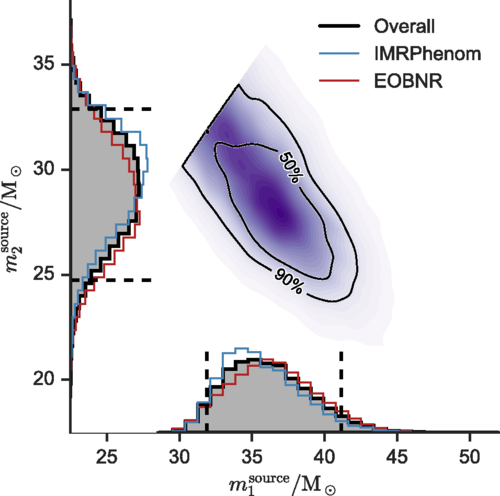}
\caption{The source frame component mass parameter posterior probability distributions for the first detected gravitational wave signal by Advanced LIGO, GW150914~\cite{Abbott:2016blz}, from~\citet{TheLIGOScientific:2016wfe}. The waveform models are IMRPhenom (blue), and EOBNR (red), with the combined posterior probability distributions (black). The 50\% and 90\% credible regions are drawn for the two-dimensional posterior probability distribution. See ~\citet{TheLIGOScientific:2016wfe} for more details.}
\label{fig:GW150914massPE}
\end{figure}

The initial observation of gravitational waves was made by the two Advanced LIGO detectors on September 14, 2015~\cite{Abbott:2016blz}. The signal detected at the LIGO Livingston Observatory (L1) and the LIGO Hanford Observatory (H1) can be seen in Fig.~\ref{fig:GW150914time}. Parameter estimation methods were then employed on this observed signal, in this case, the LALInference package of the LIGO Algorithm Library (LAL) software suite~\cite{Veitch:2014wba}, which will be described below. The physical parameters estimated for this signal include the masses, spins, luminosity distance, sky position, and other parameters which will be described in more detail below. It was from the parameter estimation results that the initial component masses $m_{1} = 36^{+5}_{-4} M_{\odot}$ and $m_{2} = 29^{+4}_{-4} M_{\odot}$ of the system in its source frame corresponding to GW150914 were obtained. The parameter estimation routines generated bivariate and univariate posterior distributions 
for these masses which are displayed in Fig.~\ref{fig:GW150914massPE}. The signal reconstruction appearing in Fig.~\ref{fig:GW150914time} is also a consequence of the parameter estimation. 
From the observation that there are two apparent point masses of $36 M_{\odot}$ and $29 M_{\odot}$ one can declare that LIGO has observed black holes, another important consequence of Einstein's theory of general relativity~\cite{1916SPAW.......189S,PhysRevLett.11.237,PhysRev.56.455}.
The observation of stellar mass black holes around $30 M_{\odot}$ also had important astrophysical implications~\cite{TheLIGOScientific:2016htt}.
Presented in Fig.\ \ref{fig:timeline}  is a timeline of important events pertaining to gravitational waves.

\begin{figure*} 
\begin{chronology}*[10]{1915}{2035}{\textwidth}[20cm]
\centering
\event[1916]{1918}{Einstein's papers on gravitational waves}
\event{1939}{Prediction of neutron stars}
\event{1957}{Chapel Hill conference}
\event{1963}{Spinning mass solution}
\event{1967}{Observation of a pulsar}
\event{1974}{Discovery of the binary pulsar}
\event{1982}{Orbital decay of the binary pulsar}
\event[1994]{1996}{LIGO and Virgo construction begins}
\event[2002]{2010}{Initial LIGO and Virgo}
\event[\decimaldate{12}{9}{2015}]{\decimaldate{19}{1}{2016}}{O1}
\event[\decimaldate{30}{11}{2016}]{\decimaldate{25}{8}{2017}}{~ O2}
\event[\decimaldate{1}{4}{2019}]{\decimaldate{27}{3}{2020}}{O3}
\event[\decimaldate{1}{10}{2022}]{\decimaldate{1}{10}{2023}}{O4}
\event[\decimaldate{1}{1}{2025}]{\decimaldate{30}{6}{2026}}{O5}
\event{2034}{LISA}
\event{2031}{Third generation detectors}
\end{chronology}
\caption{Presented here is a timeline of significant events in the history of gravitational waves. These include Einstein's papers on gravitational waves~\cite{Einstein:1916b,Einstein:1918btx}, predictions of gravitational collapse of stars and the creation of neutron stars~\cite{PhysRev.56.455,PhysRev.55.374}, and a general relativistic solution for spinning masses (such as black holes)~\cite{PhysRevLett.11.237}. The famous Chapel Hill conference in 1957 led to the acceptance by the physics community that gravitational waves truly exist~\cite{RevModPhys.29.352}. The first pulsar was discovered in 1967~\cite{1968Natur.217..709H}, followed by the binary pulsar in
1967~\cite{1975ApJ...195L..51H}, and the observation of the orbital decay of the binary pulsar by the emission of gravitational waves~\cite{1982ApJ...253..908T}. Construction for LIGO began in 1994, and 1996 for Virgo. Initial LIGO and Virgo made observations from 2002 to 2010, but with no detections. The Advanced LIGO and Advanced Virgo observing runs to date at O1, O2 and O3, while future observing runs O4 and O5 will also include KAGRA. The LISA mission is presently predicted to begin in 2034~\cite{2017arXiv170200786A}, while the third generation detector, such as Einstein Telescope~\cite{ET_Punturo} and Cosmic Explorer~\cite{Reitze:2019dyk}, should start observations in the 2030s.}
\label{fig:timeline}
\end{figure*}

\begin{figure}
\includegraphics[width=0.5\textwidth]{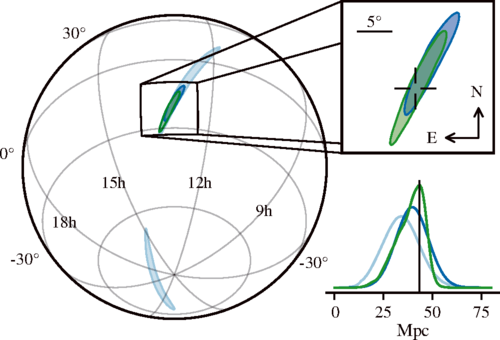}
\caption{The estimates of the sky position and luminosity distance for the source of GW170817 using the data from the two Advanced LIGO detectors and the Advanced Virgo detector. A rapid parameter estimation routine~\cite{PhysRevD.93.024013} using just the data from the two LIGO detectors (light blue contours) constrains the source sky location to 190 deg$^{2}$, while with the data from Virgo included (dark blue contours) the location uncertainty is reduced to 31 deg$^{2}$. The offline parameter estimation analysis of~\citet{Veitch:2014wba} further reduces the sky location uncertainty to 28 deg$^{2}$ (green contour). The upper right shows a zoom-in with a cross identifying the location of the galaxy NGC 4993, where the source was found~\cite{2017Sci...358.1556C}. The bottom right shows the posterior distribution functions for the luminosity distance with the vertical line showing the redshift measured distance to NGC 4993.   
See ~\citet{TheLIGOScientific:2017qsa}, the source of this figure, for more details.}
\label{fig:GW170817sky}
\end{figure}

The importance of parameter estimation was also dramatically displayed in association with GW170817~\cite{TheLIGOScientific:2017qsa}, observed by Advanced LIGO and Advanced Virgo, and produced by the merger of a binary neutron star system. This observation was the birth of gravitational wave multi-messenger astronomy. In addition to the gravitational wave signal, a gamma ray burst was observed 1.7 s after the merger time~\cite{Monitor:2017mdv}. The gravitational wave signal gave an estimate of the sky positon and distance to the source (see Fig.~\ref{fig:GW170817sky}); this was consistent with the sky position that could be inferred from the gamma ray signals detected by Fermi-GBM~\cite{2017ApJ...848L..14G} and INTEGRAL~\cite{Savchenko:2017ffs}. The localization of the emission of the signal allowed for the identification of the source, and then the discovery of the kilonova~\cite{2017Sci...358.1556C,TheLIGOScientific:2017qsa}. This was a tremendously important event, and the ability to to use parameter estimation to find the source of GW170817 led to numerous significant observations~\cite{2017ApJ...848L..12A}. It also led to important fundamental physics results, such as an independent measurement of the Hubble constant~\cite{Abbott:2017xzu}, and the demonstration that the speed of gravity is the same as the speed of light~\cite{Monitor:2017mdv}. 

The goal of this review will be to summarize the state of parameter estimation for gravitational wave observations. The review will be dedicated to applications from the ground based detectors (Advanced LIGO, Advanced Virgo, KAGRA~\cite{Aso:2013,Akutsu:2017thy,Akutsu:2019rba}). There are numerous studies addressing parameter estimation for the Laser Interferometer Space Antenna (LISA)~\cite{2017arXiv170200786A,Babak_2017}, and pulsar timing~\cite{Hobbs:2017zve,Alam:2020fjy,Alam:2020laa}. For this review we address only the ground based detector results, and encourage the reader to also explore the rich parameter estimation literature for LISA and pulsar timing. In their first three observation runs Advanced LIGO and Advanced Virgo have reported the observations of gravitational waves from $\sim 50$ compact binary mergers~\cite{LIGOScientific:2018mvr,Abbott:2020niy}. The results from the second half of the third observing run are still forthcoming. Parameter estimation routines applied to these signals have produced results in fundamental physics, astrophysics and cosmology. We present a review of these methods and results. 

\section{History}
\label{History}
We present here a review of the history of parameter estimation for gravitational waves.
Early papers that addressed the inverse problem of estimating the parameters of gravitational wave signals from ground-based laser interferometric measurements
were predominantly using maximum-likelihood (ML) based approaches. \citet{Davis1989} gave a general overview of the statistical theory of signal detection including the
classical theory of hypothesis testing, ML estimation of unknown parameters, nonlinear filtering in signal detection and pre-whitening filters to handle correlated noise.
 \citet{GurselYekta1989Nost} developed a method to estimate the source direction and the  amplitudes 
of gravitational wave  burst signals  from observations of three detectors assuming white noise using  a least squares approach and optimal filtering without any explicit model for the GW waveform.
\citet{Echeverria1989Gmot} was the first to not only estimate the mass and angular momentum of a perturbed Kerr black hole from the emitted gravitational waves  but to describe a method  for determining the uncertainty in the parameter estimates. The method is based on Wiener optimal filtering and determining the parameters that maximize the signal-to-noise ratio. However, the results were valid only for a sufficiently large signal-to-noise ratio (SNR) and provided no guidance for estimating the amplitude of the signal.
ML estimation was used by  \citet{Krolak19933451} to estimate the amplitude, phase, time of arrival and chirp mass of compact binary coalescing systems. The chirp mass is defined as
\begin{equation}
\label{eq:chirp_mass}
\mathcal{M}_c = \frac{(m_{1} m_{2})^{3/5}}{(m_{1} + m_{2})^{1/5}} ~,
\end{equation}
and it is the mass parameter that describes the evolution of a binary system as its orbit decays via gravitational wave emission. In Sec.~\ref{sec:CBC} a more complete description is given.
\citet{JaranowskiPiotr1994Ostt} extended the ML approach to data from three detectors to estimate the source direction and the strain amplitudes.
\citet{CutlerCurt1994Gwfm} determined the accuracy of ML estimates of  the  masses, spins, and the distance to earth for gravitational waves of binary inspiralling systems using lowest order Newtonian waveforms. They gave an example using the Markovi\'{c} approximation \cite{PhysRevD.48.4738} that there is no uncertainty of the location in the sky of the binary. This reduced the unknown parameters of the gravitational waveform to the distance to the binary, the cosine of the inclination angle to the line of sight, the polarization angle of the gravitational wave and the phase at collision time.  They expanded their analysis to include post-Newtonian effects on parameter estimates. 

Even though the development of inverse probability and Bayes' theorem by Thomas Bayes and Pierre-Simon Laplace  dates back to the eighteenth century and having physicists Sir Harold Jeffreys and Edwin Jaynes as  strong proponents in the twentieth century, Bayesian ideas in astronomy
did not appear until the late 1970s in the field of  image restoration~\cite{GullSF1978Irfi}. The methodology was focused on the principle of maximum entropy, finding a point estimate  based on maximizing an entropy-based prior subject to some  likelihood-based constraint \cite{HilbeJosephM2013BAAB}. This still fell short of a proper Bayesian approach using all of the posterior distribution and marginalization to determine  point estimates of  parameters and their uncertainties. 
The use of Bayesian inference in astronomy as well as most other disciplines was hindered by the difficulty of computing the posterior distribution for problems with a large number of  parameters.  Computing the joint posterior distribution using Bayes' theorem required high-dimensional integration to evaluate the normalization constant, and finding the marginal posterior distribution of each individual parameter. As numerical integration  is only feasible for dimensions up to $\approx 5$, Laplace approximation \cite{GelmanAndrew2014} works well only for unimodal and symmetric posteriors, and ordinary simulation methods based on independent random draws  such as importance sampling \cite{GelmanAndrew2014} are only efficient in low dimensions,  the progress of Bayesian inference was ultimately linked to progress in computational solutions to high-dimensional integration problems.
One such solution already came with the development of the Metropolis algorithm by~\citet{Metropolis19531087} and its extensions, the Metropolis-Hastings (MH) algorithm \cite{Hastings197097} and a modification by~\citet{BarkerA.A1965MCCO}. Instead of generating independent samples from the posterior distribution, Metropolis came up with the idea of generating {\em dependent} samples from an ergodic Markov chain, constructed such  that  the posterior distribution is its stationary distribution. However, its huge importance for Bayesian posterior computations remained unnoticed until
the seminal paper by~\citet{GemanStuart1984SRGD} that applied the MH algorithm to the problem of image restoration. This paper marked the turning point as it garnered huge interest not only from computer scientists but also statisticians. More or less at the same time, statisticians had simultaneously developed the Gibbs sampler, based on the idea of generating a sample from a multivariate distribution by generating repeatedly over cycles of full conditional distributions~\cite{TannerMartinA1987TCoP}. The Gibbs sampler was later shown to be a special case of the MH algorithm, for a proof see e.g.\ Chapter 11.5 of~\citet{GelmanAndrew2014}.
 The papers by~\citet{GelfandAlanE1990SAtC} and \citet{Gilks1996} laid out the foundations of simulation-based Bayesian posterior computation using Markov chain Monte Carlo (MCMC) methods.
Coupled with this algorithmic breakthrough, the 1990s also saw a continuous  improvement in hardware and computing power. This combination triggered revolutionary advances in Bayesian computation via MCMC. For a detailed review of the history of MCMC, see~\citet{HitchcockDavidB2003AHot}. Thus, it was in the mid to late 1990s that Bayesian computations became computationally feasible -- just at the time the LIGO Scientific Collaboration (LSC) was established in 1997 and interest in the development of parameter estimation of signals to be observed by LIGO and Virgo was growing.

In 1992, \citet{Loredo1992} had already clearly contrasted the Bayesian and frequentist approach to statistical inference and outlined its promise for astrophysics. \citet{GregoryPC1992Anmf} had advocated strongly for a Bayesian approach to statistical inference in astrophysical problems.
The paper by~\citet{PhysRevD.46.5236} laid out a Bayesian strategy for parameter estimation, using the posterior mode as a point estimate and 
normal approximation to determine a credible interval. \citet{Finn:1992xs} explored gravitational waves of binary inspirals observed by a single detector, using a quadrupole waveform approximation characterised by four parameters: amplitude, chirp mass, arrival time  and phase, again using the normal approximation for posterior computation.
\citet{Finn:1997qx} described the difference between the traditional frequentist and Bayesian approach to data analysis and pointed out the  importance of Bayesian  parameter estimation for gravitational waves. The first paper to show how computational difficulties in computing the Bayesian posterior distribution of gravitational wave parameters can be overcome by the sampling-based approach of MCMC was \citet{PhysRevD.58.082001}. It demonstrated the performance of the Gibbs sampler in the same four-parameter setup as \citet{CutlerCurt1994Gwfm}.
The MCMC-based approach to posterior computation was rapidly taken up by the LSC and applied to gravitational waves from pulsars \cite{ChristensenNelson2004Mafe, Umstatter2004,PhysRevD.72.102002,PhysRevD.76.043003}, rotating core-collapse supernovae \cite{Rover2009, PhysRevD.86.044023, Edwards2014}, and observations from the planned space-based observatory LISA \cite{Cornish:2005qw,Cornish2007,Crowder2007,Gair2010,Ali2012,PhysRevD.72.022001,Umst_tter_2005}. For binary inspirals, increasingly sophisticated MCMC techniques were developed that took more and
more accurate waveform approximations into account with growing number of parameters \cite{PhysRevD.64.022001, Pai200130, Christensen2004317,Rover:2006ni,Rover:2006bb, vanderSluys:2007st, vanderSluys:2008qx,Aasi:2013jjl}. These were further developed and  finally integrated into the LALInference adaptive MCMC routine that is in place today to estimate the 15 parameters describing the coalescence of two compact spinning binaries \cite{Veitch:2014wba}. \citet{skilling2006} developed nested sampling as an algorithm to approximate marginal likelihoods, which, as a by-product, also generates samples from the posterior distribution. Nested sampling simulates from the prior, conditional on having a likelihood value above a threshold. In many applications,  importance sampling or MCMC algorithms are required to generate these internal samples.  \citet{Veitch:2009hd} introduced nested sampling for parameter estimation of gravitational waves. Nested sampling has been integrated in LALInference  and together with the adaptive MCMC  is routinely employed for parameter estimation and evidence calculation in LIGO. Both MCMC and nested sampling were used to estimate the parameters of the very first gravitational wave signal GW150914 observed by LIGO, yielding consistent sets of parameter estimates \cite{TheLIGOScientific:2016wfe}. 

The importance that Bayesian parameter estimation played in describing the physics associated with the first direct observation of gravitational waves with GW190514 is summarized in \citet{doi:10.1111/j.1740-9713.2016.00896.x} and \citet{MeyerRenate2020Ctfp}. For the first gravitational wave multimessenger event, GW170817~\cite{TheLIGOScientific:2017qsa}, parameter estimation provided an estimation of the source position which then led to finding the optical counterpart~\cite{2017Sci...358.1556C}, observations and studies of a kilonova~\cite{Metzger:2019zeh}, estimation of neutron star radius and tidal deformability~\cite{Coughlin:2018fis,Weinberg:2018icl,Abbott:2018exr}, a speed of gravity limit~\cite{Monitor:2017mdv}, a new estimation of the Hubble constant~\cite{Abbott:2017xzu}, tests of general relativity~\cite{Abbott:2018lct}, and much more. Now with $\sim 50$ gravitational wave signal observed from compact binaries parameter estimation allows for further tests of general relativity~\cite{Abbott:2020jks} as well as detailed studies on the formation mechanisms for these systems~\cite{Abbott:2020gyp}.

\section{Methods}
\label{Methods}
We will briefly review the general Bayesian approach to statistical inference before describing the computational methods used for calculating the posterior distribution of gravitational wave parameters. For comprehensive treatments of Bayesian inference, the reader is referred to \cite{Jaynes,GelmanAndrew2014, Gregory:2005:BLD:1051497, GamermanDani2006MCMC, FeigelsonEricD2012Msmf}.

The application of Bayes' theorem, expressed in terms of probabilities  for observable events $A$ and $B$   
\begin{equation}
\p(A|B) = \frac{\p(B|A)\p(A)}{\p(B)}= \frac{\p(B|A)\p(A)}{\p(B|A)\p(A)+\p(B|A^c)\p(A^c)}
\end{equation}
is completely uncontroversial when applied for instance to medical diagnostic testing where
$A$ might be the event that a patient has a certain disease, $A^c$ its complement, i.e., that a patient does not have the disease, and $B$ the event that a diagnostic test for that disease returns a positive result.
It is purely based on probability theory and the definitions of conditional and marginal probabilities.
The conditional probability of $\p(B|A)$ refers to the {\em likelihood} and $\p(A)$ to the {\em prior} probability of $A$.
Bayes' theorem gives us a formula to update a prior probability to the {\em posterior} probability $\p(A|B)$ after observing $B$.
 Its application to the scenario where $A$ are unknown parameters of a statistical model and $B$ the observed data has been the cause for 
seemingly unresolvable disputes between frequentists and Bayesians. Whereas frequentist argue that unknown parameters are fixed quantities and thus cannot have a probability distribution, Bayesians aim to quantify the uncertainty about unknown parameters using probability and thus treat them as random quantities. Denoting observations by $\d=(d_1,\ldots,d_n)$ and unknown parameters by $\btheta=(\theta_1,\ldots,\theta_p)$, the uncertainty about the values of $\btheta$ before observing the data is expressed by the prior probability density function $\pi(\btheta)$. The likelihood function $L(\d|\btheta)$ is the conditional probability density function of the data given the unknown parameters.
Bayes' theorem in this context gives the posterior probability density of the parameters after observing the data:
\begin{equation}\label{Bayestheorem}
p(\btheta|\d) = \frac{L(\d|\btheta)\pi(\btheta)}{Z} =\frac{L(\d|\btheta)\pi(\btheta)}{\int L(\d|\btheta)\pi(\btheta)d\btheta} \propto L(\d|\btheta)\pi(\btheta).
\end{equation} 
The denominator $Z$, also called the {\em marginal likelihood, evidence, or the prior predictive distribution},  does not depend on $\btheta$ and is therefore merely a normalization constant as far as the posterior distribution of $\btheta$ is concerned. It turns out that for many sampling-based techniques, this normalization constant is not needed, therefore the posterior distribution is often written just as proportional to prior times likelihood. The normalization constant is important, though, for model comparison and is an integral part of the Bayes factor, see \ref{subsec:ModelComparison}.
Calculating the marginal posterior distribution of one of the parameters requires integrating the joint posterior over the remaining $p-1$ parameters.
Similarly, calculating the posterior mean or variance requires a further integration. Thus, the main difficulty in Bayesian posterior computation when the dimension of the parameter space is large, lies in  solving high-dimensional integration problems. These integration problems can be solved analytically only for conjugate priors,
by numerical integration only in low dimensions, using Gaussian approximations for unimodal and symmetric posteriors but, in general, require simulation-based computational techniques such as MCMC or nested sampling as described in more detail below.

For gravitational wave applications,  the data $\d$ consists of  $K$ time series of gravitational wave strain measurements $\d^{(k)}=(d^{(k)}(t))$ taken  at detector $k$ in a network of $K$ detectors which would comprise the Advanced LIGO detectors at Livingston and Hanford (USA), the Advanced Virgo detector at Pisa (Italy), the cryogenic detector KAGRA in Kamioka (Japan) and GEO 600 in Hanover (Germany). If the sampling frequency (e.g.\ 16384 Hz for LIGO)
is denoted by $f_s=\frac{1}{\Delta_t}$ and the measurements span $\tau_{obs}$ seconds, the strain time series consists of $T=f_s\times \tau_{obs}$ measurements $d^{(k)}(t), t=1,\ldots,T$ with a spacing of $\Delta_t$ seconds.
Ignoring calibration error that will be discussed in \ref{subsec:calibration}, the data $d^{(k)}(t)$ are modeled as a gravitational wave signal $h^{(k)}(t|\btheta)$
buried in interferometer noise $n^{(k)}(t)$:
\begin{equation}\label{eq:modelintime}
d^{(k)}(t) = h^{(k)}(t|\btheta) + n^{(k)}(t),\quad t=1,\ldots,T.
\end{equation}
The noise term combines a variety of noise sources such as quantum, seismic, and  thermal noise and is assumed to be mean zero, wide sense stationary and Gaussian with power spectral density (PSD) $S^{(k)}$. 
The PSD is given by the Fourier transform of the autocovariance function $\gamma^{(k)}$, i.e.,\  $S^{(k)}(f)=\sum_{\ell=-\infty}^{\infty} \gamma^{(k)}(\ell) \e^{-i\ell f}$ and uniquely characterizes the second order properties of the time series. For the moment, we also assume that $S^{(k)}$ is known.
In Sections \ref{subsec:noise} and \ref{subsec:signalplusnoise} we will explain statistical methods for estimating the spectral density for a noise-only time series and how to simultaneously estimate the spectral density and gravitational wave signal parameters, respectively.
The strain signal $h^{(k)}(t|\btheta)$ depends on, say $p$ parameters $\btheta=(\theta_1,\ldots,\theta_p)$, such as the distance of the source to the earth, the masses and spins for compact binary coalescences.
With respect to a geocentric reference frame, the strain measured at detector $k$ of a gravitational wave source with polarization amplitudes $h_+$ and $h_\times$ located in the sky at 
$(\alpha,\delta)$ where $\alpha$ is the right ascension and $\delta$ the declination of the source is 
\begin{equation}\label{eq:signalintime}
h^{(k)}(t|\btheta)=F^{(k)}_+(\alpha,\delta,\psi)h_+(t|\btheta) +F^{(k)}_\times(\alpha,\delta,\psi)h_\times(t|\btheta)
\end{equation}
where $F_{+,\times}$ denote the antenna responses as  functions of the source locations and the polarization angle $\psi$  of the gravitational waves \cite{PhysRevD.63.042003}. The time series are usually down sampled from their original sampling frequency  of 16384~Hz for LIGO and 20~kHz for Virgo to  4096~Hz,  band-pass filtered as the LIGO and Virgo detectors are sensitive and calibrated only in the frequency bands of 10~Hz to 5~kHz for LIGO and 10~Hz to 8~kHz for Virgo,  and 
notch filtered around known instrumental noise frequencies~\cite{Covas:2018,LIGOScientific:2019lzm}. Code for these pre-processing steps is included in
the LAL library \citep{lalsimulation}. The exact form of the gravitational waveform model $h_{+,\times}(t|\btheta)$ and its parameters
depends on the emitting source of the gravitational waves. We will mainly focus on binary inspiral signals for which the parameter vector $\btheta$ consists of 15 individual parameters with their prior $\pi(\btheta)$ and describe their gravitational waveforms in Section \ref{sec:CBC}. Gravitational waveform models for other sources such as pulsars, core collapse supernovae and the stochastic gravitational wave background will be presented in Section  \ref{Other-Sources}. 

For parameter estimation of ringdown-only signals, 
namely after two compact objects merge and the newly formed black hole oscillates and emits gravitational waves until it comes to equilibrium as a Kerr black hole,
the time-domain formulation of the likelihood (Eq.~\ref{eq:modelintime}) is directly used \citep{Carullo:2019flw,Isi:2019aib} to avoid the contribution of spurious frequency contributions from the pre-merger phase. However in general,  making use of the Wiener-Khinchin theorem \citep{wiener1964time}, 
the likelihood is usually specified in the frequency domain by the assumption of stationary Gaussian errors with known PSD and the independence of observations between interferometers, yielding
\begin{equation}\label{eq:Whittle}
\begin{split}
L(\d|\btheta) &= \prod_{k=1}^K L(\d^{(k)}|\btheta) \\
&\propto  \prod_{k=1}^K \e^{-\frac{1}{T}(\tilde{\d}^{(k)}-\tilde{\h}^{(k)})^*\S^{(k)^{-1}}(\tilde{\d}^{(k)}-\tilde{\h}^{(k)})} 
\end{split}
\end{equation}
where the complex vector $\tilde{\d}^{(k)}$ contains the  Fourier coefficients defined by
\[\tilde{d}^{(k)}_j=\tilde{d}^{(k)}(f_j)=\sum_{t=1}^{T} d^{(k)}(t) \e^{-itf_j}\] 
and $\S^{(k)}$ is  a diagonal matrix containing the PSD $S^{(k)}(f_j)$
at the  Fourier frequencies $f_j=2\pi j/T, j=0,\ldots,N$, $N= \lfloor(T-1)/2 \rfloor$. The likelihood in Eq.~\ref{eq:Whittle} is known as the {\em Whittle likelihood} and
provides an approximation to the exact Gaussian likelihood \citep{Contreras-CristAlberto2006ANoW, rao2020reconciling} which would have a non-diagonal covariance matrix with a Toeplitz structure given by $\Sigma^{(k)}=\left( \gamma^{(k)}(i-j)\right)_{i,j=1,\ldots,T} $ with $\gamma^{(k)}(\ell)=\int_{-\pi}^{\pi} S^{(k)}(f) \e^{i\ell f}df$ for $\ell=0,\ldots,T-1$. But it also provides a valid approximation to non-Gaussian likelihoods \citep{tang2021posterior}. The Whittle likelihood is thus an approximation in two respects: the first approximation takes place when one uses a Gaussian likelihood to approximate the potentially non-Gaussian likelihood of the observations; the second approximation stems from replacing the quadratic form corresponding to the observations in the exponent of the time-domain Gaussian likelihood with a frequency domain term by the sum of $(\tilde{d}^{(k)}(f_j)-\tilde{h}^{(k)}(f_j))^2/S^{(k)}(f_j)$.
 The Whittle likelihood approximation therefore facilitates the likelihood evaluations by avoiding matrix inversions of large covariance matrices.
Before going into more detail about  specific waveforms and prior distributions, the next subsections describe simulation-based methods to compute the posterior distribution of gravitational waveform parameters. These assume the PSD to be known and fixed for the purpose of parameter estimation. Section \ref{PSD} will review techniques for spectral density estimation.

\subsection{Markov Chain Monte Carlo}
The well-known Monte Carlo technique of rejection sampling \cite{RubinsteinReuvenY1981SatM} can be regarded as a precursor of MCMC and the Metropolis algorithm is easily understood as a generalization.

To sample from a target probability density function $p(\btheta|\d)$, rejection sampling generates a candidate $\btheta^*\sim q(\btheta)$ from a 
proposal probability density function $q(\btheta)$ which majorizes the target, i.e.,\  $p(\btheta|\d)\leq M q(\btheta)$ for all $\btheta$ and $M>0$.
Then it accepts $\btheta^*$ with probability
\[\alpha=\frac{p(\btheta^*|\d)}{M q(\btheta^*)}\]
and otherwise rejects $\btheta^*$ and draws a new candidate.
It can be shown that the
acceptance probability is $1/M$, so the rejection method is efficient, if $M$ is close to 1, i.e.,\ $q(.)$ close to $p(.)$. Note that the normalization constant of the
posterior is not needed for rejection sampling as it would cancel in the acceptance ratio. So this could easily be used for sampling from the posterior, however the main difficulty consists of finding this majorization density as this would require solving an optimization problem first. In high dimensions, it will also be difficult to find a majorization density that is close enough to the target to be efficient.

Now, the Metropolis-Hastings (MH) algorithm is an extension of rejection sampling in that it  avoids having to use a majorization proposal at the expense of generating {\em dependent} instead of independent samples. It starts with some arbitrary $\btheta_0$.
The proposal probability density function  $q(\btheta|\btheta_0)$ can now depend on $\btheta$ and does not have to majorize the target.
A  candidate $\btheta^* \sim q(\btheta|\btheta_0)$ is generated and accepted with probability
\[\alpha(\btheta_0)=\min\left\{1,\frac{p(\btheta^*|\d)}{M(\theta_0) q(\btheta^*|\btheta_0)}\right\}\]
with $\displaystyle M(\btheta_0)=\frac{p(\btheta_0|\d)}{q(\btheta_0|\btheta^*)}$. That is the new value
$\btheta_1=\btheta^*$ and otherwise, the previous value is recycled, i.e.,\ $\btheta_1=\btheta_0$. This is then iterated.
The acceptance probability of the MH algorithm can be written in its more familiar form:
\[\alpha(\btheta_{n+1})=\min\left\{1,\frac{p(\btheta^*|\d)q(\btheta_n|\btheta^*)}{p(\btheta_n|\d) q(\btheta^*|\btheta_n)}\right\}.\]
Note that just as in rejection sampling, the normalization constant of the posterior
      probability density function is not required, 
$p(\btheta^*|\d)$ and $p(\btheta_n|\d)$
      have the same normalization constant and these would cancel out
      in the acceptance probability anyway. 
In analogy to rejection sampling, the efficiency of the MH algorithm depends crucially on the choice of the proposal density.
It should be ''close" to the posterior density but easy to sample from. In practice, multivariate normal  or Student-t proposals
are often chosen. 
The original  Metropolis algorithm is
the special case where the proposal probability density function is symmetric, i.e.,\ $q(\btheta|\btheta')=q(\btheta'|\btheta)$.
This algorithm generates a Markov chain. \citet{Hastings197097}  proved the remarkable fact that with just about any choice
of the proposal density $q$, the equilibrium distribution for the Markov
chain is the posterior $p(\btheta|\d)$, as desired. \citet{TierneyLuke1994MCfE} gives the first comprehensive account of theoretical results regarding convergence rates, laws of large numbers and central limit theorems for estimates obtained by MCMC methods.
In particular, \citet{TierneyLuke1994MCfE} showed that the sample average of an ergodic Markov chain $\frac{1}{N}\sum_{n=1}^N \theta^{(n)}$ has asymptotic variance that is by a factor of 
\[\tau=1+2\sum_{\ell=1}^\infty \rho(\ell)\] 
larger than if one would average over independent samples. This factor is the so called {\em integrated autocorrelation time}
and depends on all lag-$\ell$ autocorrelations $\rho(\ell)$ of the MCMC sample and is used as a measure of the efficiency of a MCMC scheme.
A valid estimate of $\tau$ can be obtained from
CODA \citep{Plummer}, which computes a time series estimate of $\tau$ based on the spectral density estimate at zero \cite{Geweke1992}.

A special case of the MH algorithm that has found numerous useful applications in many disciplines is the Gibbs sampler.  Very flexible implementations are available in the public domain Bayesian software packages BUGS and JAGS, described in detail in \cite{LunnDavid2013TBb} and
\cite{DepaoliSarah2016JAGS}, respectively. The proposal density of the Gibbs sampler is the product of full conditional densities and it is straightforward to show that the acceptance ratio in the MH algorithm with this choice of proposal density is one, so every move is accepted.
Each iteration of the Gibbs sampler moves in a cycle through each component of $\btheta$.
In cycle $n$, the Gibbs sampler generates a new value $\theta_j^{(n)}$ for the full conditional distribution 
\[p(\theta_j|\d,\theta_1^{(n)},\ldots,\theta_{j-1}^{(n)}, \theta_{j+1}^{(n-1)},\ldots,\theta_p^{(n-1)}).\]
For many Bayesian hierarchical models, these full conditionals are conjugate distributions and can be sampled directly. 
For many nonlinear problems, as for instance in \citet{PhysRevD.58.082001}, a MH algorithm is needed to sample from the full conditionals.
The main disadvantage of the Gibbs sampler is that it can converge very slowly to the equilibrium distribution if the posterior correlations between parameters are large. Reparametrizations that reduce the posterior correlation, blocking or the introduction of auxiliary variables can improve mixing \cite{GelmanAndrew2014}.

For any application of MCMC methods, it is important to check that the algorithm has converged towards the joint posterior distribution.
Different tests for diagnosing convergence have been established, see e.g. \citet{CowlesMaryKathryn1996MCMC} and implemented in the R-package {\tt CODA} \citep{Plummer}.

Various techniques have been developed to improve mixing, respectively accelerate the convergence of the Markov chain to the posterior distribution and applied to estimating gravitational wave parameters. These include {\em  parallel tempering} \cite{SwendsenR.H.1986RMCs} for moving rapidly through a multimodal target distribution and avoiding to get stuck in local maxima. In analogy to the simulated annealing algorithm, parallel tempering samples in parallel from $L$ chains with target densities $p(\btheta|\d)^{1/T_\ell}, \ell=1,\ldots,L$ for a set of temperature parameters $T_\ell$. $T_\ell=1$ reduces to the original posterior distribution, large values of $T_\ell$ produce flatter target densities, helping to avoid local maxima.
The chains can jump from one sampler to another with a certain probability. Only the samples from the chain with $T_\ell=1$ are eventually used for posterior inference. 
{\em Evolutionary Monte Carlo} (EMC), introduced by \citet{doi:10.1198/016214501753168325} works in a similar way to parallel tempering in that a population of Markov chains at different temperatures is generated. In addition, EMC includes a genetic operator in analogy to the crossover in a genetic algorithm
 which often improves the mixing of the chains. Many variants of crossover operators have been suggested, one of the most popular ones is the snooker crossover
proposed by \citet{doi:10.1198/016214501753168325} which randomly selects a parameter say $\btheta_i$ from the current set of  parallel chain values, another one
$\btheta_j$ from the remaining ones  and a new proposal is generated by Gibbs sampling along the line connecting $\btheta_i$ and $\btheta_j$. 
Similarly, with the aim of allowing to jump between local modes when sampling from multi-modal posteriors, a differential evolution proposal \citep{BraakterC.J.F2008DEMC} has proven useful in situations where linear correlations are present \citep{Veitch:2014wba}. It draws two previous samples $\btheta_a$ and $\btheta_b$ from the Markov chain and proposes a new parameter  according to 
\[ \btheta^*=\btheta +\gamma(\btheta_a - \btheta_b)\]
where $\btheta$ denotes the current parameter and $\gamma$ is drawn from $N(0,2.38/\sqrt{2p})$.
 An alternative approach is based on the intuition that the closer the proposal distribution is to the
target, the faster convergence to stationarity is achieved. 
 {\em Adaptive MCMC} techniques \cite{RobertsGarethO2009EoAM}  aim to dynamically adjust the proposal  to the target density based on information from previously sampled values as long as the amount of adaptation is diminishing. For instance, for $p$-dimensional multivariate normal proposals and targets, the most efficient scale that determines the jump size has been shown to be $\approx 2.38/\sqrt{p}$ \cite{GelmanAndrew2014}. Moreover, the proposal  covariance matrix can be sequentially updated based on the  information from the current samples as long as the diminishing adaptation condition is satisfied. Starting with a $p$-dimensional Gaussian proposal $q_n(\btheta)=N(\btheta,0.1^2/p\, \I_p)$ for $n\leq 2p$, for later steps one uses the current empirical estimate $\Sigma_n$ of the covariance of the target based on the run so far in defining the proposal
\[ q_n(\btheta)=(1-\beta)N(\btheta,2.38^2/p\,\Sigma_n) + \beta N(\btheta,0.1^2/p\, \I_p)\]
where $\beta$ is a small positive constant, e.g.\ $\beta=0.05$. Sampling occasionally from the standard Gaussian distribution that is independent of the previous samples guarantees that the algorithm does not get stuck at problematic values on $\Sigma_n$.
 Jump sizes can also be  dynamically adjusted to obtain an optimal empirical acceptance rate of about 0.44 in one dimension, decreasing to 0.23 in high dimensions ($p>5$). An alternative strategy is {\em delayed rejection} \cite{TierneyLuke1999SaMC}. The idea behind the delayed
rejection algorithm  is that persistent rejection, perhaps in particular parts of
the state space, may indicate that locally the proposal distribution is badly calibrated to the
target. Therefore, the MH algorithm is modified so that on rejection, a second attempt to
move is made with a proposal distribution that depends on the previously rejected state. Delayed rejection 
 was also generalized to the variable dimension case \cite{GreenPJ2001Drir}. {\em Hamiltonian Monte Carlo}  avoids the random walk behavior and sensitivity to highly correlated parameters by taking a series of steps informed by first-order gradient information
\cite{2011Muhd}, simulating Hamiltonian dynamics.  Open-source software STAN that implements Hamiltonian Monte Carlo in a flexible modeling language similar to BUGS and JAGS is described in \citet{GelmanAndrew2015S}. It uses the No-U-turn sampler  \cite{HoffmanMatthew2011TNSA} and makes use of an  optimizer  that iterates to find a (local) maximum of the objective function. As both the sampler and the optimizer require gradients, a reverse-mode automatic differentiation method \cite{GriewankAndreas2000Edpa} is implemented in STAN. A method that has been very popular in physics is the {\em invariant ensemble sampler} proposed by \citet{GoodmanJonathan2010Eswa} and implemented in the open-source Python-based software package {\em emcee} by \citet{Foreman_Mackey_2013}. Similar to parallel tempering, it employs the idea of running $L$ Markov chains $\{\btheta^{(\ell)}_n\}_{n=1}^N, \ell=1,\ldots,L$  in parallel, but now the proposal for the $\ell$th chain at iteration $n$ depends on the position of the remaining ''walkers" at iteration $n$ using an affine-invariant random transformation, thus reducing the dependence of run-time  on the posterior correlations, reducing the autocorrelation time and speeding up convergence.
For problems where the dimension of the parameter space is not fixed but variable, as for instance in multiple change-point problems, variable selection in regression models, and  mixture deconvolution with an unknown number of components, the MH algorithm cannot be adopted when proposing transdimensional moves between models. Therefore,
 the MH algorithm has been extended to the {\em reversible jump MCMC} algorithm \cite{Green1995711} that constructs jumps between parameter vectors of different dimensions in such a way as to preserve the detailed balance equations which guarantee ergodicity and convergence to the posterior distribution.

\subsection{Nested Sampling}
\label{subsec:NS}
Nested sampling  was proposed by \citet{skilling2006} for computing the evidence or marginal likelihood of a Bayesian model and later shown to provide samples from the posterior distribution as a by-product. 
 It is based on a simple identity for the expectation of a positive random variable $Y$ with probability distribution function $f(y)$ and cumulative distribution function $F(y)=\int_0^y f(x)dx$:
\begin{equation}
E[Y] = \int_0^\infty y f(y)dy=\int_0^\infty(1-F(y))dy
\end{equation}
This identity can be easily shown using Tonelli's theorem \cite{Schilling2005Miam}  on switching the order of integration in a double integral as follows:
\begin{eqnarray*}
\int_0^\infty(1-F(y))dy &=& \int_0^\infty \left(\int_y^\infty f(x)dx \right) dy\\
& =& \int_0^\infty \left( \int_0^x f(x)dy \right)dx \\
&=& \int_0^\infty xf(x)dx
\end{eqnarray*}
noting that the area of integration is $\{(x,y): 0<x<\infty,x<y<\infty\}=\{(x,y): 0<y<\infty, 0<x<y\}$.
Nested sampling aims to evaluate the marginal likelihood (the denominator in Bayes theorem (\ref{Bayestheorem})):
 $Z=\int L(\d|\btheta)\pi(\btheta) d\btheta$, thus
the likelihood function $L(\btheta)=L(\d|\btheta)$ takes the role of the positive random variable $Y$. Setting \[\varphi^{-1}(y)=\p_{\pi}(L(\btheta)>y)\] one gets the representation of the marginal likelihood as a one-dimensional integral:
\[Z=\int_0^\infty \varphi^{-1}(y)dy=\int_0^1 \varphi(x)dx.\]
Such an integral can be solved numerically by the Riemann sum approximation
\[\hat{Z}=\sum_{i=1}^N (x_{i-1}-x_i)\varphi_i\]
where $\varphi_i=\varphi(x_i)$ and the grid values $0<x_N<\ldots < x_1<x_0=1$ over $[0,1]$ are chosen either deterministically, e.g. $x_i=e^{-i/N}$, or randomly.
But the function $\varphi$ is usually intractable and needs to be approximated by an iterative algorithm:
\begin{itemize}
\item At iteration 1: The nested sampling algorithm starts with $N$ walkers $\btheta_i^{(1)} \sim \pi$, $i=1,\ldots,N$, drawn independently from the prior, determines
 $\btheta_1=\mbox{arg min}_{1\leq i\leq N} L(\btheta^{(1)}_i)$, i.e.,\ the walker  with smallest likelihood out of the current walkers, and
sets $\varphi_1=L(\btheta_1)$.
\item At iteration 2: it sets $\btheta^{(2)}_i=\btheta^{(1)}_i$ but  replaces walker $\btheta^{(2)}_1$ by a sample from $\pi$ subject to the constraint $L(\btheta)>\varphi_1$, determines
$\btheta_2=\mbox{arg min}_{1\leq i\leq N} L(\btheta^{(2)}_i)$, i.e.,\ the walker  with smallest likelihood out of the current walkers, and
sets $\varphi_2=L(\btheta_2)$.
\item Iterate until a desired accuracy of $\hat{Z}$ is achieved.
\end{itemize}

\citet{skilling2006} pointed out that nested sampling  not only provides an estimate of $Z$ but also a sample $\tilde{\btheta}_i$ from the posterior distribution if one assigns appropriate importance sampling weights to $\btheta_i$, i.e.,\ defines $\tilde{\btheta}_i=(x_{i-1}-x_i)\varphi_i \btheta_i$.

Nested sampling has a convergence rate of $O(N^{-1/2})$ and computational costs $O(p^3)$ where $p$ is the number of parameters \cite{ChopinN2010Pons}.
As pointed out by \citet{ChopinN2010Pons}, the practical difficulty of nested sampling is simulation from the prior subject to the inequality  constraint $L(\btheta)>L(\btheta_i)$. \citet{skilling2006} proposed to use a finite number of MCMC steps, however convergence of  nested sampling when correlations are introduced by any embedded MCMC scheme, is an open problem.
\citet{MukherjeePia2006Asaf} consider simulating points within an ellipsoid and accept if they satisfy the constraint.
\citet{ChopinN2010Pons} propose an importance sampling method to sample from the constrained prior, nested importance sampling,
employing a similar ellipsoid strategy in scenarios where the posterior mode and Hessian at the mode are available.
Further variants of constrained sampling are nested sampling with  constrained Hamiltonian Monte Carlo \cite{BetancourtMichael2011NSwC},
Galilean nested sampling \cite{SkillingJohn2012Bcib}, slice sampling \cite{AitkenStuart2013Nsfp} and {\em diffusive nested sampling}
\cite{BrewerBrendonJ2018:DNS}.
\citet{FerozF2009Mae} and \citet{FerozFarhan2013Emdw} developed so-called {\em MultiNest},  nested sampling for multimodal distributions that has become popular in cosmological and astrophysical applications. However,  {\em MultiNest}  performs inefficiently in dimensions larger than 20 as a large part of the sampling region might  fall below the likelihood threshold and gives biased results when erroneously excluding  relevant prior volume from the sampling region \cite{BuchnerJohannes2016Astf}. 
 \citet{BuchnerJohannes2016Astf} developed a test to diagnose these types of potential failures of nested sampling.

\subsection{Model Comparison}
\label{subsec:ModelComparison}
In order to test and compare competing models within a Bayesian framework, one computes the Bayes factor, the ratio of the marginal likelihoods of the two models ${\cal M}_1$ and ${\cal M}_2$ under consideration. If $\p({\cal M}_i)$ denotes their prior probabilities, $\pi_i(\btheta_i,{\cal M}_i)$  the prior probability distribution function under model ${\cal M}_i$ with parameter vector $\btheta_i\in \Theta_i$, $i=1,2$, the Bayes factor is defined as the ratio of posterior to prior odds:
\begin{eqnarray}
B_{12}&=& \frac{\p({\cal M}_1|\d)/\p({\cal M}_2|\d) }{\p({\cal M}_1)/\p({\cal M}_2)} \label{eq:BF1}\\
&=& \frac{\int_{\Theta_1} L_1(\d|\btheta_1,{\cal M}_1)\pi_1(\btheta_1,{\cal M}_1)d\btheta_1}{\int_{\Theta_2} L_2(\d|\btheta_2,{\cal M}_2)\pi_2(\btheta_2,{\cal M}_2)d\btheta_2}=\frac{Z_1}{Z_2}
\end{eqnarray}
 This resembles a likelihood ratio statistic but unlike a likelihood ratio,  the Bayes factor is obtained by integrating  and not maximizing over $\btheta_i$. Changing to $-2\mbox{ ln\;} B_{12}$ gives the same scale as the frequentist deviance and likelihood ratio statistics.

For the special case when comparing two {\em nested} models ${\cal M}_1 \subset {\cal M}_2$ (i.e.,\ $\btheta_2=(\btheta_1,\btheta_{2-1})$ with separable  \citep{PhysRevD.76.083006} prior distributions, the Bayes factor can be written as the {\em Savage-Dickie density ratio} 
\[ B_{12}=\frac{p_2(\btheta_{2-1}=0|\d)}{\pi_2(\btheta_{2-1}=0)}\]
(where $p_2(\btheta|\d)$ denotes the posterior under the prior $\pi_2$ of model 2)
which can be estimated using the MCMC output from model 2, see e.g.\ \citet{VerdinelliIsabella1995CBFU}. Its limitation is that it only holds for a specific form of the pior for the nuisance parameters under Model 1 which is completely determined by the prior under  model 2 \citep{doi:10.1080/00031305.2020.1799861}.

In general, for non-nested models and general priors, the notoriously difficult part is the computation of the evidence
\[Z=\int_{\Theta} L(\d|\btheta,{\cal M})\pi(\btheta)d\btheta.\]
This high-dimensional integral does not have an analytic solution, in general, and numerical methods are required for its calculation.

Several computational approaches for calculating the marginal likelihood exist, for  reviews see e.g.\ \citet{RobertCP2009CmfB} and \citet{GelmanAndrew2014}, and for a comparison in the context of LISA data analysis see e.g.\ \citet{PhysRevD.76.083006}. Here we give a comprehensive overview in the hope that some of the methods that have received little attention within astrophysics so far might find more uptake and provide a valuable addition to the toolkit for model comparison of gravitational wave models.
\begin{enumerate}
\item
Under certain regularity conditions, the Bayesian information criterion (BIC), defined as 
\[BIC=-2\mbox{ ln\;} L(\d|\hat{\btheta},{\cal M}) + p \mbox{ ln\;} n\]
where $\hat{\btheta}$ denotes the maximum likelihood estimate, $p$ the number of parameters, and $n$ the sample size,
provides an approximation to $-2 \mbox{ ln\;} Z$. It combines a measure of goodness-of-fit, the log-likelihood evaluated at the maximum likelihood estimate, and a measure of complexity that penalizes the number of parameters.
Alternatively,  Akaike's information criterion (AIC) is often used for practical model selection \citep{CavanaughJosephE.2019TAic}. It differs from the BIC only in the penalty term $2p$ instead of $p \mbox{ ln\;} n$.
Frequentist analysis shows that the BIC score is an asymptotically consistent model selection procedure under weak conditions.
But there is no contribution of the prior to the BIC, so it is not a Bayesian criterion.
 A Bayesian analogue, the {\em deviance information criterion} (DIC)  was developed by 
 \citet{https://doi.org/10.1111/1467-9868.00353}. The DIC is based on the deviance $D(\btheta)=-2\mbox{ ln\;} L(\d|\btheta,{\cal M})$. While BIC uses the maximum likelihood
estimate, DIC’s plug-in estimate is based on the posterior mean of the deviance $\bar{D}(\btheta)=E_{\btheta|\d}[D(\btheta)]$:
\[DIC= \bar{D}(\btheta)+p_D\]
where the measure of complexity $p_D=\bar{D}(\btheta)-D(\bar{\btheta})$ is the difference between the posterior mean of the deviance and the deviance at the posterior mean.
 It can be justified as an estimate of the posterior predictive model performance
within a decision-theoretic framework and it is asymptotically equivalent to leave-one-out
cross-validation. The DIC has been used extensively for practical model comparison
in many disciplines as it is easy to compute when a sample of the posterior distribution is available \cite{SpiegelhalterDavidJ2014Tdic, DIC}. Another main advantage is that unlike the Bayes factor, it can be used even if improper priors are specified.

\item The Laplace formula is widely used to approximate $p$-dimensional  integrals of the form
\[I_n=\int_{\Theta} \exp\left\{-g_n(\btheta)\right\}d\btheta \]
where $g_n(\btheta)$ is a smooth  real-valued function of a $p$-dimensional vector $\btheta$, e.g.\ $g_n(\btheta)= -\mbox{ ln\;} [
 L(\d|\btheta,{\cal M})\pi(\btheta)]$ in the case of the marginal likelihood; $n$ is again the sample size.
The {\em Laplace approximation} is based on a Taylor series expansion around the posterior mode $\hat{\btheta}$ and given by
\begin{equation}
I_n\approx \exp\left\{-g_n(\hat{\btheta})\right\}(2\pi)^{p/2} |\Sigma|^{1/2}
\end{equation}
where $\Sigma$ is the inverse of the Hessian of $g_n$ evaluated at $\hat{\btheta}$. This expansion is accurate to order $O(1/n)$.
It requires finding the posterior mode, which can be done using standard optimization methods such as gradient search, and computing second derivatives for which automatic differentiation tools \cite{GriewankAndreas2000Edpa}  might be useful. However,  the Laplace approximation can be inaccurate when the integrand is far from a Gaussian density, e.g. when the posterior distribution is multimodal or severely skewed.

\item The {\em harmonic mean} estimator goes back to \citet{NewtonMichaelA1994ABIw} and is straightforward to calculate when a sample $\btheta_i$, $i=1,\ldots N$ from the posterior distribution is available.  The estimator is defined as  the harmonic mean of the likelihood values:
\begin{equation}
\hat{Z}_{HM} = \left(\frac{1}{N}\sum_{i=1}^N \frac{1}{L(\d|\btheta,{\cal  M})}\right).
\end{equation}
 It is easy to calculate but can be biased and unreliable due to its potentially infinite variance, see e.g.\ the discussion  by Radford Neal of  \citet{NewtonMichaelA1994ABIw}. An improvement of the harmonic mean estimator is given by the representation
\begin{equation}
\hat{Z}_{HM}^* = \left(\frac{1}{N}\sum_{i=1}^N \frac{\phi(\btheta_i)}{L(\d|\btheta,{\cal  M}) \pi(\btheta_i,{\cal M})}\right)
\end{equation}
in \citet{MarinJean-Michel2010Ismf}
which holds for any function  $\phi$. For $\phi=\pi$, this estimator $\hat{Z}_{HM}^*$ equals the ordinary harmonic mean estimator $\hat{Z}_{HM}$. But choosing a function $\phi$ with lighter tails than the posterior, e.g.\ for $\phi$ with constrained support given by the convex hull of sampled values within
the 10\% highest posterior density region, guarantees finite variance. 

\item An approximation of the marginal likelihood is also possible via classical {\em importance sampling}  \cite{MarinJean-Michel2010Ismf}. 
A sample $\btheta_i$, $i=1,\ldots,N$ is drawn from an importance density $q(\btheta)$ and $Z$ estimated by
\begin{equation}\label{eq:importancesampling}
\hat{Z}_{IS}=\frac{1}{N} \sum_{i=1}^N \frac{ L(\d|\btheta,{\cal  M}) \pi(\btheta_i,{\cal M})}{q(\btheta_i)}
\end{equation}
where $q$ can be chosen so that the variance of the importance sampling estimate is reduced.  This implies choosing importance functions that provide good approximations to the posterior distribution, as  for instance maximum likelihood asymptotic distributions or kernel approximations based on a pilot sample. Importance sampling distributions should have fatter tails than the target density.

\item 
{\em Thermodynamic integration} or the more general {\em path sampling} \cite{GelmanAndrew1998Sncf,Neal:2001,Xie:Lewis:Fan:Kuo:Chen:2011} make use of an auxiliary inverse temperature variable $\beta$, $0\leq \beta \leq 1$, to
define transitional distributions, namely the power posterior distributions, defined by
\begin{equation}
p_{\beta}(\btheta|\d,{\cal M}) = \frac{L(\d|\btheta,{\cal M})^{\beta}\pi(\btheta|{\cal M})}{Z_{\beta}}
\end{equation}
 providing a path from the prior ($\beta=0$) to the posterior distribution ($\beta=1$). By explicitly denoting the evidence $Z_\beta$ as a function of $\beta$ by
\[
Z(\beta)=\int_{\Theta} L(\d|\btheta,{\cal M})^{\beta}\pi(\btheta|{\cal M}) \text{d}\btheta,
\]
the log marginal likelihood has the representation as the integral over the 1-dimensional parameter $\beta$ of half the mean deviance where the expectation is taken with respect to the power posterior  \cite{Maturana-RusselPatricio2019Ssaf}:
\begin{equation}\label{eq:TI}
\mbox{ ln\;}(Z)=\mbox{ ln\;}\left( \frac{Z(1)}{Z(0)}\right)=\int_0^1 E_{\beta} \left[ \mbox{ ln\;}(p(\d|\btheta,{\cal M})\right]\text{d}\beta.
\end{equation}
The samples from  parallel tempered chains that are obtained for instance in LALInference for different values of $\beta$ provide samples from the power posteriors and the
expectation $E_{\beta}\left[ \mbox{ ln\;}(L(\bm{X}|\bm{\theta},M)\right]$ can then be estimated by the sample average. The integral
in equation (\ref{eq:TI}) is approximated by numerical integration, e.g. using the trapezoidal or Simpson's rule.

\item Another method that combines importance sampling and simulated annealing is the {\em stepping-stone sampling} algorithm, widely used in phylogenetics where it was proposed by  \citet{Xie:Lewis:Fan:Kuo:Chen:2011}.
The marginal likelihood can be seen as the ratio $Z = Z_1/Z_0$, where $Z_0 = 1$ since the prior is assumed to be proper.  The direct calculation of this ratio via importance sampling is not reliable because the distributions involved in the numerator and denominator (posterior and prior, respectively) are, in general, quite different.  To solve this problem, stepping-stone sampling expands this ratio in a telescope product of $L$ ratios of normalizing constants of the transitional distributions that~is
\begin{eqnarray*}
\label{eq:SS_ratios}
Z&=& \frac{Z_{1}}{Z_{0}} = \frac{Z_{\beta_{1}}}{Z_{\beta_{0}}}\frac{Z_{\beta_{2}}}{Z_{\beta_{1}}} \dots
\frac{Z_{\beta_{L-2}}}{Z_{\beta_{L-3}}} \frac{Z_{\beta_{L-1}}}{Z_{\beta_{L-2}}} \\
&=& \prod_{\ell=1}^{L-1} \frac{Z_{\beta_{\ell}}}{Z_{\beta_{\ell-1}}} = \prod_{\ell=1}^{L-1} r_\ell,
\end{eqnarray*}
for $\beta_0 = 0 < \beta_1 < \dots < \beta_{L-2}<\beta_{L-1} =~1$, being the sequence of inverse temperatures, where $r_\ell=Z_{\beta_\ell}/Z_{\beta_{\ell-1}}$.  These individual intermittent ratios can be estimated  with higher accuracy than
$\frac{Z_1}{Z_0}$ because the  distributions in the
numerator and denominator are generally quite similar when using a reasonable number of temperatures $L$.  In this situation the importance sampling method works well. 
The stepping-stone sampling algorithm estimates each ratio $r_\ell$ by importance sampling using $p_{\beta_{\ell-1}}$ as importance sampling distribution.  This is a suitable distribution because it has heavier tails than $p_{\beta_{\ell}}$ which leads to an efficient estimate of $r_\ell$. The estimation of each ratio is based on the identity
\begin{align*}
r_\ell = \frac{Z_{\beta_{\ell}}}{Z_{\beta_{\ell-1}}} &= \int_{\Theta}\frac{L(\d|\btheta,{\cal M})^{\beta_{\ell \quad}}}{L(\d|\btheta,{\cal M})^{\beta_{\ell-1}}} \: p_{\beta_{\ell-1}}(\btheta|\d,{\cal M}) \text{d}\btheta, 
\end{align*}
which is estimated by its unbiased Monte Carlo estimator
\begin{align*}
\widehat{r}_\ell= \frac{1}{N} \sum_{i = 1}^{N}  L( \d | \btheta_{\!\beta_{\ell-1}}^{i}, M)^{\beta_{\ell}-\beta_{\ell-1}},
\end{align*}
where $\bm{\theta}_{\!\beta_{\ell-1}}^{1}, \dots, \bm{\theta}_{\!\beta_{\ell-1}}^{n}$ are drawn from $p_{\beta_{\ell-1}}$ with $\ell = 1, \dots, L-1$. 

Therefore, the stepping-stone estimate of the marginal likelihood is defined as 
\begin{align*}
\hat{Z}_{SS} &= \prod_{\ell=1}^{L-1}\frac{1}{N} \sum_{i = 1}^{N}  L( \d| \btheta_{\!\beta_{\ell-1}}^{i}, M)^{\beta_{\ell}-\beta_{\ell-1}}.
\end{align*}

\citet{Maturana-RusselPatricio2019Ssaf}  demonstrate the performance and computational costs of the stepping-stone sampling in comparison to thermodynamic integration and nested sampling in a simulation study and a case study of computing the marginal likelihood of a binary black hole merger signal applied to simulated data from the Advanced LIGO and Advanced Virgo gravitational wave detectors.

\item For the analysis of gravitational wave signals, the {\em nested sampling} algorithm is often used for model comparison. 
Nested sampling has already been  described in detail in Sec.~\ref{subsec:NS} as it provides samples from the posterior distribution at the same time as computing the evidence. \citet{Maturana-RusselPatricio2019Ssaf} describe a block bootstrap method to determine the Monte Carlo standard error of evidence estimates and apply this both to estimates based on the stepping-stone  and nested  sampling.

\item The previous model comparison methods used Monte Carlo samples from each model to estimate each evidence.
A conceptually different approach to model comparison is given by the {\em reversible jump Markov chain Monte Carlo (RJMCMC)} 
algorithm  \cite{Green1995711}. RJMCMC constructs a single Markov chain that includes moves between both models and thus requires trans-dimensional moves. The parameter space for such a chain includes the traditional parameters and an indication of the current model. It makes use of additional random variables that enable a matching of dimensions of the models with potentially different number of parameters. The dimension-matching 
ensures that the detailed balance equations hold that are necessary to prove convergence of the MH algorithm.
The  proportion of iterations the RJMCMC algorithm spends in one model is 
a consistent estimate of its posterior probability and used via equation (\ref{eq:BF1}) to estimate the Bayes factor.
For a detailed description of the RJMCMC algorithm and an early application to the LISA source confusion problem, see~\cite{PhysRevD.72.022001}.

\end{enumerate}

\subsection{Rapid Parameter Estimation}
In order to get information to multi-messenger observing partners as rapidly as possible it is important to have methods to improve the computational speed of  parameter estimation. The LALInference~\citep{Veitch:2014wba} implementations of MCMC and nested sampling provide accurate parameter estimates but at the cost of extensive computation times that range from several hours for short signals of black hole mergers to weeks for longer neutron star coalescences. Various strategies have been employed to reduce the overall computation time which is largely dominated by the time to re-evaluate the likelihood in each MCMC or nested sampling step. To this end, \citet{CornishNeil2013FFMa,Cornish:2021wxy} and \citet{ZackayBarak2018RBaF} use the heterodyning principle and relative binning, respectively, for fast likelihood evaluation. 
With heterodyning, two signals with similar frequencies are multiplied together, producing a low-frequency signal containing the useful information, and a high frequency signal that is discarded. The relative binning concerns the subsequent greatly reduced number of frequency bins used in the analysis, hence greatly increasing the calculation speed.
Accurate approximative methods based on reduced-order models of gravitational waveforms  have been developed \citep{SmithRory,CanizaresPriscilla2015AGWP}.
Fast reduced-order quadrature allows to approximate posterior distributions at greatly reduced computational costs.  A review of waveform acceleration techniques based on reduced order or surrogate models that speed up parameter estimation is provided in \citet{Setyawati_2020}. \citet{Vinciguerra_2017} use the chirping behaviour of compact binary inspirals to sample sparsely for portions where the full frequency resolution is not required; an extension of this work for IMRPhenomXHM models~\citep{Garcia-Quiros:2020qpx} is presented in~\citet{Garcia-Quiros:2020qlt}.
An alternative approach for rapid parameter estimation  is based on highly-parallelizable grid-based techniques, see \citet{PhysRevD.92.023002} and \citet{Lange:2018pyp}, and has been successfully used for parameter estimation in the third observing runs~\citep{Abbott:2020tfl,Abbott:2020niy}. The corresponding package is called RIFT and described in Section \ref{subsec:RIFT}.  

The acceleration techniques above yield estimates of {\em all} intrinsic and extrinsic parameters but are still high-latency \citep{SideryT.2014Rtsl}. If the primary focus is on obtaining rapid estimates of the sky location and distance in order to alert electromagnetic observatories to enable follow-up searches for counterpart transient events, 
the measured time of arrival of a signal at different detectors can be used to triangulate the source position \cite{Cavalier:2006rz,FairhurstStephen2011Slwa,Fairhurst:2009tc}.
\citet{PhysRevD.89.042004} and \citet{Fairhurst:2017mvj} improve the timing triangulation by including phase consistency information. 
\citet{BerryChristopherP.L.2015Pefb} showed that timing triangulation can often be a poor approximation. 
Timing triangulation can provide the sky position in about a minute (or maybe less) but at the expense of accuracy. A rapid as well as accurate Bayesian sky localization algorithm known as
Bayesian triangulation and rapid localization (BAYESTAR) was developed by \citet{PhysRevD.93.024013}. It avoids expensive post-Newtonian model waveforms and MCMC iterations by simplifying the likelihood function.  BAYESTAR conditions on the intrinsic parameters and uses the maximum likelihood estimates of  the  time delay, amplitude and phase on arrival at each of the network detectors as data to construct an approximate likelihood. Numerical quadrature is used to obtain the marginal posterior distribution of the sky location and  parallelization at pixel level yields  estimates of the sky position $10^4$ times faster than LALInference, within a minute of detection, at an accuracy comparable to the full coherent MCMC-based posterior inference \cite{BerryChristopherP.L.2015Pefb}.

\subsection{Machine Learning}

To speed up parameter estimation, deep
learning approaches, particularly variational autoencoders and
convolutional neural networks, have recently been explored
\citep{GeorgeDaniel2018DLfr,GabbardHunter2019Bpeu,ShenHongyu2019DaBN,ChuaAlvin2020LBpw,Green2020,Alvares2020exploring}. Deep
learning approaches train neural networks to learn the posterior
through stochastic gradient descent to optimize a loss function. The
training samples require only sampling from the prior and the
likelihood which is fast. It also has the advantage that training can
be performed offline and the estimation of parameters from an observed
gravitational wave signal becomes almost instantaneous.  A recent summary of how machine learning might aid gravitational wave signal
searches is presented in~\citet{Cuoco:2020ogp}. These methods
are still in their infancy and are being further developed to be able
to handle the full parameter space of binary inspirals \cite{green2020complete} and longer
duration waveforms from multiple detectors. They hold great promise
for low-latency parameter estimation and a fast electromagnetic
follow-up. Machine learning approaches have also been explored to produce a fast approximation to waveform generation \citep{Khan_2021,Schmidt_2021}, and much quicker parameter estimation~\cite{Dax:2021tsq}.

\section{Noise Power Spectral Density Estimation}
\label{PSD}
It is critical for the accurate estimation of the physical parameters to also accurately estimate the noise PSD for each of the detectors.
In the previous section, the noise PSDs were assumed to be known so that the factors relating to the determinant of the PSD matrices in the Whittle likelihood approximation 
\begin{equation}\label{eq:Whittlefull}
L(\d|\btheta)\approx 
 \prod_{k=1}^K \frac{1}{\det( \pi T\S^{(k)})}\e^{-\frac{1}{T}(\tilde{\d}^{(k)}-\tilde{\h}^{(k)})^*\S^{(k)^{-1}}(\tilde{\d}^{(k)}-\tilde{\h}^{(k)})} 
\end{equation}
could be ignored, yielding the simpler likelihood in Eq.~\ref{eq:Whittle}. In practical LIGO data analyses, the noise PSD has often been estimated {\em off-source} by the Welch method \cite{1161901} which   averages  the periodograms of several time segments of  noise close to but not including the signal and with the same length as the signal segment. This averaging  reduces the variance and provides a consistent estimate of the PSD. For the purpose of parameter estimation, this Welch estimate or the median of the periodograms \cite{Veitch:2014wba} has then often been substituted for the true PSD in Eq.~\ref{eq:Whittle} and assumed known and fixed. But as demonstrated by \citet{ZackayBarak2019DGWi}, the noise PSD of LIGO data drifts slowly over timescales of dozens of seconds to minutes and failures to take this into account results in a loss of sensitivity in matched filter searches for graviational waves. Similarly, \citet{Chatziioannou2019} show the potential effects of this so-called {\em off-source} PSD estimation on parameter estimates in simulation studies. They find that the fractional change in the  width of posterior credible intervals ranges from a few percent to 25\%.
 Their work demonstrates the superiority of {\em on-source} PSD estimation, i.e.,\ using the signal segment for simultaneously estimating the PSD and waveform parameters. 
 Furthermore, assuming the noise PSD to be known does not take the full uncertainty into account and might yield too narrow credible bands for the physical parameters, as shown  in \citet{PhysRevD.102.023008}.

 In this section, we review methods for spectral density estimation.
In Sec.~\ref{subsec:noise} we first consider techniques for estimating the noise PSD for time series that consist of pure instrumental noise before combining these with parameter estimation methods for time series consisting of signal and noise in Sec.~\ref{subsec:signalplusnoise}.

\subsection{Estimation of the Power Spectral Density Estimation With No Signals Present}
\label{subsec:noise}
For ease of exposition, let  us consider a data stream in just one detector. Furthermore, we assume that the data does not contain any gravitational wave signal, i.e.,\ $d(t)=n(t), t=1,\ldots,T$. Under the usual  assumption of stationarity and Gaussianity, the Whittle likelihood takes the form
\begin{eqnarray*}
L(\n|\S)&\approx&\frac{1}{\det(\pi T \S)}\e^{-\frac{1}{T}\tilde{n}^*\S^{-1}\tilde{n}}\\ 
&=& \exp\left\{-\sum_j \left[\frac{\tilde{n}(f_j)^2}{T S(f_j)} + \log(\pi T S(f_j)) \right]\right\}
\end{eqnarray*}
A parametric method to estimate the PSD was introduced by   \citet{RoverChristian2011Mcrn} by putting conjugate inverse Gamma
distributions on the
unknown spectral densities $S(f_j)$ which yielded Student-t marginal distributions for the errors. Though fast to compute, this estimate is not consistent in that the posterior distribution of the PSD estimate will not concentrate around the true spectral density with increasing sample size. A closely related approach that also attempts to account for the uncertainty in the PSD is given by \citet{LittenbergTyson2013Ftco} and \citet{Veitch:2014wba} that treats $S(f_j)$ as fixed but  introduces an additional scale factor
$\eta_j$ for each frequency bin, i.e.,\ replacing $S(f_j)$ by $\eta_j S(f_j)$ and placing  Normal priors on $\eta_j$ with mean 1.
\citet{TalbotColm2020Gawa} derived the likelihood after marginalization over the uncertainty in the median PSD estimate and show that their analysis is robust with respect to outliers. They also investigate the impact of mean- and median-based noise PSD estimation methods on the astrophysical inference of GW151012.
\citet{Cuoco:2000gv} employed the classical parametric
spectral density estimation methods based on fitting autoregressive and autoregressive moving average models.

\begin{figure}
\centering
\includegraphics[width=0.5\textwidth]{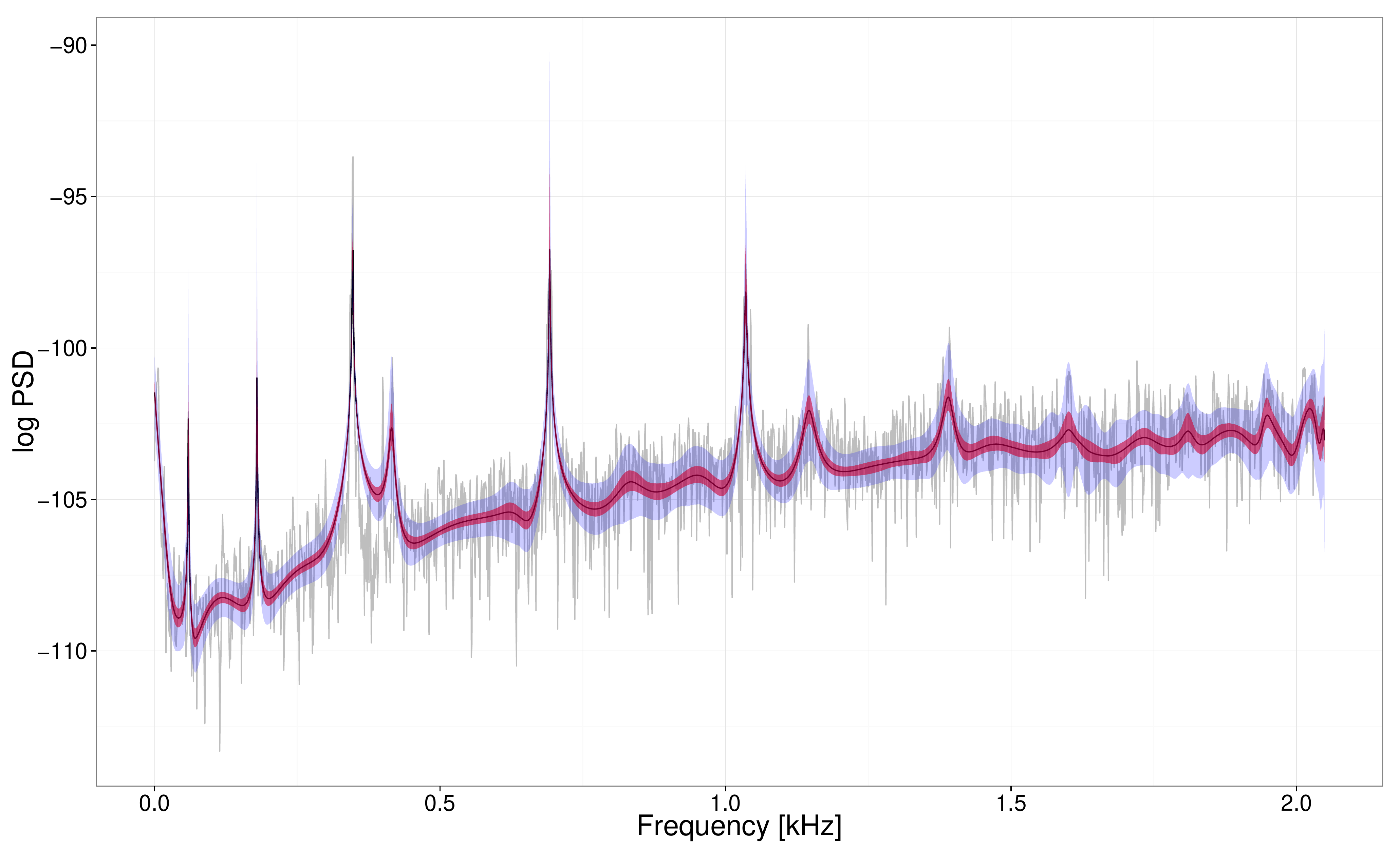}
\caption{Estimated log spectral density for a 1 s segment of Advanced LIGO S6 data.
The posterior median log spectral density estimate  using the corrected likelihood with  an AR(35) working
model (solid black thin line), pointwise 90\% credible region (shaded red thicker line), and uniform 90\%  credible
region (shaded violet band) are overlaid with the log periodogram (grey data)  \citep{KirchClaudia2019BWNC}.}
\label{fig:LIGOS6}
\end{figure}

But even though parametric models are efficient when the parametric model is correctly specified, they can give biased results under model misspecification. Therefore, considering the presence of glitches (short duration noise transients) and slow adiabatic drift in LIGO/Virgo/KAGRA noise, robust Bayesian  {\em nonparametric} approaches to PSD estimation have been developed. These  can then be combined with Bayesian estimation of the model parameters when a gravitational wave signal is present, as illustrated in Sec.~\ref{subsec:signalplusnoise}. 
The BayesLine algorithm \cite{Littenberg:2014oda} uses a mixture of Lorentzians and cubic splines to model  spectral lines and the smooth part of the PSD, respectively. BayesLine is a component of the BayesWave analysis~\cite{Cornish:2014kda,Cornish:2020dwh}. This is a very flexible approximation that treats the number and location of the knots of the cubics splines and also the number, location and line width of the spectral lines as unknown parameters.
This approach then uses a RJMCMC procedure to allow for the change in dimension of the parameter space. \citet{EdwardsMatthew2019Bnsd} modified the
Bernstein-Dirichlet process
prior on the PSD specified by \citet{ChoudhuriNidhan2004BEot}. The Bernstein-Dirichlet process prior models the PSD by a mixture of Bernstein polynomials where the number of mixture components is variable and the mixture weights are induced by a Dirichlet process. To improve on the approximation properties of Bernstein polynomials, \citet{EdwardsMatthew2019Bnsd} suggested to replace these by B-splines and put a second Dirichlet process prior on the knot spacings. They demonstrate the method's ability to pick up sharp spectral peaks and lines in the data
from the LIGO S6 science run.
To speed up this algorithm, \citet{Maturana2021} use P-splines, i.e.,\ B-splines with fixed number and location of knots where the smoothness prior is derived from the difference-based penalty of penalized splines, avoiding the computationally intensive Dirichlet process prior and reversible jump type simulations. These algorithms are implemented in the R-packages {\tt bsplinePsd}~\citep{Edwards:bsplinePsd:2018} and {\tt psplinePsd}~\citep{psplinepackage}.
Instead of modifying the Bernstein-Dirichlet process prior, \cite{KirchClaudia2019BWNC}  improve the Whittle likelihood approximation by making use of an autoregressive working model. They prove posterior consistency, superior small sample performance measured by integrated absolute error and  frequentist coverage probability, and achieve a better estimation of sharp peaks in the PSD as demonstrated using LIGO S6 noise and shown in Fig.~\ref{fig:LIGOS6}.  The implementation of this PSD estimation method  is available in the R package {\tt beyondWhittle}~\citep{Meier:beyondWhittle}.

\subsection{Estimation of the Power Spectral Density Estimation with Signals Present}
\label{subsec:signalplusnoise}
 \citet{PhysRevD.102.023008} quantified the effect of PSD uncertainty on waveform parameter estimates by first sampling from the PSD using BayesWave and then obtaining samples from the compact binary coalescence waveform parameters conditional on each sampled PSD, still not simultaneously sampling from the joint posterior distribution of waveform parameters and PSD.
Currently, the most often used method to simultaneously estimate the PSD and the signal waveform is BayesWave \cite{Cornish:2014kda,Cornish:2020dwh}. BayesWave combines the BayesLine PSD estimation method with simultaneously estimating a gravitational wave transient signal and glitches, both modeled as a mixture of Morlet-Gabor wavelets. While this gives estimates of the Morlet-Gabor wavelet coefficients and can thus reconstruct the signal taking the full uncertainty in the PSD estimate into account, it is not intended for estimating the astrophysical waveform parameters.

Most of the Bayesian noise PSD estimation techniques discussed in Sec.~\ref{subsec:noise} lend themselves to a simultaneous estimation of interferometer-specific PSDs  $\S^{(k)}$ and astrophysical waveform parameters $\btheta$ via the Whittle likelihood Eq.~\ref{eq:Whittlefull}. In general, this can be done by combining the respective  MCMC algorithms used for parameter and PSD estimation in a Gibbs sampling step. One iterates between
\begin{itemize}
\item conditioning on a PSD estimate and performing one iteration of the parameter estimation routine,
\item then conditioning on the just obtained new estimate of $\btheta$, calculating the residuals $\tilde{\d}^{(k)}-\tilde{\h}^{(k)}(.|\btheta)$  and performing one iteration of the PSD MCMC routine. 
\end{itemize}
This Gibbs sampling procedure has been applied by \cite{PhysRevD.92.064011} 
to  simultaneously fit rotating
core collapse supernova gravitational wave burst signals embedded in simulated
Advanced LIGO -- Advanced Virgo noise and the PSD using the Bernstein-Dirichlet process prior and  by \citet{EdwardsMatthew2020IaAN} to simultaneously estimate the parameters of a galactic binary signal and the noise PSD in simulated LISA data using the B-spline prior.  Similarly, in a blocked Gibbs sampler \citet{Chatziioannou2021} simultaneously estimates the compact binary coalescence waveform parameters and the noise PSD using BayesLine and extends this to the simultaneous estimation of instrumental glitches using sine-gaussian wavelets. Thus, in the applications in \citet{PhysRevD.92.064011},  \citet{EdwardsMatthew2020IaAN} and  \citet{Chatziioannou2021},
 the marginal posterior distributions of
the signal parameters take the full uncertainty of the PSD estimates
into account. This is in contrast to \citet{Chatziioannou2019} who used the BayesWave PSD estimate that was obtained from the same stretch of data as the signal but assumed fixed for the purpose of parameter estimation of  injected compact binary coalescence signals in real LIGO O2 noise.

\section{Parameter Estimation for Gravitational Waves from Coalescing Compact Binaries}
\label{sec:CBC}
The gravitational wave observations from LIGO and Virgo to date have all been produced by compact binary systems~\cite{Abbott:2020niy}. This is a system where there is an exellent ability to use a waveform model for its expected emitted signal that can be compared with  the observed data. Black holes and neutron stars are compact objects, and behave like point particles, hence simplifying the analysis. As we will see below, tidal effects will deform a neutron star at some level, and this will change the inspiral behavior. But to start, one can assume the validity of general relativity and derive the evolution of a binary system of point particles, with the system losing energy due to the emission of gravitational waves. General relativity is a non-linear theory, so it will ultimately be impossible to produce a closed form or exact functional description for the gravitational waves, and approximations with different levels of accuracy will be used. The methods used to derive waveforms will be summarized below. There are excellent descriptions of the physics of coalescing compact binary mergers, and what information can be extracted from their observation with gravitational waves, to be found in~\citet{Sathyaprakash:2009xs} and~\citet{Abbott:2016bqf}.

We start with the lowest level of approximation, which is essentially coupling Newtonian orbital mechanics and gravitational wave emission via linearized general relativity. This  presentation is for the dominant quadrupolar multipole moment of radiation. 
We follow the presentation of \citet{moore2013general}, and use this as a means to develop an understanding of the physical parameters before moving on to more complex waveform theories.

Consider two point masses, $m_{1}$ and $m_{2}$, orbiting around one another and separated by a distance of $D$. The total mass is $M = m_{1} + m_{2}$, while the dimensionless mass ratio is $\eta = m_{1} m_{2}/(m_{1} + m_{2})^2$. The orbital frequency of the system is $f$. An observer directly above the orbital plane of the system a distance $R$ away would observe the gravitational wave strain 
\begin{equation}
\begin{aligned}
{\mathbf h} = \frac{16 \pi^{2} G M \eta D^{2} f^{2}}{R c^{4}} [ {\mathbf h_{+}} cos(4 \pi f t + \phi_{0}) \\
+ {\mathbf h_{\times}} sin(4 \pi f t + \phi_{0}) ] ~ ,
\end{aligned}
\label{eq:h0}
\end{equation}
where the two polarization states are represented by ${\mathbf h_{+}}$ and ${\mathbf h_{\times}}$.  This is a circular polarization. $\phi_{0}$ is a phase shift depending on the initial conditions.
If one observes the binary system edge on from a distance $R$, then the gravitational waves received will be
\begin{equation}
{\mathbf h} = \frac{8 \pi^{2} G M \eta D^{2} f^{2}}{R c^{4}} {\mathbf h_{+}} cos(4 \pi f t + \phi_{0}) ~ ,
\label{eq:h90}
\end{equation}
so linearly polarized, and smaller in magnitude by a factor of 2.
In general the observer will be at an angle of $\iota$ with respect to the normal of the orbital plane. In this case the amplitude of the ${\mathbf h_{+}}$ polarization
component will be diminished by a factor of $(1 + cos^{2}\iota)/2$, while the amplitude of the ${\mathbf h_{\times}}$ polarization component will be diminished by $cos\,\iota$~\cite{Usman:2018imj}.
Note that the gravitational wave signal frequency is twice the orbital frequency, $f_{gw} = 2f$; this is the frequency of the gravitational
wave that will be detected. 
Since the system is losing energy via gravitational wave emission, the orbit will decay. This will cause the orbital
frequency to increase,
\begin{equation}
\frac{df}{dt} = \frac{192 \pi \eta}{5 c^{5}} (2 \pi f G M)^{5/3} f^{2} \propto f^{11/3} ~ .
\label{eq:fdot1}
\end{equation}
This defines the frequency evolution of the system and the gravitational wave signal. The frequency evolution is what results in the {\it chirp} like signal. As the binary system's orbit decays the frequency of the resulting gravitational wave signal
will increase, and consequently the amplitude will increase as well. 

One can then calculate the time it will take for the two masses to spiral into one another. Starting from a separation distance of $D_{0}$ the coalescence time, namely when the masses collide at $D = 0$ is
\begin{equation}
t_{c} = \frac{5 D_{0} c^{5}}{256 \eta (GM)^{3}} ~ .
\end{equation}
This simple calculation essentially assumes an orbit described by Kepler's Laws, with energy loss from the emission of gravitational waves. However, the general evolution of the system is evident. It is also now possible to see which parameters pertaining to the source can be described using parameter estimation methods on the detected gravitation wave signal.

Gravitational wave signals from compact binary coalescence have been detected by LIGO and Virgo~\cite{LIGOScientific:2018mvr}. These detectors are L-shaped, hence quadrupole detectors. A gravitational wave descending directly downward (normal incidence) on such a detector will register the maximal response for the  ${\mathbf h_{+}}$ polarization and no response for the ${\mathbf h_{\times}}$ polarization. Of course gravitational waves can come from any direction. Let us define a Cartesian coordinate system with the arms of the detector defining the  ${\mathbf x}$ and ${\mathbf y}$ directions, with ${\mathbf z}$ normal to those in a right hand sense.
The units for the sky location of an astronomical source are right ascension (RA) and declination (Dec). But in terms of the Cartesian coordinate system, we can say that the angle between the propagation direction of the gravitational wave, ${\mathbf k}$, and the ${\mathbf z}$ axis is $\theta$. The projection of ${\mathbf k}$ onto the ${\mathbf x}$-${\mathbf y}$ plane makes an angle $\phi$ with respect to the ${\mathbf x}$ axis. Finally, $\psi$ defines the angle of polarization about the ${\mathbf k}$ direction of propagation. 
A quadrupole detector will respond to the incoming gravitational wave with a detected signal of 
\begin{equation}
h = F_{+}(\theta,\phi,\psi) h_{+} + F_{\times}(\theta,\phi,\psi) h_{\times} ~ ,
\end{equation}  
where $h_{+}$ and $h_{\times}$ are the amplitudes of the two polarizations and
\begin{equation}
\label{eq:Fplus}
F_{+}(\theta,\phi,\psi) = \frac{1}{2}(1 + cos^{2}\theta) cos2\phi cos2\psi - cos\theta sin2\phi sin2\psi ~ ,
\end{equation}
and
\begin{equation}
\label{eq:Fcross}
F_{\times}(\theta,\phi,\psi) = \frac{1}{2}(1 + cos^{2}\theta) sin2\phi cos2\psi - cos\theta sin2\phi cos2\psi ~ .
\end{equation}

While it is difficult to determine the sky location of the source with just one interferometer,
with two detectors one can use the interferometer responses and delay in the arrival time of the signal to constrain the source location~\cite{Rover:2007ij,Rover:2006bb}. The addition of more detectors improves the ability to resolve a source~\cite{Aasi:2013wya,Pankow:2019oxl}. The current LIGO-Virgo network of three detectors has already succeeded in estimating the sky-location of GW170817 to 28 deg$^2$.
The long-term goal is to have a world-wide network of 5 detectors, which in addition to the two LIGO detectors and Virgo would include the Japanese KAGRA detector~\cite{Aso:2013,Akutsu:2017thy} and a third LIGO detector in India~\cite{Unnikrishnan:2013qwa}. With such a network it should be possible to constrain the location of a compact binary source to a few deg$^2$ on the sky. 

One can now see the parameters that can be estimated using the data from gravitational wave detectors. With even just one detector it is possible (with sufficient SNR) to produce estimates of the two mass parameters, $m_{1}$ and $m_{2}$, the time of coalescence $t_{c}$, and the inspiral phase $\phi_{0}$. 
With two or more detectors, estimates can be made of the sky position (RA and Dec), the distance to the source $R$, orbital plane inclination angle $\iota$, and the polarization angle $\psi$. More complicated orbits, for example including eccentricity, would consquently require additional parameters for the model.

In terms of parameter estimation, the mass parameter which can be estimated most accurately is the chirp mass, $\mathcal{M}_c$ (Eq.~\ref{eq:chirp_mass}).
This is because it is the mass term that directly influences the frequency derivative. For example, Eq.~\ref{eq:fdot1} can be rewritten as 
\begin{equation}
\frac{df}{dt} = \frac{192 \pi}{5 c^{5}} (2 \pi f G \mathcal{M}_c)^{5/3} f^{2} ~ .
\label{eq:fdot2}
\end{equation}
Hence $\mathcal{M}_c$ is the only mass term that contributes to the frequency evolution. 
The orbital frequency $f$ is certainly an important parameter for describing the physical system producing the gravitational waves, but the detectors will be directly observing the gravitational wave frequency $f_{gw} = 2f$. Hence the time evolution of the phase terms in Eq.~\ref{eq:h0} and Eq.~\ref{eq:h90} will vary like $2 \pi f_{gw} t$.

The signal model presented above has neglected to address the fact that the universe is expanding. Gravitational waves that have been emitted at a large distance will experience a redshift, and the detected gravitational wave frequency will be less than the emitted frequency by a factor of $1 + z$, where $z$ is the cosmological {\it redshift} of the signal source. With the expanding universe the gravitational wave signal will provide an estimate of the luminosity distance~\cite{moore2013general}, $D_{L}$, which we will henceforth use instead of $R$. When LIGO and Virgo detect a gravitational wave signal from a compact binary coalescence, the masses derived via parameter estimation will be the {\it detector frame} masses. The physical parameters that are needed to understand the origin of the signal are the {\it source frame} masses~\cite{1987GReGr..19.1163K}. One must divide the detector frame masses by $1 + z$ to produce those of the source frame. This requires the use of additional cosmological information. Objects in the expanding universe move away from us at velocity $v$ that is related to the distance $D$ via the Hubble constant, $H_{0}$, by 
\begin{equation}
v = H_{0}\, D ~ .
\label{eq:Hub}
\end{equation}
The coalescing binary gravitational wave signal provides an estimate for the luminosity distance to the source. Converting distance to redshift requires the assumption about the cosmology of the universe. LIGO and Virgo have been using the cosmological parameters derived from the Planck observations of the cosmic microwave background~\cite{Ade:2015xua}. The cosmological calculation converts the luminosity distance to a redshift.

Ultimately one must use general relativity to describe everything about the evolution of the binary system. While Newtonian equations of motion emerge as an approximation to general relativity, it should be no surprise that it is necessary to use the full theory of general relativity to describe the orbital mechanics and gravitational wave emission. As an example of the difference between Newtonian orbits and those from general relativity, consider a point particle in an orbit about a central mass $M$. For Newtonian mechanics, the radial acceleration is
\begin{equation}
\frac{d^{2}r}{dt^{2}} = \frac{- G M}{r^{2}} + \frac{\ell^{2}}{r^{3}} ~ ,
\label{eq:orbit_Newton}
\end{equation}
where $t$ is coordinate time, and $\ell$ is the anglular momentum per unit mass. On the other hand, the radial acceleration from general relativity is
\begin{equation}
\frac{d^{2}r}{d\tau^{2}} = \frac{- G M}{r^{2}} + \frac{\ell^{2}}{r^{3}} - \frac{3 G M \ell^{2}}{c^{2} r^{4}}~ ,
\label{eq:orbit_GR}
\end{equation}
where $\tau$ is the proper time. With Newtonian gravity it is possible to have a stable circular orbit for any radius. However the effect of the additional term in Eq.~\ref{eq:orbit_GR} creates an {\it innermost stable circular orbit}. For $r < 6 G M/c^{2}$ it is not possible to sustain a circular orbit, and the particle will fall toward $r = 0$~\cite{moore2013general}. This has important implications for predicting the gravitational wave signal for a coalescing compact binary. The derivation above considered the gravitational wave emission from the inspiral part of the signal, but an accurate description must also consider this {\it plunge} part of the signal.

The {\it no-hair theorem} says that black holes in general relativity can be completely described by three parameters: their mass, their spin (angular momentum), and their electric charge~\cite{Misner1973}. Astrophysical black holes will likely be uncharged (the surrounding plasma will quickly neutralize any net charge)~\cite{Narayan:2005ie}, and will be described by their mass and spin, namely Kerr black holes~\cite{PhysRevLett.11.237}. For a binary black hole coalescence the final part of the process, after the plunge, will be the merger of the two initial black holes to form a final black hole. By the no-hair theorem the final product of the merger must be a Kerr black hole. This means that the merger process will be the initial formation of a black hole that undergoes a {\it ringdown} to a final stable configuration. This merger-ringdown process will result in gravitational wave emission as well. In fact, the observed gravitational waves from the ringdown of a newly formed black hole can be used to test the no-hair theorem; see~\citet{Dreyer:2003bv}, \citet{Kamaretsos:2011um} and \citet{Isi:2019aib}.


The spin of the initial component masses will add six additional parameters. Essentially, these are the total spin magnitude and the spin direction (being a vector) for each initial mass. While general relativity is the correct theory of gravity, one can consider the limit when the theory is well approximated by Newtonian gravity. The gravitational field $\vec{g}(\vec{r})$ at $\vec{r}$ created by a point mass $M$ at the origin is 
\begin{equation}
\vec{g}(\vec{r}) = \frac{-G M \hat{r}}{r^2} ~ .
\end{equation}
Similarly, the electric field $\vec{E}(\vec{r})$ at $\vec{r}$ created by a point charge $Q$ at the origin is (Coulomb's Law)
\begin{equation}
\vec{E}(\vec{r}) = \frac{k_{e} Q \hat{r}}{r^2} ~ ,
\end{equation}
with $k_{e}$ the Coulomb constant. The form of the equations is similar. And while we know that a moving charge will create a magnetic field $\vec{B}$, it should not come as a surprise that a moving mass would create a so-called gravitomagnetic field $\vec{B}_{G}$~\cite{moore2013general}. The effect of gravitomagnetism would cause a spinning gyroscope orbiting the (rotating) Earth to be torqued through Lense-Thirring precession and geodetic precession. The measurement of these effects was the goal of NASA's Gravity Probe B mission~\cite{PhysRevLett.106.221101}. 

For compact binary systems we can make a comparison with atoms and spin-orbit coupling. For the binary system there will be a coupling between the gravitomagnetic field created by the orbital angular momentum, and the gravitomagnetic field created by the spins of the compact objects. This will affect the orbital dynamics, and hence the generation of the gravitational waves. The spin angular momentum, $\vec{S}_{j}$, of a black hole of mass $m_{j}$ will vary in magnitude between $0$ and a maximum allowed by general relativity of $G m_{j}^{2}/c$. One often uses the dimensionless spin vector, $\vec{\chi}_{j} = \vec{S}_{j} c/(G m_{j}^2)$, whose magnitude varies from 0 to a maximum of 1 by cosmic censorship (a proposition to avoid naked singularities)~\cite{Wald:1997wa}. 

LIGO and Virgo report their results in terms of two global effective spin parameters, in addition to the individual component spins. The {\it effective inspiral spin parameter}, $\chi_{\rm eff}$, is a mass averaged sum of the spins of the two initial masses aligned with the orbital angular momentum, as presented in~\citet{PhysRevD.82.064016}, \citet{Ajith:2009bn}  and \citet{LIGOScientific:2018mvr}. The {\it effective precession spin parameter}, $\chi_{\rm p}$, is a measure of the component masses spin parallel to the orbital plane~\cite{Hannam:2013oca,Schmidt:2014iyl}. The presence of $\chi_{\rm p}$ will change the dynamics of the orbit, creating for example, a precession of the orbital plane~\cite{PhysRevD.49.6274,PhysRevD.54.2438}. A waveform model that includes spins is more complex, but it is also representative of how we expect to observe black holes and neutron stars. The first step in complexity will be models that consider only the presence of spin in the direction of the obital angular momentum, the so-called spin-aligned models with a $\chi_{\rm eff}$. The more complete and more complex model also considers spin in the orbital plane, hence with orbital precession, and a non-zero $\chi_{\rm p}$. These models are described below in Sec.~\ref{subsec:BBH}.

\subsection{Binary Black Hole}
\label{subsec:BBH}
The gravitational waves from a binary black hole are the simplest binary system to predict, as black holes have no internal structure that must be taken into account. That said, it will still require 15 parameters to describe the gravitational waves produced by a binary black hole system in a circular (namely non-eccentric) orbit. The research into the prediction of the gravitational waveforms goes back many years. In \citet{Landau:1951} the dynamics of the orbit of a binary system, the production of gravitational waves, and the decay of a circular orbit via energy loss from gravitational wave emission are addressed. 

In the early 1960s the complexity of the binary orbit was increased. The evolution of a binary in an eccentric orbit was studied by \citet{PhysRev.131.435}, and it was shown that gravitational wave emission would cause the orbit to circularize. This then led to a subsequent study where the emission of gravitational waves from a binary system is derived via approximative solutions derived using expansions of the field equations in terms of the gravitational coupling constant and the velocity of the masses with respect to the speed of light, $v/c$~\cite{PhysRev.136.B1224}. The expansion methods have become  critically important in the evolution of gravitational waveform development. The study of \citet{PhysRev.136.B1224} presented a derivation of both the energy and the angular momentum carried away from a binary by gravitational waves; this was demonstrated for a binary system with eccentricity. 

A strong motivation for the study of binary systems was provided by \citet{1975ApJ...195L..51H} with the discovery of the binary pulsar PSR 1913+16. It was quickly realized that this system would provide an excellent opportunity to test general relativity~\cite{1974CRASM.279..971D}, and especially the existence of gravitational waves~\cite{1975ApJ...196L..63W}. The decay of the orbit of PSR 1913+16 was convincingly confirmed by  \citet{1982ApJ...253..908T}, showing that the energy loss was exactly what one would expect with general relativity and the emission of gravitational waves. Years of subsequent observations have even more convincingly supported the initial observations~\cite{taylor:1989,2010ApJ...722.1030W,0004-637X-829-1-55}.

By the 1970s the activity associated with the development of gravitational wave detectors was well underway. These detector projects included the resonant bar detectors~\cite{PhysRevLett.22.1320,PhysRevLett.24.276,PhysRevLett.38.454}, and laser interferometers~\cite{weiss:1972,PhysRevD.17.379,0022-3735-12-11-010}. Simultaneously there was a rapid development of the theoretical physics methods to more accurately describe the gravitational wave signals that these detectors would hopefully soon observe. 
A method to approximate the solutions to the non-linear Einstein equation is to derive order-by-order deviations from Newtonian gravity, the so-called post-Newtonian formalism. These approximative solutions are expansions in terms of small parameters, similar to a Taylor series.
\citet{1975ApJ...197..717E} used a {\it post-Newtonian} approximation expansion in general relativity to describe the gravitational radiation field and the energy flux from sources. 
The expansion was done in powers of the velocity of the signal source, namely powers of $v/c$; the initial derivation for gravitational waves was done to 3/2 post-Newtonian order.
This post-Newtonian method had been introduced by \citet{1965ApJ...142.1488C} to derive hydrodynamics equations in general relativity. The method was then applied to a system of two masses for circular orbits, gravitational bremsstrahlung, and head-on collisions~\cite{1976ApJ...210..764W}.
And so began the derivation of gravitational wave solutions for the inspiral phase of binary systems, to higher and higher post-Newtonian order. Soon 2.5 post-Newtonian solutions were presented in \citet{1981PhLA...87...81D} and \citet{Itoh:2001np}, and then 3.0~\cite{Blanchet_1998}. The effects from the spin of the masses must be addressed, and have been included in the post-Newtonian waveforms; these include spin-orbit and spin-spin coupling. Post-Newtonian waveforms up to order 4.0 have been derived. For an excellent and comprehensive review of post-Newtonian descriptions of gravitational wave signals see \citet{Blanchet:2013haa}.

\subsubsection{Compact Binary Parameter Estimation}
\label{subsubsec:bbhpe}
After the initial proposal by \citet{PhysRevD.58.082001} to use MCMC methods for Bayesian gravitational wave data analysis, these parameter estimation methods were then tested by \citet{PhysRevD.64.022001} with 2.5 order post-Newtonian waveforms as expressed in the frequency domain~\cite{PhysRevD.62.082001}. The MCMC was implemented with a Gibbs sampler~\cite{Gilks1996}. This initial demonstration considered the data from one detector, and five parameters: the two masses, the time and phase at coalescence, and a signal amplitude. The demonstration was also concerned with displaying rapid and efficient
sampling from a complex posterior distribution for the steps of the Gibbs  sampler; this was accomplished with a Metropolized type of adaptive rejection sampling~\cite{doi:10.2307/2347565,doi:10.2307/2986138}.   

These MCMC methods were then expanded by \citet{Christensen2004317} with the use of a Metropolis-Hastings MCMC~\cite{Metropolis19531087,Hastings197097}. This was again a demonstration for a single detector and a signal described by five parameters~\cite{PhysRevD.62.082001}, also with the 2.5 post-Newtonian waveforms of \citet{PhysRevD.62.082001}. The code was developed to be compatible with LIGO data and data analysis.
Starting with initial LIGO's second science run, S2, these MCMC parameter estimation methods were applied to signals injected into the data~\cite{PhysRevD.72.082001}. Further improvements to this initial method also involved the development of importance resampling~\cite{GelmanAndrew2014,doi:10.1080/00031305.1992.10475856} to improve the convergence of the Markov chains, and informative priors to better correspond to the expected conditions expected with  LIGO and Virgo observations~\cite{Rover:2006ni}. 

The application of Bayesian parameter estimation for gravitational wave signals observed by a network of detectors was demonstrated in \citet{Rover:2006bb}. This was a presentation of a MCMC routine for  coherent  parameter  estimation for binary compact objects with multiple interferometric  gravitational  wave  detectors, for example the three detectors in the LIGO-Virgo network. This increased the number of parameters to nine, including the distance and sky position; spin was neglected in this initial study. As described in Sec.~\ref{Methods} (see Eq.~\ref{eq:Whittle}), for interferometer $k$, the likelihood function takes the form
%
\begin{eqnarray*} \label{eqn:detectorlikelihoodInt}%
L^{(k)}(\btheta)&\propto&\\
&\exp&\left\{-\frac{1}{T}(\tilde{\d}^{(k)}-\tilde{\h}^{(k)}(\btheta))^*\S^{(k)^{-1}}(\tilde{\d}^{(k)}-\tilde{\h}^{(k)}(\btheta)) \right\} 
\end{eqnarray*}
%
where the
detector data is~$\d^{(k)}$, which is the sum of the detector noise and the gravitational wave signal~$\h^{(k)}(\btheta)$, 
which is described by the parameters~$\btheta$.
The Fourier transforms of the data
$\tilde{\d}^{(k)}$ and $\tilde{\h}^{(k)}(\btheta)$, 
appear in the likelihood along with the noise PSD $\S^{(k)}$. The parameters $\btheta^{(k)}$ that describe the signal in detector $k$ are:  masses $m_1$ and $m_2$; luminosity distance $D_L$; inclination angle $\iota$; coalescence phase $\phi_0$; polarization $\psi$; and detector-specific parameters:
local coalescence time $t_c^{(k)}$;
local sky altitude $\vartheta^{(k)}$;
and local sky azimuth $\varphi^{(k)}$. 
Making the assumption that the noise in the detectors is independent, the product of the individual likelihoods gives
the network likelihood, Eq.~\ref{eq:Whittle}.
The study of \citet{Rover:2006bb} implemented a prior for the gravitational signal amplitude that considerd the combined effects of the distance and masses, which was then multiplied by the prior for the inclination and sky position. 
For this study the MCMC  was implemented as a Metropolis-sampler~\cite{Gilks1996,GelmanAndrew2014}.
With the large number of parameters simulated annealing was used to better sample the parameter space. This study demonstrated effects that are important for present LIGO-Virgo observations, especially GW170817~\cite{TheLIGOScientific:2017qsa}. There is a correlation between the estimation of the luminosity distance $d_{L}$ and the inclination angle $\iota$. Also, if the signal strength (SNR) in a particular detector is small it still adds information and can improve the sky position localization. Again, this was what was observed with GW170817 where the signal was weak in Virgo, yet the inclusion of the Virgo data in the parameter estimation improved the sky position estimation and was crucial for identifying the source via electromagnetic observations~\cite{2017ApJ...848L..12A}. 

The multi-detector, coherent MCMC for gravitational wave signals from compact binary systems was further expanded and presented in \citet{Rover:2007ij}. Here a more advanced waveform model was used that was 3.5 post-Newtonian order in phase and 2.5 order in amplitude~\cite{PhysRevD.65.061501,Arun_2004}.
This study used sophisticated MCMC methods for the Metropolis sampler, such as evolutionary MCMC~\cite{doi:10.1198/016214501753168325}, a generalization motivated by genetic algorithms~\cite{10.5555/534133}. These methods yielded an enhancement over parallel tempering, with faster and more dependable convergence toward the correct posterior distribution. 

The natural evolution of the parameter estimation for compact binary coalescence produced gravitational wave signals was the inclusion of spin for the two initial component masses. In the work of \citet{vanderSluys:2007st,vanderSluys:2008qx} the first step in this process was to consider a neutron star ($1.4 M_{\odot}$) -- black hole ($10 M_{\odot}$) binary; the black hole is assumed to be spinning while the spin of the neutron star can be ignored. This brings the total number of parameters considered to 12. Such a parameter estimation routine is demanding due to the 
sizable parameter number and large correlations between them; this creates a parameter space with much structure.
The waveform used was 1.5 post-Newtonian in phase, but Newtonian order in amplitude; in this way the signal could be computed analytically, which was convenient for this initial demonstration with large parameter number.
The spin of the black hole can couple with the orbital angular momentum (an analogy to atomic spin-orbit coupling); this will lead to modulation of the signal amplitude and phase. In addition, such coupling will create a precession of the orbital plane~\cite{PhysRevD.49.6274}. However, the modulations in the observed signal can actually benefit the parameter estimation and sometimes remove parameter degeneracies. While the MCMC was running,
the correlations between the Markov chains for the different parameters were measured. The covariance matrix was calculated. The subsequent samples for the Markov chains were then drawn from the new multivariate normal distribution. This increased the efficiency of the MCMC. Parallel tempering was also employed.
The study showed that for reasonable SNRs, the sky position of the source could be estimated to tens of square degrees, thereby displaying the potential for gravitational wave multimessenger astronomy. The subsequent study of \citet{Raymond:2008im} demonstrated that the use of high post-Newtonian templates with spin provided improved sky position estimation. The application of the 12 parameter MCMC was successfully demonstrated for hardware-injected signals in the LIGO S5 data, thereby showing the efficacy of the method in the presence of real interferometer detector noise~\cite{van_der_Sluys_2009}.

The parameter estimation for a compact binary coalescence where the spin of both component masses is considered is presented in \citet{Raymond:2009cv}. For this the total number of parameters is 15. Parallel tempering is again used for the MCMC~\cite{Rover:2006bb}. For the study the  waveform  was  3.5 post-Newtonian order in phase, Newtonian  amplitude, and spin related terms that are 2.5 post-Newtonian order in phase; these are waveforms in the adiabatic circular orbit inspiral regime that are driven by radiation reaction~\cite{PhysRevD.67.104025}. Signals were injected into synthetic Gaussian noise and also actual LIGO data. These injected signals had an SNR of 11.3, and with that accurate parameter estimation was achieved. This study also made the comparison between injected signals and short duration noise transients, glitches. Bayes factors were calculated for the comparison between the model where a signal is present, as compared to just Gaussian noise. A harmonic mean method was used for the calculation of the evidence, and hence the Bayes factor~\cite{NewtonMichaelA1994ABIw}, see also \ref{subsec:ModelComparison}.

The possibilities for Bayesian parameter estimation for compact binary coalescence produced gravitational wave signals was greatly expanded with the important paper of \citet{Veitch:2009hd} which introduced nested sampling as a possible method for use. Nested sampling, described in Sec.~\ref{Methods}, offers a parameter estimation method that is potentially faster than MCMC. This is important for signals with a large number of parameters, such as the 15 parameters associated from compact binary coalescence. In addition to the generation of posterior probabability functions for the parameters, the study also addresses model selection and the generation of Bayes factors. Nested sampling and MCMC both became integral for the LIGO-Virgo parameter estimation software LALInference~\citep{Veitch:2014wba}, which is described in Sec.~\ref{sec:LALInference}. 

\subsubsection{Compact Binary Waveform Modeling}
\label{subsubsec:waveforms}
As demonstrated by the observations of numerous binary black hole produced gravitational wave signals by LIGO and Virgo~\citep{LIGOScientific:2018mvr,Abbott:2020niy}, it is critical to have a model for more than just the inspiral part of the signal. The decades of work on the development of waveform models that include the merger and ringdown of the newly formed black hole have become essential in this era where gravitational waves from binary black holes are commonly observed and accurate parameter estimation is required~\cite{Mandel:2014tca}. The limitations of the post Newtonian expansion, in powers of $v/c$, become evident in the final few orbits of a binary black hole as the relative velocity approaches the speed of light. This is well illustrated in the observation of GW150914, where the two black holes reached a relative velocity of $v/c \sim 0.6$ at merger~\citep{Abbott:2016blz}. In order to address the limitations of the post-Newtonian approach, the {\it Effective One-Body} (EOB) formalism was developed~\citep{Buonanno:1998gg,Buonanno:2000ef,Damour:2000we}.
The post-Newtonian expansion in powers of $v/c$ is replaced by a non-polynomial function of $v/c$ that addresses the non-perturbative characteristics of the true signal~\citep{Damour-SP}.
The EOB method was quickly expanded to include spin for the initial component black hole in \citet{Damour:2001tu}.

A critically important point in the development of gravitational wave signal waveforms was the numerical calculation of the final orbit, the plunge and then the ringdown from the newly formed black hole system; this achievement was presented in \citet{Pretorius:2005gq}. This breakthrough motivated many groups to develop numerous different numerical solutions to general relativity and the prediction of gravitational wave signals from binary black hole systems. See for example, the works of \citet{Baker:2005vv}, \citet{Lindblom:2005qh}, and \citet{Campanelli:2005dd}. A subsequent important development was the ability to combine analytical relativity with numerical relativity, as demonstrated by the work of \citet{Buonanno:2006ui}.
With this work numerical relativity completed EOB waveforms were constructed.
This has subsequently led to the use of templates constructed via these methods for use by LIGO and Virgo, see~\citet{Taracchini:2013rva} and \citet{Ossokine:2020kjp}.

The Numerical INjection Analysis (NINJA) project was initiated in order to study the sensitivity of gravitational wave analysis pipelines to numerical simulations of waveforms. The Bayesian parameter estimation methods were used to help verify the validity of numerical relativity waveforms as part of the NINJA project~\citep{Cadonati_2009,Aylott:2009ya,Aasi:2014tra}.

A frequency domain phenomenological model for the generation of gravitational wave signals has also been developed and used by LIGO and Virgo~\cite{Khan:2015jqa,Pratten:2020ceb}. The phenomenological models use a
combination of analytic post-Newtonian and EOB methods to describe the inspiral, merger and ringdown. Numerical relativity simulations are then used
to calibrate EOB coefficients that could not be defined otherwise, and also free parameters associated with the merger and ringdown.
These models have considered a spin-aligned configuration where the black holes spins are parallel to the orbital angular momentum. However, they have also been extended to account for in-plane spin and orbital precession by twisting up non-precessing waveforms to simulate the precessional motion via the addition of one parameter, the effective precession spin parameter $\chi_{\rm p}$, as described in \citep{Hannam:2013oca,Schmidt:2014iyl}.

For the first detected gravitational wave signal GW150914~\cite{Abbott:2016blz} the waveforms that were used for the parameter estimation
included the numerical relativity completed EOB waveforms~\cite{Taracchini:2013rva,Purrer:2014fza}, as well as the phenomenological model with aligned spin~\citep{Khan:2015jqa,Pratten:2020ceb}. 
A phenomenological model that included spin precession was also used~\citep{Hannam:2013oca}. The analysis was redone in order to correct a transformation of coordinates relating the nonprecessing and precessing systems; this analysis used the spin aligned EOB waveforms and the precessing phenomenological waveforms, and produced parameter estimates equivalent to the initial study~\citep{TheLIGOScientific:2016wfe}. Finally, a revised and improved analysis using fully precessing waveforms (EOB and phenomenological) was presented in \citet{Abbott:2016izl}.

Even from his orginal calculations, Einstein knew that gravitational waves would be at least quadrupolar~\citep{Einstein:1918btx}. But just like electromagnetic radiation (where the lowest multipole radiation is dipole), it is possible to have higher order multipoles.
For gravitational waves, extensions past quadrupolar multipole moments are what are referred to as higher-multipole emission. 
Note that the presence of higher order modes and precession does not actually increase the number of physical parameters to estimate; it is just a more accurate waveform. 
Higher multipoles will be detectable in gravitational wave signals observed from compact binary systems with large inclination
angles, hence the absence of the higher multipoles would allow for the exclusion of those large angles.
This  can  break  the orbital plane inclination angle -- distance degeneracy, which will then improve the constraints on the inferred source inclination and luminosity distance~\cite{Chatziioannou:2019dsz}.

The recent O3 observations by Advanced LIGO and Advanced Virgo have displayed more complicated signals that have shown the importance of having more complex waveforms that encompass higher multipoles and orbital precession. The binary systems responsible for the gravitational wave signals GW190412~\citep{LIGOScientific:2020stg} and GW190814~\citep{Abbott:2020khf} had larger mass ratios, and the presence of higher multipoles was confirmed. For GW190412~\citep{LIGOScientific:2020stg} the system was a binary black hole with initial component masses estimated to be $m_{1} = 29.7^{+5.0}_{-5.3} M_{\odot}$ and $m_{2} = 8.4^{+1.8}_{-1.0} M_{\odot}$, for a mass ratio of $q = 0.28^{+0.13}_{-0.06}$. The inclination angle (folded to $[0,\pi/2]$) was estimated to be $\theta_{JN} = 0.73^{+0.34}_{-0.24}$. Between the large mass ratio and the relatively large inclination angle it should not be a surprise that there was evidence for higher order modes, as given by a log$_{10}$ Bayes factor $\geq 3$. The estimate for the orbital precession was $\chi_{p} = 0.30^{+0.19}_{-0.15}$, not providing strong evidence.
Numerous waveforms were used that incorporated higher multipole modes and orbital precession. These include EOB waveforms which are
created  by  doing  an  analytical inspiral-merger-ringdown description that are based on post-Newtonian, black-hole perturbation theory, numerical-relativity results; see \citet{Pan:2013rra}, \citet{Babak:2016tgq} and \citet{Ossokine:2020kjp}. There were also the phenomenological waveforms which included higher multipoles with no precession~\cite{London:2017bcn}, or with precession~\citep{PhysRevD.100.024059,PhysRevD.101.024056}; further advances have been implemented 
with the IMRPhenomXHM models~\citep{Garcia-Quiros:2020qpx,Garcia-Quiros:2020qlt}.
In addition, a numerical relativity surrogate model with higher multipole modes, but only with spin aligned with the angular momentum~\citep{PhysRevD.99.064045}, was used to verify the parameter estimation results for GW190412~\citep{LIGOScientific:2020stg}. A similar analysis, with comparable results, was conducted for GW190814~\citep{Abbott:2020khf}. The binary black hole system that produced this signal also had a large mass ratio: $m_{1} = 23.2^{+1.1}_{-1.0} M_{\odot}$ and $m_{2} = 2.59^{+0.08}_{-0.09} M_{\odot}$, and $q = 0.112^{+0.008}_{-0.009}$. There was also a relatively large inclination angle of $\theta_{JN} = 0.8^{+0.3}_{-0.2}$. See Section~\ref{subsubsec:DetGW190814} for more discussion about these events.

Finally, the very massive (total mass of $150 M_{\odot}$) binary black hole merger signal, GW190521~\citep{Abbott:2020tfl,Abbott:2020mjq}, shows some evidence of orbital precession. As such it was necessary to have waveforms that accounted for this effect. And while GW190521 did not show evidence for the presence of higher order modes, it is important to also have these effects in the waveforms too.  
For the analysis of GW190521, the LIGO and Virgo used 
the numerical relativity surrogate model NRSur7dq4~\cite{Varma:2019csw}, the effective-one-body model SEOBNRv4PHM~\cite{Ossokine:2020kjp,Babak:2016tgq} and the phenomenological model IMRPhenomPv3HM \cite{PhysRevD.101.024056}. See Section~\ref{subsubsec:DetGW190521} for more discussion about this event.

\subsection{Binary Neutron Star}
\label{subsec:BNS}
The gravitational waves from a neutron star binary are different to those of a binary black hole. The potential of inducing tidal deformations in the neutron stars must be included; see the discussion in Sec.~\ref{subsubsec:DetGW170817} and Eq.~\ref{Eq:Love}. This was the case for the analysis of GW170817~\cite{TheLIGOScientific:2017qsa}.
As the orbital frequency increases, the neutron star tidal effects begin to affect the phase and become significant for orbital frequencies above about 300 Hz; these effects could be observable in the gravitational wave signal~\cite{PhysRevD.81.123016}.
These tidal deformations produce a mass-quadrupole moment that will advanced the coalescence~\cite{PhysRevD.77.021502}. 
The initial analysis of GW170817 used waveforms incorporating
the effects of spin aligned with the orbital angular momentum~\cite{Boh__2013}, spin-spin interactions~\cite{Boh__2015,PhysRevD.93.084054}, and tidal interactions~\cite{PhysRevD.83.084051,Dietrich:2018uni,Nagar:2018zoe}. Subsequent analysis of GW170817 used waveforms that incorporated different theoretical  predictions for the equation of state of the neutron stars~\cite{LIGOScientific:2019eut}, and also potential coupling between $p$ and $g$ modes from within neutron stars~\cite{Weinberg:2018icl}. See also \citet{LIGOScientific:2018hze} for more details on GW170817.

\subsection{Neutron Star - Black Hole Binary}
\label{subsec:NSBH}
The gravitational waves from a neutron star - black hole binary will encode the large disparity between the masses, with the neutron star mass being around $2 M_{\odot}$ or less, and the black hole mass of course being unconstrained. This will allow for the presence of higher order modes to be observable in the waveform. In addition, if the black hole mass is not too large, possibly $10 M_{\odot}$ as an upper limit, then tidal distortion of the neutron star could be observable before the merger. As such, LIGO and Virgo will use waveforms that incorporate tidal effects~\cite{Nagar:2018zoe,Matas:2020wab,Thompson:2020nei}, similar to the situation for binary neutron stars, as described in Sec.~\ref{subsec:NSBH}. Recently LIGO and Virgo announced the discovery gravitational waves coming from two neutron star - black hole mergers, GW200105 and GW200115~\cite{LIGOScientific:2021qlt}. These events will be futher discussed in Sec.~\ref{subsubsec:DetGW200105_GW200115}.

\section{Detections of Gravitational Waves by LIGO and Virgo}
\label{sec:Detections}
In this section we review the detection of gravitational waves made by LIGO and Virgo to date, and describe the use of parameter estimation methods to extract physical information. LIGO and Virgo have announced the detection of 11 gravitational wave events during O1 and O2, ten from binary black hole mergers and one from a binary neutron star merger~\cite{LIGOScientific:2018mvr}. These events have provided numerous opportunities to use parameter estimation methods to extract physics information, test general relativity, and predict the expansion of the universe and the population of compact binary systems. LIGO and Virgo have recently announced another 39 gravitational wave events from compact binary mergers in catalog GWTC-2 \citep{Abbott:2020niy}. GWTC-2 includes gavitational-wave events from the first half of O3, namely the events from O3a. The observation of two neutron star - black hole coalescenses by LIGO and Virgo have recently been announce~\cite{LIGOScientific:2021qlt}.

\subsection{Interferometer Calibration}
\label{subsec:calibration}
Before starting with a detected signal one must first take into account the calibration of the gravitational wave detectors, and the associated uncertainty on the calibration parameters. The calibration of Advanced LIGO is explained in detail in \citet{Cahillane:2017,Abbott:2016jsd,Sun:2020wke}, and for Advanced Virgo in \citet{Acernese:2018bfl}. These calibration uncertainties will ultimately affect the parameter estimation routines that are attempting to extract the physical parameters associated with the detected gravitational wave signals. Clearly the gravitational wave detectors like LIGO and Virgo are complicated instruments, and hence the calibration of their sensitivity to gravitational waves is a necessary but difficult procedure. 

Interestingly MCMC methods are used to conduct the statistical analysis on the LIGO interferometer response functions. The detector parameters used for constructing a strain signal from the phase changes in the interferometer light and the servo-loop signals controlling the interferometer performance are estimated with an MCMC. From this analysis the ultimate uncertainties in the calibration are extracted~\cite{Cahillane:2017}.

A clear and brief presentation of errors in the calibration of gravitational wave detectors and how they will affect parameter estimation can be found in \citet{SplineCalMarg-T1400682}. Following their explanation, a gravitational wave of amplitude $\tilde{h}(f)$ (as expressed in the frequency domain) arrives at the detector. The data for the recorded signal, $\tilde{d}(f)$, is
\begin{equation}
\label{eq:FFT_signal}
\tilde{d}(f) = \tilde{h}_{obs}(f) + \tilde{n}(f) ~ ,
\end{equation}
where $\tilde{n}(f)$ is the noise in the detector, while $\tilde{h}_{obs}(f)$ is the apparent gravitational wave signal observed by the detector. There are always unavoidable uncertainties in the calibration of the detectors. A way of expressing this is through the frequency dependent uncertainty in the calibration of the magnitude of the gravitational wave signal, $\delta A(f)$, and the phase uncertainty, $\delta \phi(f)$. These uncertainties are frequency dependent, but the assumption is that they are also continuous as a function of frequency. One can then express the observered gravitational wave signal with respect to the {\it real} gravitational wave impinging on the detector as
\begin{equation}
\tilde{h}_{obs}(f) = \tilde{h}(f) \left(1 +  \delta A(f)\right) e^{i \delta \phi(f)} ~ .
\end{equation}
There will be different calibration uncertainties for the different detectors. The calibration uncertainties are not large; for example, for the time of the GW150914 detection, LIGO reported calibration uncertainties of less than 10\% in magnitude and 10$^{o}$ in phase for the frequency band 20 Hz to 1 kHz~\cite{Abbott:2016jsd}. It is also the case that when the observing run ends LIGO and Virgo re-do their calibration, which often results in diminishing the uncertainties. For the O1 and O2 gravitational wave observations reported in \citet{LIGOScientific:2018mvr} the final calibration uncertainties were 3.8\% for magnitude and 2.1$^o$ in phase for LIGO Livingston, 2.6\% for magnitude and 2.4$^o$ for LIGO Hanford, and 5.1\% for magnitude and 2.3$^o$ for Virgo.
In the study \citet{SplineCalMarg-T1400682} an approximation for the phase terms is introduced that will simplify the actual computer based computations, namely
\begin{equation}
 e^{i \delta \phi} \sim \frac{2 + i \delta \psi(f)}{2 - i \delta \psi(f)} ~ ,
\end{equation}
and the phase term $\delta \psi(f)$ is used instead. 

A spline interpolation is used to model the calibration errors, and this is the method currently used for LIGO-Virgo parameter estimation studies with LALInfernece~\cite{Veitch:2014wba}. At the nodes for the splines, $f_{i}$, are the magnitude errors, $\delta A_{i}$, and phase errors, $\delta \psi_{i}$.
The nodal points are selected to be distributed uniformly in log $f$.
There are posterior probability distribution functions generated for the calibration errors at the nodal points, and this is done as part of the overall parameter estimation calculation along with the physical parameters of the gravitational wave source.
This calibration uncertainty procedure is conducted independently for each of the gravitational wave detectors.
Calibration uncertainty is also modeled in a similar way with Bilby~\cite{Romero-Shaw:2020owr}.

With the small uncertainties it is reasonable to use a prior distribution that is Gaussion for the calibration uncertainties, namely,
\begin{equation}
p(\delta A_{i}) = N(0,\sigma_{A_i}) ~ ,
\end{equation}
and
\begin{equation}
p(\delta \psi_{i}) = N(0,\sigma_{\psi_i}) ~ ,
\end{equation}
and the $\sigma$s are the calibration uncertainties. The prior distributions for the calibration uncertainties are then used as part of the parameter estimation process, and for example, are part of the parameter estimation routines in LALInference~\cite{Veitch:2014wba}. When estimating the physical parameters of a gravitational wave source one marginalizes over these calibration uncertainties~\cite{SplineCalMarg-T1400682}. 
Note that the absolute timing accuracy of data between the detectors, 10 $\mu$s, is so small that its potential contribution to affects on parameter estimation are much less than those of the calibration uncertainty~\cite{Abbott:2016jsd}. 

The study of \citet{Payne:2020myg} applied a calibration model that was physically motivated and then used it as part of a comprehensive inference strategy for compact binary mergers. In order to make the analysis more efficient importance sampling was applied. Events from LIGO-Virgo catalog GWTC-1~\cite{LIGOScientific:2018mvr} were analyzed. It was found that the estimation of the calibration error was not the limiting factor for the estimation of the physical parameters from the gravitational wave signals. 

While the current LIGO and Virgo calibration strategies involve injecting electrical signals at appropriate parts of the control system, or using photon actuators to push the mirrors, new gravitational methods are currently under investigation.
In the studies of \citet{Estevez:2018zdr,Estevez:2020ulq} spinning masses were used to conduct a {\it Newtonian} calibration of Virgo. This initial demonstration of Newtonian calibration gave results consistent with the standard calibration results of Virgo, however more work is necessary to reduce noise with this method.
In the study of \citet{Essick_2019} the authors use the belief in the correctness of general relativity, the relative amplitude and phase measurements of the gravitational wave in multiple detectors, and external electromagnetic observations that provide constraints on the distance to the source and angle of inclination of the orbital plane through the information from the jet observation. The study was able to show, using Bayesian parameter estimation methods~\cite{Veitch:2014wba}, that with the observations of GW170817 one could calibrate the amplitude calibration of the LIGO detectors to $\pm 20\%$ and $\pm 15\%$ for the phase~\cite{Essick_2019}. 
Accurate calibration will also have important implications for measuring the Hubble constant with gravitational waves from compact binary mergers; in fact, gravitational wave sources can themselves be used to help calibrate the detectors~\citep{Schutz:2020hyz}.

\subsection{Binary Black Holes}
\label{subsec:DetBBH}
When the first detection was made, GW150914~\cite{Abbott:2016blz}, it truly was the birth of a new type of astronomy. Since the time of Galileo, every time a new type of telescope has been used new and often unexpected discoveries have been made. This was certainly the case with the first detected gravitational wave signal, GW150194. The two LIGO detectors simultaneously detected this signal at 09:50:45 UTC on September 14, 2015. Since then LIGO and Virgo have detected a further 49 gravitational wave signals from compact binaries, with the majority being binary black holes, during O1, O2 and O3a, with even more coming from the O3b (final 5 months of O3) observations. It is informative to start with how much was learned from the first event alone, especially through the use of parameter estimation methods.

\subsubsection{GW150914}
\label{subsubsec:DetBBHGW150914}
GW150914, the first detection of a gravitational waves, was made right at the beginning of the first observational run for Advanced LIGO. The signal was confidently detected with 24 for the SNR and 1 event in 203 000 years for the false alarm rate~\cite{Abbott:2016blz}. 

The first important result from parameter estimation would be the sky position estimate. By providing a possible location for the source it is possible for other observers (electromagnetic radiation, high energy neutrinos) to look for a counterpart signal. The simplest way to create a sky map is to use the difference in the arrival times of the signal in the different detectors. For GW150914 the time delay of 
$6.9^{+0.5}_{-0.4}$
ms between the Livingston and Hanford detectors produced a location patch in the sky.
This is what is typical done with the signal search pipelines. However, Bayesian parameter estimation routines take into account the nature of the signal as defined by the model, and also consider the response of the different interferometric detectors to the polarization state of the gravitational wave. 

\begin{figure}
\includegraphics[width=0.5\textwidth]{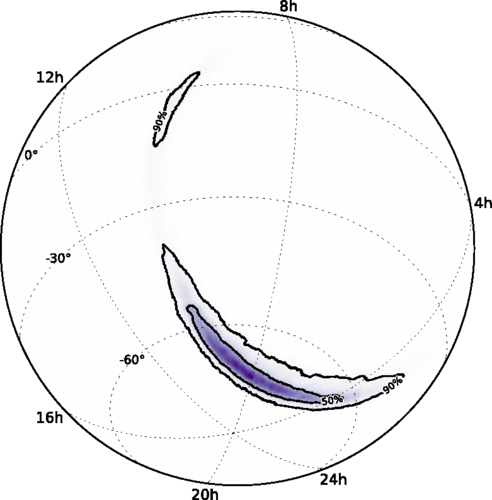}
\caption{The two-dimensional (right ascension $\alpha$ in hours, declination $\delta$ in degrees) probability distribution function for the position of the source of GW150914 on the sky. 
The 50\% and 90\% contours of the for the credible regions are given. This probability distribution function was made with LALInference~\cite{Veitch:2014wba}, which is futher described in Sec.~\ref{sec:LALInference}.
See ~\cite{TheLIGOScientific:2016wfe} for more details.}
\label{fig:GW150914sky}
\end{figure}

For GW150914 there was an initial sky position estimate released 2 days after the event that reported a 50\% credible region of $\sim 200$ deg$^2$ in size, and a 90\% region of $\sim 750$ deg$^2$~\cite{ligo18330virgo}. This initial circulation used the output of two signal search routines, Coherent WaveBurst~\cite{Klimenko:2015ypf} and Omicron+LALInference Burst~\cite{Lynch:2015yin}. Coherent WaveBurst conducts a limited maximum likelihood (using the antenna response of the detectors') estimate of the reconstructed signal on a grid of the sky, while Omicron+LALInference Burst makes the assumption of a sinusoidally modulated Gaussian signal and then uses Bayesian inference. On January 16, 2016, some 4 months after the event, LIGO and Virgo released an update to the sky position estimate~\cite{ligo18858virgo}. This new sky position estimate used LALInference~\cite{Veitch:2014wba}, which is further described in Sec.~\ref{sec:LALInference}, and BAYESTAR~\cite{PhysRevD.93.024013}, which is further described in Sec.~\ref{sec:BAYESTAR}. The BAYESTAR sky maps are made using the information from the signal search pipelines: merger times, signal amplitudes, and signal phases. The LALInference sky map was created from Bayesian MCMC and Nested Sampling analyses, and was considered to be the most accurate sky map, at the expense of the computational time. The LALInference sky map for GW150914 has a 90\% credible region of 630 deg$^{2}$.
See~\cite{Abbott:2016gcq} for a full description of the methods used to produce the sky position estimates for GW150914, and the efforts that were subsequently done by astronomical observers to try and find a counterpart.
Note that the sky localization for GW150914 was later improved from 230 deg$^2$ due to improved calibration uncertainty~\cite{TheLIGOScientific:2016pea}. This also improved the inclination estimation. See \citet{LIGOScientific:2018mvr} for the  most recent estimates of the parameters.
Fig.~\ref{fig:GW150914sky} displays the LALInference generated sky map for GW150914. 

While numerous observing systems looked for a counterpart signal, there was only one possible associated observation. Fermi/GBM reported a possible gamma ray event of energy above 50 keV of duration 1 s that occurred 0.4 s after the GW150914 merger time, with a false-alarm probability of 0.0022. The sky location determination for the gamma ray event was not well localized but was consistent with part of the gravitational wave localization from the Advanced LIGO~\cite{Connaughton:2016umz}. This observation was not confirmed by other observers.

\begin{figure}
\includegraphics[width=0.5\textwidth]{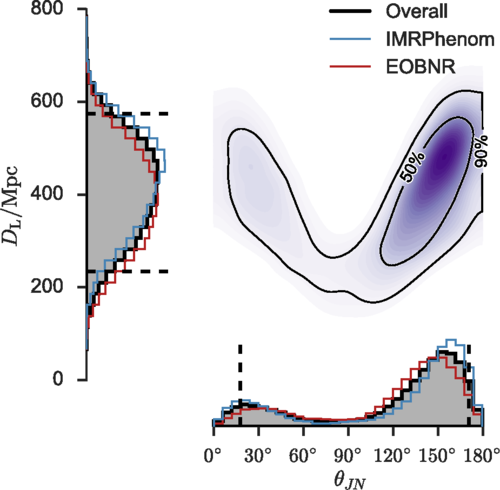}
\caption{The two-dimensional probability distribution function for the luminosity distance $D_{L}$ and the orbital plane inclination angle $\theta_{JN}$ for GW150914.
Two different model waveforms were used: the effective-one-body numerical relativity (EOBNR, in red)~\cite{Taracchini:2013rva,Mroue:2013xna}, the inspiral-merger-ringdown phenomeno-logical formalism (IMRPhenom, in blue)~\cite{Hannam:2013oca,Schmidt:2014iyl,Husa:2015iqa,Khan:2015jqa,Bohe:PPv2,Ajith:2007kx,Pan:2007nw,Ajith_2007}, and the combined total (in black).
The 50\% and 90\% credible regions are also presented.
See~\cite{TheLIGOScientific:2016wfe} for more details.}
\label{fig:GW150914dist-incl}
\end{figure}

In addition to the right ascension and declination, the distance to the source is another critical parameter in the attempt to locate a gravitational wave source. However, there is a degeneracy between the luminosity distance, and the angle of incidence for the orbital plane of the binary system. If the normal to the orbital plane points directly to the observer, then the amplitude of the gravitational wave signal will be bigger, mimicking a closer source. As the angle between the normal to the orbital plane and the line of sight of the observer departs from 0 to $\pi/2$ radians, the amplitude diminishes, mimicking a source farther away. This effect and its implications for parameter estimation is explained in~\citet{Rover:2006bb,Rover:2007ij}. The parameter estimation results for GW150914, as generated by LALInference, produce posterior distribution functions for the luminosity distance and the orbital plane inclination that can be seen in Fig.~\ref{fig:GW150914dist-incl}. The source of GW150914 was evidently at a distance of 
$410^{+160}_{-180}$ Mpc, or a redshift of 
$0.09^{+0.03}_{-0.04}$~\cite{TheLIGOScientific:2016wfe} using the Planck's estimated cosmology parameters~\cite{Ade:2015xua}.

For a binary black hole merger the information on the inclination of the orbital plane does not offer much useful information. However, for binary neutron stars, or a black hole - neutron star binary, there could be a jet produced, and the subsequent gamma ray observations would depend on that angle~\cite{Burns:2019byj}. The importance of the orbital plane inclination angle for binary neutron star mergers will be addressed in Sec.~\ref{subsubsec:DetGW170817}.

The masses for the binary system that produced GW150914 were critical in explaining the nature of the initial and final objects. From the character of the signal it appeared that the objects were point masses, and parameter estimation provided the estimate of the masses of the two intial objects to be $m_{1} = 36^{+5}_{-4} M_{\odot}$ and $m_{2} = 29^{+5}_{-4} M_{\odot}$~\cite{Abbott:2016blz,TheLIGOScientific:2016wfe}. The posterior distribution functions for the two initial masses can be seen in Fig.~\ref{fig:GW150914massPE}. 
Note that for the initial analysis of GW150914 the prior distribution for each initial component masses was uniform between 10 and 80 $M_{\odot}$~\cite{TheLIGOScientific:2016wfe}.
The amplitude of GW150914 reached a maximum at about 150 Hz, which implies that the orbital frequency of the binary system was about 75 Hz. Using Newtonian mechanics and neglecting the effects from a small redshift of $z \sim 0.1$ this implies and orbital separation of about 210 km for the orbital frequency of 75 Hz. The $\sim 30 M_{\odot}$ initial component masses are far in excess of what is possible for a neutron star. Also, a pair of stars would not sustain their spherical shapes and act like point particles, and in fact, would be far larger than this implied separation distance. The only reasonable explanation, aside from exotic and new physics, is that these two masses are black holes. This is further supported by the fact that the full general relativistic analysis of this merger is consistent with point particles meeting together at a relative velocity of $\sim 0.6 c$~\cite{Abbott:2016blz}. Further support for the black hole hypothesis also comes from the merger and ringdown part of the signal, which is further discussed below in the context of tests of general relativity in Sec.~\ref{sec:TGR}.

As explained in Sec.~\ref{sec:CBC}, the mass parameter that is most most accurately described by parameter estimation is the chirp mass. For the system that produced GW150914 this was estimated to be $\mathcal{M} = 28^{+2.0}_{-1.7} M_{\odot}$. And while the total initial mass was $M = 65^{+4.5}_{-4.0} M_{\odot}$, the final mass was estimated to be $M_{f} = 62^{+4.1}_{-3.7} M_{\odot}$, implying that about $3 M_{\odot} c^{2}$ of energy was converted into the production of gravitational waves. This peak gravitational wave luminosity was $3.6^{+0.5}_{-0.4} \times 10^{56}$ erg/s, or $200^{+30}_{-20} M_{\odot} c^{2}$/s~\cite{Abbott:2016blz,TheLIGOScientific:2016wfe}. On the Earth, for this instant GW150914 was about 10 times brighter than the full-moon!

The angular momentum components of the system are also critically important parameters to estimate, as they might provide some clues to the formation history of the black holes. As described above, the parameter estimation routines for LIGO-Virgo estimate the initial spins of the component black holes, and with the orbital angular momentum, an estimate of the spin of the remnant black hole is also estimated. GW150914 provided the first measurement of the spin of black holes, although not all the information can be extracted for a signal of finite SNR, and observable only for a limited time in the observation frequency band of the detectors.

LIGO and Virgo use two different models for the coalescing compact binary parameter estimation with spin for the initial masses. The simpler model only considers the spin of the initial masses to be aligned or anti-aligned with the orbital angular momentum vector. As such, for circular orbits this model consists of 11 parameters describing the physical system. When the spin directions for the initial masses can be in any direction it is possible to induce precession of the spins and the orbital plane. This more complex mode has 15 parameters for circular orbits. The non-precessing analysis used an effective-one-body model~\cite{Taracchini:2013rva} 
that was adjusted in consideration of numerical relativity simulation results~\cite{Mroue:2013xna}. 
These were the considerations for the initial parameter estimation analysis of GW150914~\cite{TheLIGOScientific:2016wfe}.

\begin{figure}
\includegraphics[width=0.5\textwidth]{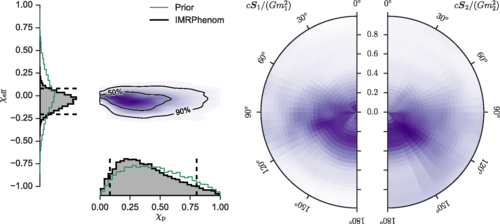}
\caption{The spin parameters estimated for GW150914. On the left are the posterior probability distribution functions for the spin parameters $\chi_{\rm eff}$ and $\chi_{\rm p}$. The 50\% and 90\% credible regions overlap the two-demensional posterior probability. On the right are the posterior probability functions for the dimensionless spins of the initial black holes with respect to the orbital angular momentum vector. There is no evidence for significant initial spin.
See~\cite{TheLIGOScientific:2016wfe} for more details.}
\label{fig:GW150914spin}
\end{figure}

\begin{figure}
\includegraphics[width=0.5\textwidth]{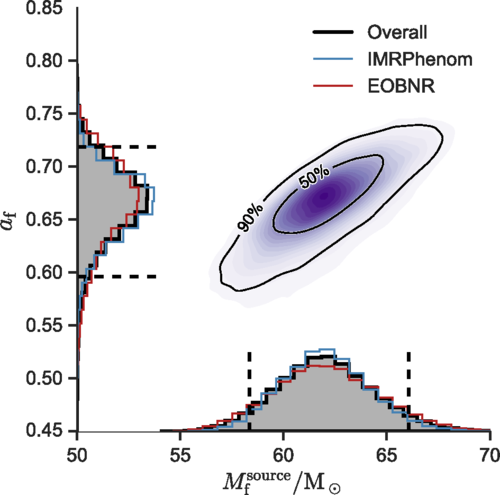}
\caption{Estimates for the spin and mass parameters in the source frame for the black hole remnant associated with GW150914. The one-dimensional probability distribution functions are calculated for the spin aligned EOBNR waveform (red), the spin precessing IMRPhenom waveform (blue), and the {\it Overall} average (black). The 90\% credible intervals are indicated by the dashed lines. The two-dimensional probability distribution function is overlaid by the 50\% and 90\% credible regions. Note that the spin label in the figure, $a_{f}$, is equivalent to what is called $\chi_{f}$ in the main text.
See~\cite{TheLIGOScientific:2016wfe} for more details.}
\label{fig:GW150914finalmassspin}
\end{figure}

The initial analysis of the Advanced LIGO GW150914 data showed that there was no appreciable spin for the two initial black holes; however, the posteriors for the spins are essentially uninformative. The mass weighted linear combination of the initial spins aligned with the orbital angular momentum was estimated to be $\chi_{\rm eff} = -0.07^{+0.16}_{-0.17}$, consistent with zero, but slightly negative. The posterior distribution for the effective in-plane spin parameter essentially corresponds to the prior, and hence a 90\% constraint of $\chi_{\rm p} < 0.71$ is set. Estimates were made on the total spins of the initial black holes: $\chi_{1} = 0.32^{+0.49}_{-0.29}$ with an upper bound of $0.69\pm 0.08$, and $\chi_{2} = 0.44^{+0.50}_{-0.40}$ with an upper bound of $0.89\pm 0.13$. Conservation of angular momentum converts the initial spins, the orbital angular momentum, and the angular momentum carried away in gravitational waves, into a spin for the remnant black hole of $\chi_{f} = 0.67^{+0.05}_{-0.07}$~\cite{TheLIGOScientific:2016wfe}. See Fig.~\ref{fig:GW150914spin} for a summary of the posterior distrubution functions for the initial spins, and Fig.~\ref{fig:GW150914finalmassspin} for the posterior distributions for the final mass and final spin of the remnant black hole.

For the initial analysis of GW150914 the prior distributions for the spins were uniform between 0 and 1 for $\chi_{1}$ and $\chi_{2}$. For the nonprecessing spin model the prior is such, with that the initial black hole spin vectors could be aligned or anti-aligned with the orbital angular momentum vector, with magnitudes between 0 and 1. For the precessing spins model the initial spin angular momentum priors are uniform across all directions, again with spin magnitudes between 0 and 1~\cite{TheLIGOScientific:2016wfe}.

Subsequent analyses of GW150914 continue to show that the spin parameters for the two initial black holes are not large, and that the inclusion of precession in the model does not affect the results. There is no indication of precession for GW150914~\cite{Abbott:2016wiq,LIGOScientific:2018mvr}.

\begin{figure}
\includegraphics[width=0.5\textwidth]{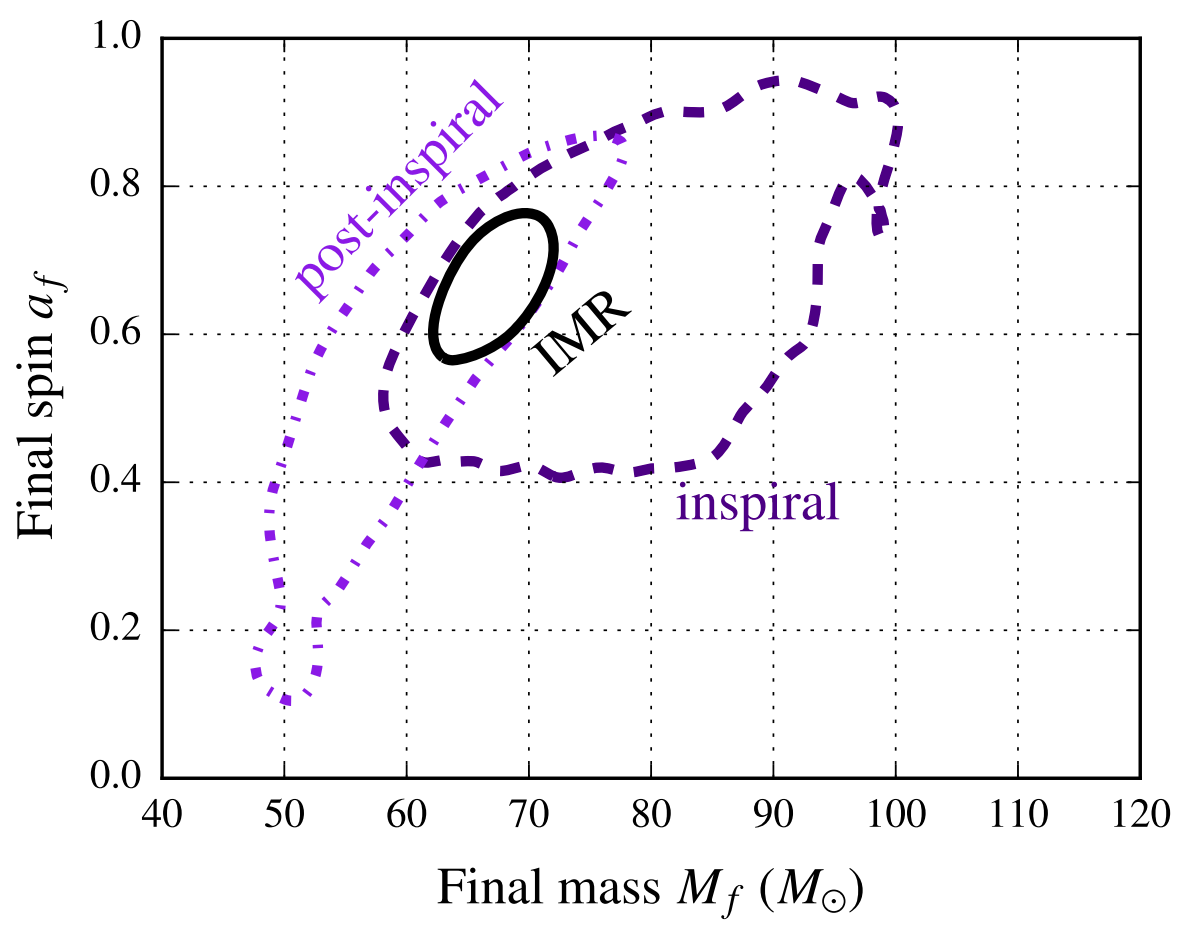}
\caption{A display of the 90\% credible regions for the final spin and final mass for the black hole remnant for GW150914 based on the parameter estimation of the signal from the low-frequency inspiral part of the signal, and the high-frequency merger-ringdown part of the signal (labled post-inspiral). The black line represents the estimate from the fill inspiral-merger-ringdown analysis. 
Note that the spin label in the figure, $a_{f}$, is equivalent to what is called $\chi_{f}$ in the main text.
See~\cite{TheLIGOScientific:2016src} for more details. Figure courtesy of LIGO Laboratory~\cite{GW150914-IMR}.}
\label{fig:GW150914-TGR-IMR}
\end{figure}

Tests of general relativity have been made with GW150914. General relativity should describe everything about the merger of a binary black hole system, from the orbital inspiral, to the merger of the two black holes, and then the ringdown of the newly formed black hole. A test has been done to subtract the most probable waveform from the gravitational wave data, using the procedure described in Sec.~\ref{subsec:TGRSRT}. After the subtraction of the waveform the residuals from the two LIGO data streams are more probable to represent Gaussian noise than residual gravitational wave energy~\cite{TheLIGOScientific:2016src,LIGOScientific:2019fpa,LIGOScientific:2019hgc}. A comparison was done for the estimation of the final black hole mass and spin, using the data from the inspiral part of the signal (low frequency), and then the data from the merger and ringdown (high frequency). Fig.~\ref{fig:GW150914-TGR-IMR} presents a display of the 90\% credible regions for the final spin and final mass for the black hole remnant for GW150914 using the low frequency (inspiral) and high frequency (merger and ringdown) parts of the signal.

\begin{figure}
\includegraphics[width=0.5\textwidth]{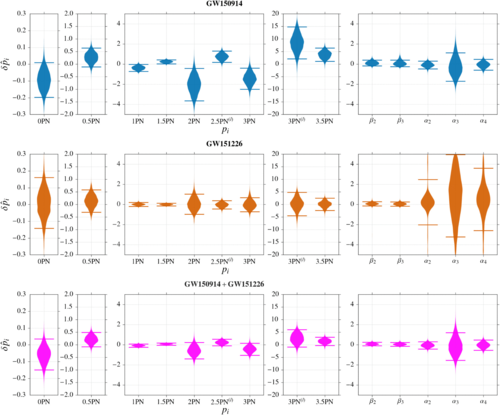}
\caption{The {\it violin plots}, or the posterior density distributions and 90\% credible intervals for the deviations from the post-Newtonian inspiral parameters $\phi_{i}$, the intermediate regime parameters $\beta_{i}$, and the merger-ringdown parameters $\alpha_{i}$. The results are given for GW150914, GW151226, and a combined result. Some parameters for GW150914 diverge slightly from zero. The parameters for GW151226 are consistent with zero, as are the combined results.
See~\cite{TheLIGOScientific:2016pea} for more details.}
\label{fig:GW150914violin}
\end{figure}

The methods used to conduct a parametrized test of general general relativity via an examination of the post-Newtonian and phenomenological numerical relativity parameters, as described in Sec.~\ref{subsec:TGRParamTests}, were also applied for GW190514. Fig.~\ref{fig:GW150914violin} diplays the 90\% credible intervals for the post-Newtonian inspiral parameters $\phi_{i}$, the intermediate regime parameters $\beta_{i}$, and the merger-ringdown parameters $\alpha_{i}$. The results are presented for the two O1 events GW150914 and GW151226, as well as the combined result; these results are consistent with general relativity for all the parameters. 
The study of~\citet{Yunes:2016jcc} addresses the implications of these observations for theoretical physics.

\subsubsection{GW170814 and GW170818}
\label{subsubsec:DetGW170814}
The first three-detector gravitational wave detection GW170814, produced by a binary black hole merger, provided the first opportunity to test whether the polarization of the gravitational waves was consistent with the predictions of general relativity~\cite{Abbott:2017oio}. GW170818 was another gravitational wave from a binary black hole merger that was detected with three detectors with sufficient SNR to allow for a polarization test~\cite{TheLIGOScientific:2016pea}. See Sec.~\ref{subsec:TGRPolarTests} for a description of the methods used.
For GW170814 the modal comparison analysis gave a Bayes factor of 30 for tensor polarization as opposed to vector polarization, and 220 for tensor versus scalar~\cite{LIGOScientific:2019fpa}. With GW170818 there was a Bayes factor of 12 for tensor polarization versus vector, and 407 for tensor versus scalar~\cite{LIGOScientific:2019fpa}.

\subsubsection{GW190412 and GW190814}
\label{subsubsec:DetGW190814}
The observations of GW190412~\citep{LIGOScientific:2020stg} and GW190814~\citep{Abbott:2020khf}, both seen in observing run O3, displayed new and important effects. All three LIGO-Virgo detectors were in observational mode for these two events. These two gravitational wave events were produced from binary black hole systems where the mass ratio for the initial constituent masses, $q = m_{1}/m_{2}$, displays a significant asymmetry. For GW190412 the parameter estimation gave $m_{1} = 30.1^{+4.6}_{-5.3} M_{\odot}$ and $m_{2} = 8.3^{+1.6}_{-0.9} M_{\odot}$, or $q = 0.28^{+0.12}_{-0.07}$. While for GW190814 the parameter estimation provided $m_{1} = 23.2^{+1.1}_{-1.0} M_{\odot}$ and $m_{2} = 2.59^{+0.08}_{-0.09} M_{\odot}$, or $q = 0.112^{+0.009}_{-0.008}$. This low mass for $m_{2}$ with GW190814 has provoked much discussion and research as it is not certain whether this object is a black hole or a neutron star; it has even been proposed that this could be a strange quark star~\cite{PhysRevLett.126.162702}.

When such significant mass ratios are present, it is possible to observe the effects of higher order multipoles, namely past the dominant quadrupole mode. 
One can describe the emitted gravitational waves in terms of a series of spin-wighted sperical harmonics~\citep{RevModPhys.52.299}. For example, the two polarizations would take the form~\citep{LIGOScientific:2020stg}
\begin{equation}
\label{eq:sph-harm}
h_{+} - i h_{\times} = \sum_{l \geq 2} \sum_{-l \leq m \leq l} \frac{h_{lm}}{D_{L}} Y_{lm}(\theta,\phi) ~ , 
\end{equation}
where the direction of propagation to the observer is defined by the angles $(\theta,\phi)$, $D_{L}$ is the luminosity distance to the source, $Y_{lm}(\theta,\phi)$ are the spherical harmonics, and $h_{lm}$ is the amplitude of each multipole. For binary systems the quadrupole $l = m = 2$ mode is expected to dominate, but when the mass ratio becomes different from one the contribution of the $l = m = 3$ mode can become important. 

The waveforms used in the analyses accounted for both orbital precession and higher order multiples. These were the effective-one-body numerical relativity waveform SEOBNRv4PHM~\citep{Babak:2016tgq,Ossokine:2020kjp}, the phenomenological IMRPhenomPv3HM~\citep{PhysRevD.100.024059,PhysRevD.101.024056}, as well as the numerical-relativity surrogate NRHybSur3dq8 model (for GW190412) which includes higher order multipoles~\citep{Varma:2019csw,PhysRevD.99.064045}..
LALInference~\citep{Veitch:2014wba} was used to generate the parameter estimation results.

\begin{figure}
\includegraphics[width=0.5\textwidth]{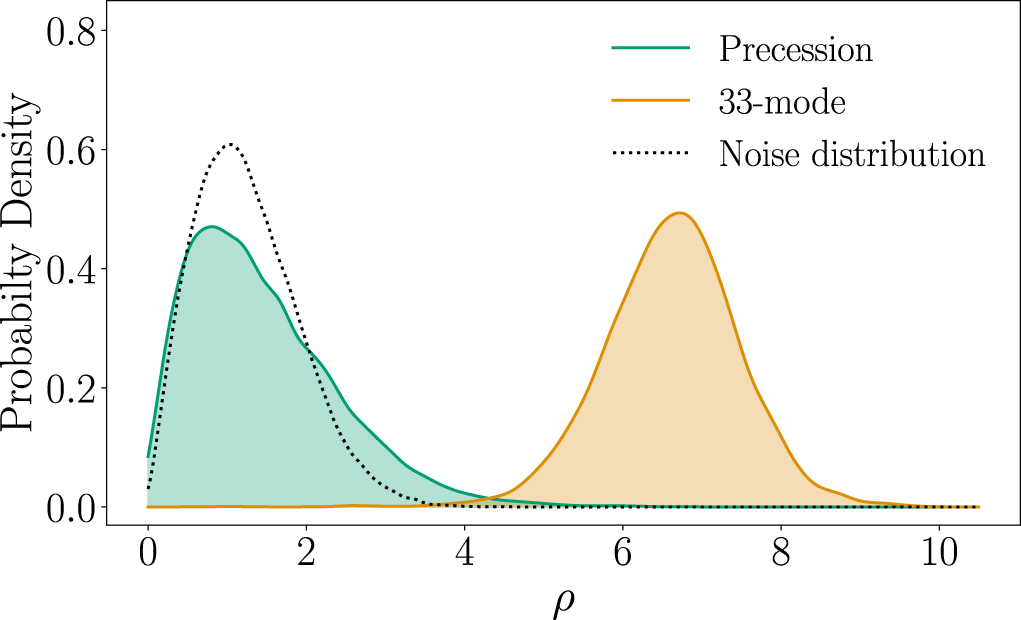}
\caption{For the observed gravitational wave signal GW190814, the posterior distribution for the $l = m = 3$ mode SNR is in orange, while that for precession is in green~\citep{Fairhurst:2019vut,Fairhurst:2019srr}.
The expected distribution for Gaussian noise is displayed with the dotted line.
See ~\citet{Abbott:2020khf} for more details.}
\label{fig:GW190814_SNR}
\end{figure}

As these two events displayed, signal models which include higher multipoles are more effective in constraining the parameters, especially the initial component masses. Consequently, the waveforms used for parameter estimation need to include these effects. The Bayes factor, $\mathcal{B}$, which compares the presence of high order multipoles to a pure quadrupole model in the waveforms, consistently had log$_{10} ~ \mathcal{B} > 3$ for various different signal models~\citep{LIGOScientific:2020stg}. 
This was also the case for GW190814, where the evidence was more significant, with log$_{10} ~ \mathcal{B} > 9.6$~\citep{Abbott:2020khf}. The presence of higher multipoles is sufficiently strong for this event that it can be determined that the $l = m = 3$ mode is the dominant higher order multipole, with log$_{10} ~ \mathcal{B} > 9.1$  
in support of the signal containg both the $l = m = 2$ and $l = m = 3$ multipole modes, as opposed to just the quadrupole $l = m = 2$ alone. Further evidence for the presence of the $l = m = 3$ mode in the GW190814 signal can be found in Fig.~\ref{fig:GW190814_SNR} where the inferred SNR of this mode is observed, whereas the inferred SNR for orbital precession is not significant~\citep{Fairhurst:2019vut,Fairhurst:2019srr}. 
For GW190412, an important result is that the effective spin parameter of the primary (most massive) black hole primary can be measured, $\chi_{\rm eff} = 0.25^{+0.08}_{-0.11}$. This is distinct from other events, and is discussed in detail in \citet{Zevin:2020gxf}.
For GW190412 there was no evidence for precession.

\subsubsection{GW190521}
\label{subsubsec:DetGW190521}
One of the most important events observed in O3 was GW190521~\citep{Abbott:2020tfl,Abbott:2020mjq}. This was the most massive binary black hole produced gravitational wave event observed to date by LIGO and Virgo. The initial binary system had black holes of masses $m_{1} = 85^{+21}_{-14} M_{\odot}$ and $m_{2} = 66^{+17}_{-18} M_{\odot}$ (90\% credible intervals). The final black hole has a mass of $142^{+28}_{-16} M_{\odot}$, making this an observation of the formation of an intermediate mass black hole~\citep{doi:10.1142/S021827181730021X,2017mbhe.confE..51K}. It is also difficult to explain the formation of the initial $85 M_{\odot}$ by stellar processes as it falls within the $\sim 64 - 135 M_{\odot}$ mass gap from (pulsational) pair-instability supernova processes~\citep{Spera:2017fyx,Farmer_2019,Woosley:2021xba}. The luminosity distance was estimated to be $D_{L} = 5.3^{+2.4}_{-2.6}$ Gpc, or a redshift of $z = 0.82^{+0.28}_{-0.34}$.

The observed GW190521 signal provides an indication for the effects of orbital precession. If the initial black holes have significant spin in the orbital plane there will be an induced precession of the orbital plane; this is a consequence of a gravitational spin-orbit coupling~\citep{PhysRevD.52.821}. There was no evidence for higher order multipoles in the signal. However, it is still important to include both orbital precession and higher order modes in the waveforms because this can help to break the inclination angle -- distance degeneracy and produce better parameter estimates~\citep{Chatziioannou:2019dsz}. As such, for the study of GW190521 the numerical relativity surrogate model NRSur7dq4~\cite{Varma:2019csw} was used in the LIGO-Virgo discovery presentation in \citet{Abbott:2020tfl}. Two other waveforms, namely the effective-one-body model SEOBNRv4PHM~\cite{Ossokine:2020kjp,Babak:2016tgq} and the phenomenological model IMRPhenomPv3HM~\citep{PhysRevD.101.024056}, were also used and gave consistent results, as presented in \citet{Abbott:2020mjq}.

\begin{figure}
\includegraphics[width=0.5\textwidth]{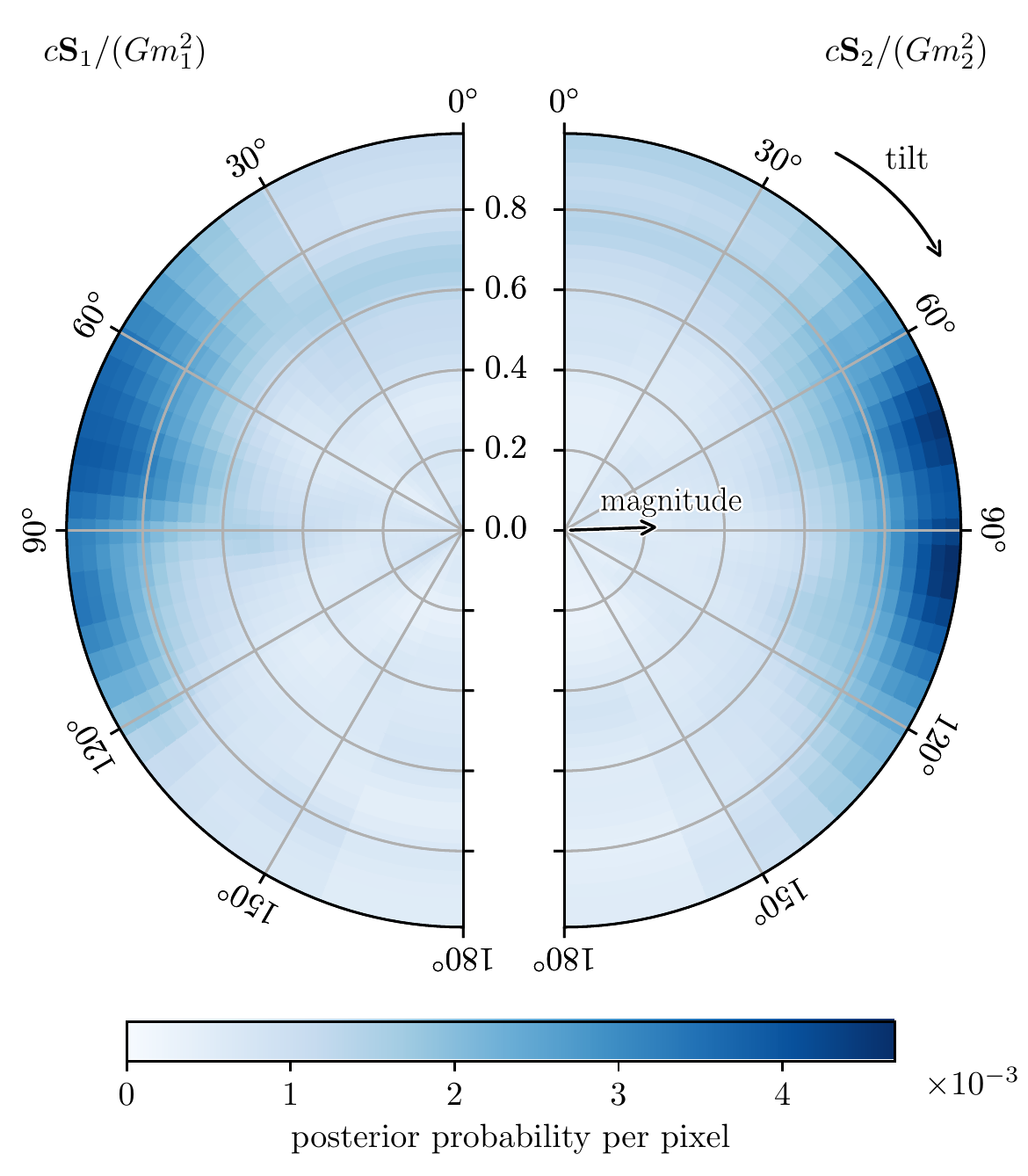}
\caption{Displayed are the estimated posterior distributions for the spins of the two initial black holes which produced GW190521. Having significant spin in the orbital plane would induce obital precession. These distributions show weight for the posteriors for large spin and near to 90$^o$, namely large spin in the orbital plane. Spins of 0$^o$ would be aligned with the orbital angular momentum.
See ~\citet{Abbott:2020tfl} for more details.}
\label{fig:GW190521_disk}
\end{figure}

The parameter estimation for GW190521 provided values for the dimensionless spin vectors of $\chi_{1} = 0.69^{+0.27}_{-0.62}$, and $\chi_{2} = 0.73^{+0.24}_{-0.64}$. The precession spin parameter was estimated at $\chi_{\rm p} = 0.68^{+0.25}_{-0.37}$, while
the effective spin parameter was $\chi_{\rm eff} = 0.08^{+0.27}_{-0.36}$. The Bayes factor for the presence orbital precession was calculated to be log$_{10} ~ \mathcal{B} = 1.06^{+0.06}_{-0.06}$, thereby showing slight evidence. Fig.~\ref{fig:GW190521_disk} displays the estimated posterior distributions for the spins of the two initial black holes, showing weight for the posteriors at large spin and near the orbital plane at 90$^o$~\citep{Abbott:2020tfl}.

\subsubsection{O1 and O2 Catalog, GWTC-1}
\label{subsubsec:DetO1O2}
The totality of the observations from the O1 and O2 observing runs were reported by LIGO and Virgo in their first gravitational wave transient catalog, GWTC-1~\cite{LIGOScientific:2018mvr}. This corresponds to the confident detections of 10 binary black hole produced gravitational wave signals, and one signal produced by a binary neutron star. The catalog presents the basic information about the sources of these signals, as derived via the Bayesian parameter estimation routines. This information includes the estimates of the initial and final masses, the effective aligned spin, the final spin, peak luminosity, total radiated energy, luminosity distance and the sky position. The tests of general relativity from the binary neutron star merger were presented in \citet{Abbott:2018lct}. Similarly, the tests of general relativity from the gravitational waves from the 10 binary black hole mergers are presented in \citet{LIGOScientific:2019fpa}.

\subsubsection{O1, O2 and O3a Catalog, GWTC-2}
\label{subsubsec:DetO1O2O3a}
The catalog encompassing the O3a results, GWTC-2 \citep{Abbott:2020niy}, has added another 39 compact binary produced gravitational wave events, including a binary neutron star produced signal (GW190425). One event could be from a neutron star - black hole merger, although this event has the lowest significance in the catalog. From this list of 50 compact binary produced signals further studies into the population properties of compact binaries~\citep{Abbott:2020gyp}. Additional tests of general relativity have also been conducted using the events from GWTC-2~\citep{Abbott:2020jks}.


\subsection{Binary Neutron Stars}
\label{subsec:detBNS}
The observation of gravitational waves from the merger of a binary neutron star system provides numerous additional scenarios over what can be done with the observation of a binary black hole system. Neutron stars are made of matter. As such, with a binary neutron star merger electromagnetic radiation will be emitted. In fact, it has been theorized that a source for short gamma ray bursts could be from binary neutron star mergers~\cite{Berger:2013jza}. 

Black holes essentially 
behave as point particles as they spiral into one another in a binary merger, and while neutron stars are incredibly dense, the tidal gravitational field will eventually distort the shape of the neutron stars as they approach each other. The tidal gravitational field induces mass quadruple moments in the neutron stars~\cite{PhysRevD.45.1017}, and this has the effect of increasing the rate at which the binary system coalesces~\cite{PhysRevD.77.021502}. This effect can be calculated and incorporated into the waveform. As such, it could be possible to extract information on the nuclear equation of state from the observation of gravitational waves from a binary neutron star merger~\cite{PhysRevD.83.084051,Damour:2012yf,Chatziioannou:2020pqz}.

While it is likely that the end product of a binary neutron star merger will be a black hole, it is possible that some of the matter will be ejected from the collision, and could be observable. 
This remnant is called a kilonova. Rapid neutron capture would form heavy elements, and with them being radioactive, the activity could be visible.
This ejected material could also be the source of heavy elements, such as gold and platinum, in the universe~\cite{1974ApJ...192L.145L,Metzger:2019zeh}

A unique feature about coalescing compact binary gravitational wave signals is that it is possible to use Bayesian inference to estimate the luminosity distance to the source. If the redshift of the source can also be measured then the expansion of the universe, or the Hubble constant, could be measured~\cite{1986Natur.323..310S,Nissanke:2009kt}. This would be a new and unique way to measure the expansion of the universe, and very different from the use of cosmic microwave background data~\cite{Christensen_2001,Ade:2015xua} or supernovae observations~\cite{Riess:2016jrr}.

All of these effects are of great importance for astrophysics and fundamental physics. Hence there is tremendous interest in observing gravitational waves and electromagnetic radiation from a binary neutron star merger. Many of the studies necessitate finding the source, so this requires good parameter estimation for the sky position and the distance. In such a way, multi-messenger astronomy can be conducted.

\subsubsection{GW170817}
\label{subsubsec:DetGW170817}
The birth of gravitational wave multimessenger astronomy occured with the simultaneous observations of a gravitational wave signal GW170817~\cite{TheLIGOScientific:2017qsa} followed 1.7s by the observation of the short gamma ray burst GRB 170817A~\cite{2017ApJ...848L..14G,Savchenko:2017ffs}. With the subsequent detection of the kilonova counterpart~\cite{2017Sci...358.1556C}, the source was observed across the electromagnetic spectrum, from X-ray to radio~\cite{2017ApJ...848L..12A}. This event provided the means to study numerous important astrophysical and fundamental physics effects.

The success of GW170817 for multimessenger astronomy came from the ability to identify the location of the source. The gravitational wave detection was not trivial, however. A strong noise transient, a {\it glitch}, occured during the time that the signal was being recorded at LIGO Livingston, 1.1 s before the coalescence time. 
For the initial rapid response the data containing the glitch was removed with a Tukey window function.
The SNR was large in the data for LIGO Hanford and LIGO Livingston. The SNR in the Virgo data was relatively small, however as noted in \citet{Rover:2006bb}, this is still information that can be used in a parameter estimation routine to improve the localization of the source. The data from all three gravitational wave detectors was crucial for the signal source identification. The rapid parameter estimation routine BAYESTAR~\cite{PhysRevD.93.024013} (see also Sec.~\ref{sec:BAYESTAR}) was able to estimate that position of the source in the sky to 31 deg$^{2}$, and an initial estimation of the luminosity distance of $40 \pm 8$ Mpc~\cite{GCN21513}. The sky position estimation can be seen in Fig.~\ref{fig:GW170817sky}. This is what allowed for the location of the source to be identified in the galaxy NGC 4993 10.9 hours after the detection of the gravitational wave and the gamma ray burst~\cite{GCN21529}.

In order to subsequently conduct effective parameter estimation, the glitch in the LIGO Livingston data was modeled in time and frequency with wavelets, namely with the BayesWave algorithm~\cite{Cornish:2014kda} as described in Sec.~\ref{subsec:signalplusnoise}, and then subtracted from the data. Note that this event has subsequently provoked the development of even more sophisticated techniques in glitch subtraction~\cite{Pankow:2018qpo}. The waveform~\cite{PhysRevD.44.3819} model used for the Bayesian parameter estimation~\cite{Veitch:2014wba} incorporated the effects of spin aligned with the orbital angular momentum~\cite{Boh__2013}, spin-spin interactions between the two initial masses~\cite{Boh__2015,PhysRevD.93.084054}, and tidal interactions on the neutron stars~\cite{PhysRevD.83.084051,Bernuzzi:2012ci}. With that the parameter estimation could report that the initial component masses were consistent with what are expected for neutron stars. However, there was some dependency on the prior for the effective spin parameter. Assuming that the initial system had low spin the two initial masses were estimated to be in the range of 1.17 to 1.6 $M_{\odot}$, with a total initial mass estimate of $2.74^{+0.04}_{-0.01} M_{\odot}$, and a chirp mass estimate of $1.188^{+0.004}_{-0.002} M_{\odot}$~\cite{TheLIGOScientific:2017qsa}. 

The parameter estimation for GW170817 considered the possible distortion of the spherical shape of the neutron stars from tidal forces. The tidal deformity is defined as the ratio of the induced mass quadruple moments to the tidal gravitational field, and is given by
\begin{equation}
\label{Eq:Love}
\Lambda = \frac{2 k_{2}}{3} \left(\frac{c^{2} R}{G m}\right)^{5} ~ ,
\end{equation}
where $R$ is the neutron star radius, $m$ the mass, and $k_{2}$ the second Love number, which is theorized to be in the range of 0.05 to 0.15 for neutron stars~\cite{PhysRevD.81.123016}. The initial analysis constrained the tidal deformity to $\Lambda \leq 800$~\cite{TheLIGOScientific:2017qsa}. A subequent analysis assumed that the equation of state for the two neutron stars was the same, that the equation of state could allow for neutron stars in excess of $1.97 M_{\odot}$ (to be consistent with the observed mass of for the pulsar J0348+0432 of $\sim 2 M_{\odot}$~\cite{Antoniadis1233232}), and that the spins of the neutron stars were consistent with the observed spins of binary neutron stars in our galaxy. With such assumptions LIGO and Virgo were able to produce the estimate of $\Lambda = 190^{+390}_{-120}$, as well as to estimate the radii of the neutron stars to be $11.9^{+1.4}_{-1.4}$ km~\cite{Abbott:2018exr}. LIGO and Virgo have used the gravitational wave data for GW170817 and Bayesian parameter estimation methods to investigate other characteristics of the neutron stars and their equations of state~\cite{Weinberg:2018icl,LIGOScientific:2019eut}.

The ability to measure the expansion rate of the universe, the Hubble constant, was another significant byproduct of the observation of GW170817 and the use of Bayesian parameter estimation routines on the gravitational wave data. The ability to use gravitational wave data to measure the Hubble constant was a long anticipated reward for gravitatational wave astrophysics~\cite{1986Natur.323..310S}. Recall Eq.~\ref{eq:Hub}, $v = H_{0} D$.
The parameter estimation for a compact binary coalescence provides an estimate for the luminosity distance $D_{L}$. However, the estimation of another parameter is correlated with the distance, namely the angle of inclination of the normal to the orbital plane of the system with respect to the line of sight, $\iota$. Variations in $D_{L}$ and $\iota$ both affect the amplitude of the detected gravitational wave, hence their correlation when they are estimated. In order to generate a posterior probability distribution function for the luminosity distance $D_{L}$ one must marginalize over the inclination angle $\iota$ which adds uncertainty to the distance estimate. The Hubble constant parameter estimation effort was significantly improved by the fact that the source of GW170817 was found to be in the galaxy NGC 4993. Because of this the two sky position parameters could be fixed, and with that constraint the distance was estimated to be $D_{L} = 43.8^{+2.9}_{-6.9}$ Mpc, where the error bars correspond to the 68.3\% credible interval. The velocity of the source to the line of sight could be measured from the observed velocity of NGC 4993; this is done via redshift measurements. An allowance was also made for the peculiar velocity of NGC 4993 within its local cluster of galaxies. With this the Hubble constant was estimated to be $H_{0} = 70^{+12}_{-8}$ km s$^{-1}$ Mpc$^{-1}$, where again the error bars represent the 68.3\% credible interval~\cite{Abbott:2017xzu}. This measurement is independent of the other methods used to measure the Hubble constant, but the result is consistent. For example, measurements of the cosmic microwave background data give the estimate 
$H_{0} = 67.74 \pm 0.46$ km s$^{-1}$ Mpc$^{-1}$~\cite{Ade:2015xua}, while type Ia supernovae observations give $H_{0} = 73.24 \pm 1.74$ km s$^{-1}$ Mpc$^{-1}$~\cite{Riess:2016jrr}. 
LIGO and Virgo have subsequently added binary black hole merger events and information from galaxy catalogs to produce an estimation of
$H_{0} = 69^{+17}_{-8}$ km s$^{-1}$ Mpc$^{-1}$~\cite{LIGOScientific:2019zcs}. This has also spurred other groups to use gravitational wave data to estimate the Hubble constant~\cite{Finke:2021aom}.

Binary neutron star mergers were thought to be a source of short gamma ray bursts~\cite{Berger:2013jza} with the gamma rays ejected in a jet perpendicular to the orbital plane of the binary, namely, parallel to the orbital angular momentum of the the system~\cite{PhysRevLett.96.031102}. Hence, the observation of the gamma rays and the estimation of the observation angle with respect to the jet, $\iota$, will provide much important information for understanding the formation of jets from the binary neutron star mergers. GW170817 and GRB 170817A provided such critical data~\cite{Monitor:2017mdv}. The gravitational wave data indicate that the viewing angle to the source is anti-alinged, namely the angular momentum vector of the system is pointing away from us. The estimation of the cosine of the inclination, $cos\iota$, is that it is constrained to the range [-1.00, -0.81], at 68.3\% confidence, or equivalently [-144$^o$, 180$^o$]~\cite{TheLIGOScientific:2017qsa}. The Hubble constant question can also be inverted, namely to use the previously determined Hubble constant measurements as a prior and produce an improved estimate on the inclination angle $\iota$. Using the Hubble constant from the cosmic microwave background measurement of the Planck mission~\cite{Ade:2015xua} the 68.3\% confidence band for $cos\iota$ is  [-1.00, -0.92], or [157$^o$, 177$^o$] for the inclination angle $\iota$. Using the supernova produced value for the Hubble constant ~\cite{Riess:2016jrr} the similar constraints are  [-0.97, -0.85] for $cos\iota$ and [148$^o$, 166$^o$] for $\iota$~\cite{Abbott:2017xzu}. The parameter estimates for GW170817 have been further updated in~\citet{LIGOScientific:2018mvr}.

\subsubsection{GW190425}
\label{subsubsec:DetGW190425}
LIGO and Virgo have announced the detection of gravitational waves from another possible binary neutron star merger GW190425~\citep{Abbott:2020uma}.
This was observed in the the third Advanced LIGO - Advanced Virgo observing run, O3. The event was only confidently detected in the LIGO Livingston detector. Virgo was taking data, but the SNR was too low for it to contribute to the detection. LIGO Hanford was not on-line during this event. As a consequence there was a very large uncertainty in the sky position of the source, $\sim 8300$ deg$^{2}$. The luminosity distance was estimated to be $159^{+71}_{-69}$ Mpc, much farther than the $\sim 40$ Mpc distance for GW170817. Because of the large sky position uncertainty and large distance, no electromagnetic counterpart to GW190425 was observed. 
For the parameter estimation a phenomenological waveform~\citep{Hannam:2013oca} is used, namely  
PhenomPv2NRT \citep{Dietrich:2018uni}; this model incorporates spin precession and tidal interactions~\citep{PhysRevD.96.121501}. 
We quote here the results using the high-spin prior (dimensionless spin magnitudes for the two initial neutron star of $\chi < 0.89$).
An interesting consequence of the the observation of GW190425 was the relative large masses for what are assumed to be a pair of neutron stars. The chirp mass was estimated to be $1.44^{+0.02}_{-0.02} M_{\odot}$, and total mass $3.4^{+0.3}_{-0.1} M_{\odot}$; for comparison, GW170817 had, when also using a high-spin prior, estimates for the chirp mass of $1.188^{+0.004}_{-0.002} M_{\odot}$ and total mass of $2.82^{+0.47}_{-0.09} M_{\odot}$. For GW190425 no tidal effects were observed, and a limit for the combined dimensionless tidal deformability was set at $\Lambda < 1100$~\citep{Abbott:2020uma}.

\subsection{Neutron Star -- Black Hole Binaries}
\label{subsec:detNSBH}
The merger of a neutron star - black hole binary is another source of gravitational waves. These events are interesting for a number of reasons. If the black hole is not too massive the neutron star could be tidally disrupted before crossing the event horizon~\cite{Stachie:2021noh} and could be a source of gamma rays~\cite{2013PhRvD..87h4053S,Berger:2013jza} or a kilonova~\cite{Kawaguchi:2016ana,Metzger:2019zeh,Mochkovitch:2021prz,Foucart:2020ats}. This could also provide information on the equation of state for the neutron star material~\cite{Harry:2018hke}. The formation mechanisms for neutron star - black hole binaries are also an important area of study~\cite{Broekgaarden:2021hlu}.

\subsubsection{GW200105 and GW200115}
\label{subsubsec:DetGW200105_GW200115}
LIGO and Virgo have detected gravitational waves from two neutron star - black hole binary mergers~\cite{LIGOScientific:2021qlt}. These were both detected in January, 2020, during the second half of observing run O3 (O3b). GW200105 was detected at LIGO Livingston, while LIGO Hanford was not observing; the SNR for the event in Virgo was very low, implying that this was essentially a single detector observation. Virgo data was used for paramter estimation, yielding mass estimates (low spin prior) of $8.9^{+1.1}_{-1.3} M_{\odot}$ (90\% credible intervals) for the presumed black hole, and $1.9^{+0.2}_{-0.2} M_{\odot}$ for the presumed neutron star. The estimated luminosity distance is $280^{+110}_{-110}$ Mpc. GW200115 was detected by all three LIGO-Virgo detectors. The mass estimates were (low spin prior) are $5.9^{+1.4}_{-2.1} M_{\odot}$ for the presumed black hole, and $1.4^{+0.6}_{-0.2} M_{\odot}$ for the presumed neutron star. The estimated luminosity distance is $310^{+150}_{-110}$ Mpc. No electromagnetic or neutrino counterpart to these events was detected.

For parameter estimation pBilby was used~\cite{10.1093/mnras/staa2483} (see Sec.~\ref{sec:Bilby}), as well as RIFT~\cite{Lange:2018pyp} (see Sec.~\ref{subsec:RIFT}). In order to verify the results LALInference was also used~\cite{Veitch:2014wba}. The primary parameter estimation analysis did not assume that tidal effects were present. The phenomenological model IMRPhenomXPHM~\cite{Pratten:2020ceb} and the EOBNR model SEOBNRv4PHM~\cite{Ossokine:2020kjp} were used. These models included the effects of orbital precession and higher order modes, although the presence of these effects was not observed for either event. The possible tidal deformation of the neutron stars were investigated using models that include such an effect; these assume that spins are aligned with the orbital angular momentum. These are the phenomenological IMRPhenomNSBH~\cite{Thompson:2020nei} and the EOBNR SEOBNRv4\_ROM\_NRTidalv2\_NSBH~\cite{Matas:2020wab} models. Tidal deformation was not observed.

\section{Testing General Relativity}
\label{sec:TGR}
The observations of gravitational waves by LIGO and Virgo now present a possibility to test general relativity in a way that has never been possible before. 
The parameter estimation methods used to examine the gravitational wave signals are inherently model dependent. For the analyses conducted by LIGO and Virgo the basic assumption is that general relativity is correct. However the same parameter estimation methods can be extended to encompass alternatives to general relativity. The general relativistic models can be extended and parameter estimation can then be conducted, and if the additional parameters produce non-zero estimates it could be evidence for a violation of general relativity. Model comparison methods such as those described in Sec.~\ref{subsec:ModelComparison} can be applied directly between general relativity and the alternative model. In this section we summarize the methods used by LIGO and Virgo to test general relativity. We discuss the specific results in Sec.~\ref{sec:Detections} where the summary of some of the observed results from LIGO-Virgo parameter estimation are presented. The first detection, GW150914~\cite{Abbott:2016blz}, provided the first opportunity to conduct numerous tests of general relativity~\cite{TheLIGOScientific:2016src}. The first three-detector observation, GW170814, allowed for an examination of the polarization of gravitational waves and to test their consistency with general relativity~\cite{Abbott:2017oio}. Subsequently in their first three observing periods, O1, O2 and O3a (O3a is the first 6 months of O3), LIGO and Virgo observed a total of $\sim 50$ gravitational wave signals from compact binary coalescence~\cite{LIGOScientific:2018mvr,Abbott:2020niy}, and these observations have provided further opportunities to test general relativity~\cite{LIGOScientific:2019fpa,Abbott:2020jks}. 
Not all candidates were included in the analyses testing general relativity; this is reserved for the best candidates.
Hierarchical analyses in tests of general relativity were also employed in~\citet{Carullo:2021yxh} and~\citet{Ghosh:2021mrv}.

For the tests of general relativity based on the LIGO-Virgo O1 and O2 results, presented in~\citet{LIGOScientific:2019fpa}, it was assumed that inconsistencies in general relativity would occur in the same fashion for all events, regardless of the properties of the source. In the latest LIGO-Virgo study of~\citet{Abbott:2020jks}, which includes O3a, this approach was loosened and a hierarchical inference technique from~\citet{Zimmerman:2019wzo} and \citet{Isi:2019asy} was used for some tests.
For every particular gravitational wave event that is tested, the parameters associated with a violation of general relativity are selected from some universal distribution that has been created from inference on all of the events.
While this distribution is initially not known, it could be resolved with an appropriate description of gravity (beyond general relativity) and the data from numerous events. One can then do tests by comparing the derived distribution with what one expects from general relativity. 

The observation of gravitational waves from a binary neutron star merger, GW170817~\cite{TheLIGOScientific:2017qsa}, provided further tests of general relativity, especially due to the long period of time that the signal was observable in the LIGO-Virgo operating frequency band and its large SNR~\cite{Abbott:2018lct}. The observation of a gamma ray signal, 1.7 s after the binary neutron star merger, allowed for unique tests of general relativity and Lorentz invariance~\cite{Monitor:2017mdv}.   

As with any parameter estimation routine, the model for the signal is of critical importance. For the gravitational waves produced by binary black holes the LIGO-Virgo analyses have used the effective-one-body SEOBNRv4 waveforms~\cite{Bohe:2016gbl} (specifically, the frequency domain SEOBNRv4\_ROM) for non-precessing spins for the black holes. To account for precessing spins the phenomenological waveforms IMRPhenomPv2~\cite{Khan:2015jqa,Husa:2015iqa,Hannam:2013oca} were used~\cite{LIGOScientific:2019fpa}. 
For some of the events, higher order modes were taken into account, and for these the SEOBNRv4HM model~\citep{Ossokine:2020kjp,Babak:2016tgq} and the IMRPhenomPv3HM model~\citep{PhysRevD.100.024059,PhysRevD.101.024056} were employed.
These waveforms for binary black holes also address the inspiral, merger and ringdown for the black holes. For gravitational waves from binary neutron stars the effects of the tidal deformations of the neutron stars must be taken into account. As such LIGO and Virgo have used the NRTidal models~\cite{PhysRevD.96.121501,Dietrich:2018uni} for the necessary additional phase factor.
The presence of eccentricity has been ignored in the binary black hole and binary neutron star models used by LIGO-Virgo to date~\cite{LIGOScientific:2019fpa,Abbott:2018lct}. As described in Sec.~\ref{subsec:calibration}, uncertainties in the calibration of the detector data are introduced as part of the overall parameter estimation, and the introduced parameters associated with detector response function; in the end there is a marginalization over these parameters~\cite{TheLIGOScientific:2016wfe}.

\subsection{Signal Residual Test}
\label{subsec:TGRSRT}
The parameter estimation methods used by LIGO and Virgo are run on detected signals. This is typically done using LALInference~\cite{Veitch:2014wba}. From this a best fit (in terms of maximum likelihood) waveform is produced.
After subtracting the best fit waveform from the observed gravitational wave data,  a test is then performed to see if the remaining residual is consistent with Gaussian noise. This method has been applied to all 10 gravitational wave signals detected during O1 and O2 that were produced by binary black holes~\cite{LIGOScientific:2019fpa}; the method is further explained in~\cite{LIGOScientific:2019hgc}. 
The waveform model was IMRPhenomPv2~\cite{Khan:2015jqa,Husa:2015iqa,Hannam:2013oca}. 
This method was also applied for tests of general relativity for the events in the second LIGO-Virgo catalog~\citep{Abbott:2020niy},
and is further discussed in~\citet{Ghonge:2020suv}.
For O3a the signal residual test was conducted on a further 24 binary black hole produced gravitational wave signals~\cite{Abbott:2020jks}.

\begin{figure}
\includegraphics[width=0.5\textwidth]{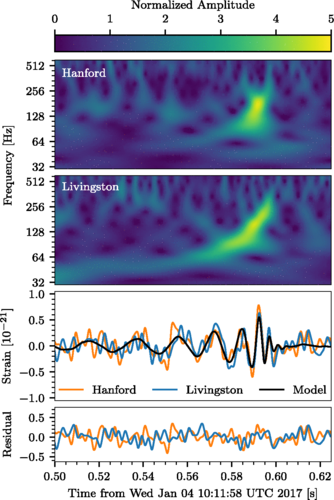}
\caption{The detected gravitational wave signal of GW170104 by in the two Advanced LIGO detectors. The top panel shows the time-frequency expression of the data from LIGO-Hanford, while the second panel does the same for LIGO-Livingston. The time series for the data from the two detectors is displayed in the third panel. The LIGO-Livingston data had been inverted to account for a sign difference with respect to LIGO-Hanford, and has been adjusted by 3 ms because of the difference in the arrival times. This time series data has been filtered with a 30-350 Hz bandpass. The black line is the waveform corresponding to the maximum likelihood from parameter estimation using the precessing spin model. The maximum likelihood waveform is subtracted from the gravitational wave data, and the residuals are then displayed in the bottom panel. A statistical analysis of the residuals shows that they are more consistent with Gaussian noise than the presence of remaining coherent gravitational wave energy.
See~\cite{Abbott:2017vtc} for more details.}
\label{fig:GW170104TS}
\end{figure}

For the 10 binary black hole produced gravitational wave signals reported in~\cite{LIGOScientific:2018mvr}, the best fit waveforms were produced and subtracted from a 1 s stretch of the
LIGO and Virgo data, with the merger time in the middle. Since the model waveform was constructed as to adhere to general relativity, by substracting the best fit model waveform one should be left with a residual that resembles the noise of the detector. The residual is tested to confirm this assumption. LIGO and Virgo have used the BayesWave~\cite{Cornish:2014kda,Littenberg:2014oda} algorithm to conduct these tests; see Sec.~\ref{subsec:BayesWave} for a description of BayesWave. With the signal substracted, BayesWave analyzes the residual data streams from the two or three detectors involved in the detection. Three models are considered: the data contain an elliptically polarized gravitational wave signal that is coherent in the different data streams plus Gaussian noise; there are uncorrelated noise transients (glitches) present and Gaussian noise; there is only Gaussian noise. BayesWave then calculates Bayes Factors for model comparison. 
A p-value is computed to take into account the variable background.
A consequence of this analysis is also a network SNR for the presence of a coherent gravitational wave signal in the different residual data streams. 
Fig.~\ref{fig:GW170104TS} displays this process for the gravitational waves detected from the binary black hole merger GW170104~\cite{Abbott:2017vtc}. Further examples from the LIGO-Virgo observations are summarized in Sec.~\ref{sec:Detections}.

\subsection{Inspiral - Merger - Ringdown Consistency for Binary Black Holes}
\label{subsec:TGRBHR}
The estimated parameters from the inspiral part of a binary black hole system can be compared with the parameters estimated from the signal pertaining to the merger and ringdown of the final black hole, see \citet{Ghosh:2016qgn} and \citet{Ghosh:2017gfp}. The basic assumption for this test is that the underlying theory that describes the inspiral, merger, and ringdown is general relativity. As such, one should be able to conduct parameter estimation on these different parts of the gravitational wave signal and recover the same physical parameters that describe the system. 
For the binary black hole systems the goal is to estimate the mass and spin of the final black hole remnant from the low frequency inspiral part of the signal, and then again for the high frequency merger-ringdown part of the signal~\cite{LIGOScientific:2019fpa}. 
An important goal will be to observe overtones in addition to the fundamental ringdown mode of the remnant and test the no-hair theorem~\cite{PhysRevX.9.041060}.
The study of~\citet{Isi:2019aib} analyzed LIGO data for GW150914 and claimed that there was evidence at the 3.6$\sigma$ level for the presence of the $l=m=2$ quasinormal mode and one overtone. The LIGO-Virgo analysis of~\citet{Abbott:2020jks} also searched for the presence of overtones and found evidence for GW190521 (Bayes factor of 19.5), and GW150914 (Bayes factor of 4.3).

A separation between the inspiral and merger-ringdown regimes must be chosen for the analysis.
The final black hole mass and spin are obtained from the initial masses and spins using numerical relativity fits~\cite{PhysRevD.90.104004,PhysRevD.95.024037,PhysRevD.95.064024,Hofmann_2016}, and then the Kerr innermost stable circular orbit frequency is computed from the final mass and spin using the expressions in~\citet{1972ApJ...178..347B}. This frequency is used as the transition from inspiral to merger-ringdown. 
With that, the remnant's final spin and mass are estimated with the low frequency inspiral part of the signal, and then again for the high frequency merger-rigndown part of the signal.
The Bayesian parameter estimation is done with LALInference~\cite{Veitch:2014wba}, with the spin precessing phenomenological IMRPhenomPv2~\cite{Khan:2015jqa,Husa:2015iqa,Hannam:2013oca}
and IMRPhenomPv3HM~\citep{PhysRevD.100.024059,PhysRevD.101.024056} waveforms, 
plus the effective-one-body SEOBNRv4 waveform for binary black holes with spins that are nonprecessing~\cite{Bohe:2016gbl}. For the two frequency regions, the source parameters are estimated, and then comparisons are done with simulations from numerical relativity~\cite{PhysRevD.95.024037,Hofmann_2016,PhysRevD.95.064024}. In this way the final remnant's mass $M_{f}$ and dimensionless spin
$\chi_{f} = c \lvert \vec{S}_f \rvert / (G M_{f}^2)$ 
are calculated from the data before and after the innermost stable circular orbit, and compared for consistency. This is done by calculating the overlap of the posterior distributions for these parameters. The test also calculates the posteriors on the final mass and spin deviation parameters and quotes the quantile of this distribution at which the general relativity predicted value of is recovered~\citep{Ghosh:2017gfp}.
Finally, this analysis tests the emission of energy and angular momentum predicted by general relativity, especially in the non-linear phase of the merger-ringdown.


This test was done for 7 of the 10 binary black hole mergers observered during O1 and O2 that had sufficient SNR for both parts of the signal. 
Using uniform priors for the masses and the magnitude of the spins, plus priors for the spin directions that are isotropic, the parameter estimation and analysis for the final black hole remnant and spin magnitude were found to be consistent between the inspiral part of the signal, and the merger-ringdown part of the signal. 
This analysis was then repeated for 12 events in O3a, also using uniform priors for the deviation parameters.
No deviations from general relativity were observed~\cite{LIGOScientific:2019fpa,Abbott:2020jks}.
See Sec.~\ref{subsubsec:DetBBHGW150914} for an example with GW150914.

The inspiral-merger-ringdown test has also been done including higher order modes with the IMRPhenomPv3HM waveform~\citep{PhysRevD.100.024059,PhysRevD.101.024056} using {\tt pBilby}~\citep{10.1093/mnras/staa2483} for parameter estimation; see Sec.~\ref{sec:Bilby} for more information on {\sc Bilby}. This was done for GW190412 and GW190814~\citep{LIGOScientific:2020stg,Abbott:2020jks}.
Similarly, the inspiral-merger-ringdown analysis has also been demonstrated using the NRSur7dq2 waveform~\citep{Blackman:2017pcm} and the RIFT package~\citep{Lange:2018pyp} for parameter estimation~\citep{Breschi:2019wki}; see Sec.\ref{subsec:RIFT} for more information on RIFT.  

\subsubsection{Remnant Properties}
\label{subsebsec:remanants}

LIGO-Virgo have also examined just the ringdown signal from binary black hole mergers~\citep{Abbott:2020jks}. The remnant after the merger will initially be a non-spherical object, but by the no hair theorem, it must come to equilibium as a Kerr black hole. The excited remnant ringdown results in the emission of different damped sinusoidal signals, quasi-normal modes, that depend only on the final mass and spin of the Kerr black hole, plus the integer indices of the modes. The information from the observation of gravitational waves from the ringdown would hence describe the final state of the remnant. A comparison can then be made to the energy and angular momentum emitted through gravitational waves during the inspiral. For the parameter estimation of signals only containing the ringdown a time-domain formulation of the likelihood is used. This method avoids the contribution of spurious frequency contributions from the pre-merger phase, or an abrupt windowing around the peak of the signal which would give Gibbs phenomena~\cite{Carullo:2019flw,Isi:2019aib}.
Another ringdown analysis is based on the EOB waveforms and concentrates on the gravitational wave signal and damping time for a particular mode (220)~\citep{Brito:2018rfr}. All of the results from both analyses show consistency of the ringdown parameter estimation with those derived from the full inspiral-merger-ringdown signal.

Another possible effect associated with the remnant, and outside of the theory of general relativity, concerns the effect of echoes on the generated gravitational waves. This could pertain to exotic compact objects like fuzzballs~\cite{Lunin:2002qf,Mathur:2008nj} and gravastars~\cite{Mazur:2001fv,Mazur:2004fk}. With such an object there may be a surface, between the light ring and the location where the event horizon would be, that would reflect gravitational waves. In such a case, when two compact objects merge and emit gravitational waves, some of the signals created would be reflected off this surface, producing the so-called gravitational wave echoes. In fact, a cavity like structure could be created, and a series of echo signals would be emitted, successively smaller in amplitude. This effect was claimed to have been observed in the LIGO data for GW150914 by~\citet{Abedi:2016hgu}.
As presented in~\citet{Abbott:2020jks}, LIGO and Virgo have implemented a signal template search for echoes of ringdown signals from binary black hole mergers. A Bayes factor is calculated between the presence of inspiral-merger-ringdown-echo signals, and the general relativity predicted inspiral-merger-ringdown signal. The data from 31 binary black hole mergers were analyzed. No evidence for the presence of echoes was found. 

\subsection{Parametrized Tests of Gravitational Waveforms}
\label{subsec:TGRParamTests}
Since general relativity is a non-linear theory, simple or closed form solutions are rare. Binary orbital systems, losing energy through the emission of gravitational waves, are described through post-Newtonian approximations, namely expressing the orbit in terms of varying orders of $v/c$~\cite{PhysRevLett.74.3515,PhysRevLett.93.091101,PhysRevD.71.124004,Blanchet:2013haa}. For the standard gravitational wave parameter estimation for a LIGO and Virgo detected signal the different post-Newtonian approximants are summed together to form the model of the signal. However, if general relativity is not the correct theory to describe gravitational phenomena, a discrepency might arise between the observation and the general relativistic model. Tests are done on general relativity by allowing for a phase shift for the different post-Newtonian approximate terms. 
Parametric deviations are also added to phenomenological parameters in the merger-ringdown phases.
In terms of parameter estimation, this is the introduction of an additional phase parameter for each term in the post-Newtonian expansion~\citep{Meidam:2017dgf}.

\begin{figure}
\includegraphics[width=0.5\textwidth]{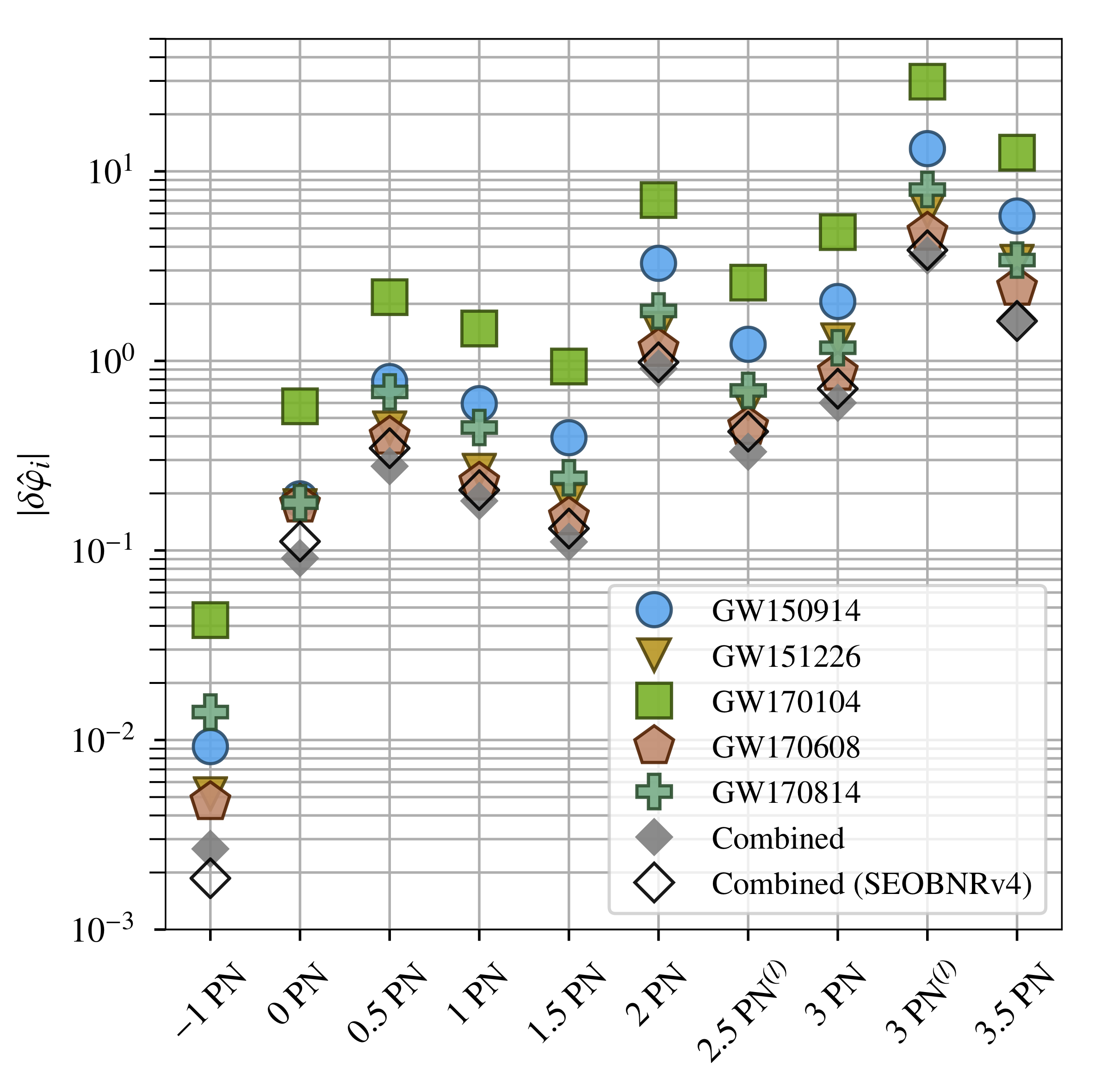}
\caption{The upper bounds at 90\% confidence for the post-Newtonian binary inspiral parameters $\delta \phi_{i}$ that are testing general relativity.
Results are presented for the five loudest binary black hole produced gravitational wave signals in O1 and O2. For the five individual events the upper bounds are given based on the IMRPhenomPv2 waveform model. Then a final bound is calculated from the five events together by combining the individual posteriors for $\delta \phi_{i}$. A combined result is also presented for the SEOBNRv4 waveform. 
See~\cite{LIGOScientific:2019fpa} for more details. Figure courtesy of LIGO Laboratory~\cite{GW150914-IMR}.}
\label{fig:GWTC1-PN_TEST}
\end{figure}

This approach to test general relativity is divided amongst the three parts of the signal: inspiral, intermediate stage, and merger-ringdown. In the study of the 10 binary black hole events from O1 and O2 reported in~\cite{LIGOScientific:2018mvr}, the IMRPhenomPv2 waveforms~\cite{Khan:2015jqa,Husa:2015iqa,Hannam:2013oca} are used, and the modifications to the various phase terms are added to the expansion terms for this model~\cite{LIGOScientific:2019fpa}. 
In addition, the SEOBNRv4\_ROM waveform~\citep{Bohe:2016gbl} was applied to search for parametrized modifications in the inspiral. 
For the inspiral part of the signal $\delta\phi_{i}$ represents the additional phase for the $i^{th}$ Newtonian or post-Newtonian term. The intermediate stage has two possible perturbations to the phenomenological coefficients, $\beta_{2}$ and $\beta_{3}$. For the final merger-ringdown part of the signal the three phase perturbations to the phenomenological coefficients are denoted by $\alpha_{2}$, $\alpha_{3}$ and $\alpha_{4}$. With the general notation that $p_{i}$ represents the waveform coefficients $\phi_{i}$, $\alpha_{i}$ and $\beta_{i}$, the parameter estimation code LALInference~\cite{Veitch:2014wba} is modified to make the adjustment $p_{i} \rightarrow (1 + \delta p_{i}) p_{i}$. If the parameter estimation produced $\delta p_{i}$ is consistent with 0, then there is no evidence for a violation of general relativity. The LIGO-Virgo studies conduct these tests by varying one $\delta p_{i}$ parameter at a time, and calculate their posterior distribution functions~\cite{LIGOScientific:2019fpa}. A theory for gravity that is different than general relativity would probably cause a deviation in all of the $\delta p_{i}$ parameters.
This simplifying choice is made because if all of the parameters vary together, the correlations are  so strong that at current SNR, no meaningful constraints can be made; this was displayed with GW150914 in~\citet{TheLIGOScientific:2016src}.
Fig.~\ref{fig:GWTC1-PN_TEST} uses the data from the five loudest gravitational wave signals during O1 and O2 from binary black hole mergers to set upper bounds on the magnitude of the post-Newtonian parameters corresponding to the inspiral part of the gravitational wave signal, $\delta \phi_{i}$. This study was repeated for 24 binary black produced gravitational wave signals from O3a~\cite{Abbott:2020jks}.

\subsection{Spin Induced Quadrupole Moment}
\label{subsec:TGRSIQM}
A new test of general relativity by LIGO-Virgo for their binary black hole merger events concerns the possible inducement of a quadrupole moment due to the spin of the black hole~\citep{Abbott:2020jks}. The spin of a black hole $\chi$ will create a quadrupole moment $Q$ given by
\begin{equation}
\label{eq:spin-quad}
Q = - \kappa \chi^{2} m^{3} ~,
\end{equation}
where $m$ is the black hole mass and $\kappa$ depends on the object. With general relativity, a black hole will have $\kappa = 1$. A neutron star could have $\kappa$ in the range of $\sim 2$ to $14$~\cite{Harry:2018hke,Abbott:2020jks}, while for a boson star it could be several hundred~\citep{Chia:2020psj}. A spin induced quadrupole moment will change the phase of the inspiraling compact binary. The LIGO-Virgo study makes a simplifying assumption that for a binary black hole, each object will have equal $\kappa$ parameters. The spin induced phase variation is included in the waveforms, and parameter estimation is done in estimating the deviation of $\kappa$ from 1 for the binary black hole mergers observed in O1, O2 and O3a. The resulting distribution for the deviation of $\kappa$ is consistent with zero, hence supporting the validity of general relativity~\cite{Abbott:2020jks}.

\subsection{Polarization Tests}
\label{subsec:TGRPolarTests}
General relativity predicts only two tensor polarization modes for gravitational waves, while alternative theories of gravity could could produce vector or scalar modes~\cite{Chatziioannou:2012rf}; see also \citet{Will:2005va}, \citet{Berti:2015itd} and \citet{Callister:2017ocg} which give many references for alternative theories of gravity that can produce different combinations of these polarizations. 
The two LIGO detectors are relatively well-aligned with respect to one another with only the curvature of the Earth over the 3000 km displacement causing a slight misalignment. As such, it is not possible for just the two LIGO detectors to observe and test the polarization of the detected gravitational waves. However the addition of a third detector to the network, Virgo, and especially because of its orientation with respect to the LIGO detectors, provides for the means to probe the polarization state of the observed gravitational waves. When Advanced Virgo joined O2 it quickly participated in three gravitational wave observations that could be used to test the polarization content of the signals, specifically the two binary black hole produced signals GW170814~\cite{Abbott:2017oio} and GW170818~\cite{LIGOScientific:2018mvr}, and the binary neutron star produced signal GW170817~\cite{TheLIGOScientific:2017qsa}. 

For the first three-detector observation of gravitational waves, GW170814, a model comparison was done between the scenarios where the gravitational waves were entirely of a tensor polarization, entirely of a vector polarization, or entirely of a scalar polarization. The response of interferometric gravitational wave detectors is different for the tensor, vector and scalar polarizations~\cite{Callister:2017ocg,Isi:2017fbj}. The LIGO-Virgo Bayesian parameter estimation software, LALInference~\cite{Veitch:2014wba}, is used to analyze the three-detector data. The particular interferometer responses for the different polarizations change the estimates for the sky location and luminosity distance for the source, however the estimates for the masses and spins remain unchanged. For GW170814 LIGO and Virgo 
calculated a Bayes factor exceeding 200 for a pure tensor polarization model as compared to a pure vector tensor model; the Bayes factor exceeded 1000 for the preferences of a pure tensor polarization model as compared to the pure scalar polarization model~\cite{Abbott:2017oio}. A subsequent reanalysis of GW170814 using cleaned and recalibrated data reported a Bayes factor of 30 for tensor compared to vector polarization, and 220 for tensor versus scalar~\cite{LIGOScientific:2019fpa}. A similar study of the binary black hole produced event GW170818
gave a Bayes factor of 12 for tensor polarization compared to vector, and 407 for tensor compared to scalar~\cite{LIGOScientific:2019fpa}. 

A further 17 events have been studied in this way in O3a~\cite{Abbott:2020jks}. The analysis done for O3a is distinct from previous analyses as it uses a Bayesian version of the null-stream test of polarizations suggested in \citet{GurselYekta1989Nost}.

The binary neutron star merger event, GW170817, produced even stronger evidence in support of general relativity and tensor gravitational wave polarization. 
The pure tensor polarization model compared to the pure vector polarization model was favored with a log$_{10}$ Bayes factor of 20.81, while the pure tensor polarization model compared to the pure scalar polarization model was preferred with a log$_{10}$ Bayes factor of 23.09. These substantially larger Bayes factor results, as opposed to those from the binary black hole observations, are due to a variety of factors. The location on the sky of GW170817 could be determined extremely well from electromagnetic observations. The network SNR of GW170817 was large, and the position of the source in the sky was beneficial for a polarization test given the orientation of the detectors~\cite{Abbott:2018lct,Takeda:2020tjj,LIGO-T2000405}.  

LIGO and Virgo have searched for alternative polarizations in continuous gravitational wave signals~\cite{Abbott:2017tlp}, and in the stochastic gravitational wave background~\cite{LIGOScientific:2019vic,Abbott:2021xxi}. Note that with these studies there is a search for every possible combination of the different polarizations.
See Secs.~\ref{subsec:stochastic} and \ref{subsec:CW} for more details.

\subsection{Gravitational Wave Propagation}
\label{subsec:TGRGWP}
According to general relativity the speed at which gravitational waves propagate should be equal to the speed of light. This can now be tested. The mergers of a binary neutron star systems were thought to be a source of short gamma ray bursts. This was confirmed with the coincident observation of gravitational waves from a binary neutron star merger, GW170817~\cite{TheLIGOScientific:2017qsa}, and a gamma ray burst, GRB 170817A 1.7 s after the merger time~\cite{Monitor:2017mdv}.

\begin{figure}
\includegraphics[width=0.5\textwidth]{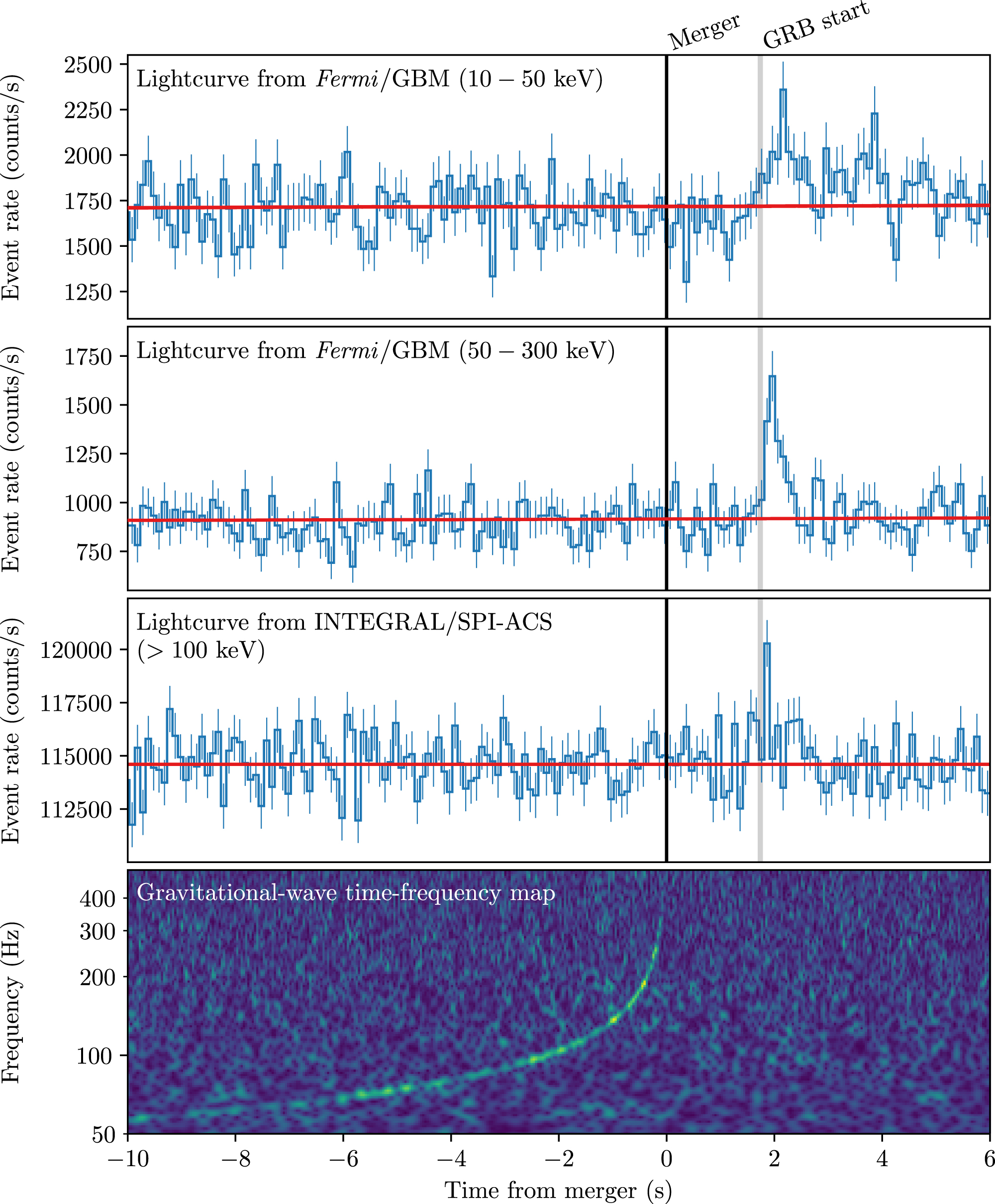}
\caption{The observation of GW170817 by LIGO, and GRB 170817A by Fermi/GBM~\cite{2017ApJ...848L..14G} and INTEGRAL~\cite{Savchenko:2017ffs}. The top row is a Fermi/GBM lightcurve for gamma ray energies between 10 and 50 keV, while the second is similar but for 50 to 300 keV. The third row is the lightcurve from INTEGRAL for 100 keV to 80 MeV. The bottom row displays the time-frequency map from the coherent sum of the data from LIGO-Livingston and LIGO-Hanford. See~\cite{Monitor:2017mdv} for more details.}
\label{fig:GW170817_GRB}
\end{figure}

The parameter estimation for the GW170817 signal provided the time at which the two neutron stars coalesced. In addition, the observations of a short gamma ray burst were made by the Fermi Gamma-ray Burst Monitor (Fermi/GBM)~\cite{2017ApJ...848L..14G} and the Anti-Coincidence Shield for the Spectrometer for the International Gamma-Ray Astrophysics Laboratory (INTEGRAL)~\cite{Savchenko:2017ffs}. The recorded signals from LIGO, Fermi/GBM and INTEGRAL are dispayed in Fig.~\ref{fig:GW170817_GRB}, and the 1.7 s delay is apparent.

The gravitational wave parameter estimation routines for this binary neutron star merger produce an estimate of the luminosity distance to the source. With the gravitational wave data the luminosity distance was estimated to be $40^{+8}_{-14}$ Mpc~\cite{TheLIGOScientific:2017qsa}. However, the follow-up multi-messenger observing campaign identified the source of GW170817 and GRB 170817A to be in the galaxy NGC 4993. Redshift measurements of the galaxy and the use of the cosmological expansion Hubble constant then provided a comparable estimation of the distance to the source to be $42.9^{+3.2}_{-3.2}$ Mpc~\cite{2017ApJ...848L..12A}. 

In principle, once one has a measurement of the arrival time difference between the gravitational waves and the gamma rays, plus an estimate of the distance, that will allow for a calculation of the difference in the speeds. 
To compare the speed of light, $c$, with the speed of gravity, $v_g$, one should use the most conservative estimate for the distance, namely the lower limit, 26 Mpc.
One does not know, however, the time of the emission for the gamma rays relative to the time of the merger of the neutron stars. While it could very likely correspond to the 1.7 s observation, one can assume that the gravitational waves and gamma rays were emitted simultaneously. 
In this case, the speed of gravity would be faster than the speed of light and the difference in the speeds would be $\Delta v = c - v_g$, $\Delta v/c \approx -3 \times 10^{-15}$.
Another conservative assumption could be that the gamma emission happened 10 s after the merger~\citep{Ciolfi:2014yla,Rezzolla:2014nva}. In this case $\Delta v/c \approx 7 \times 10^{-16}$. As such, the GW170817 and GRB 170817A data from LIGO, Virgo, Fermi/GBM and INTEGRAL were used to constrain the difference between $v_g$ and $c$ to be~\cite{Monitor:2017mdv}
\begin{equation}
-3 \times 10^{-15} \leq \frac{\Delta v}{c} \leq +7 \times 10^{-16} ~ .
\end{equation}  

Another method to probe the validity of general relativity is to conduct parametrized tests of the propagation of gravitational waves. From GW170817 and GRB 170817A one knows that $v_g$ is extremely close to $c$. From relativity a particle of mass $m$, momentum $p$, and energy $E$ will have this quantities related by
\begin{equation}
E^{2} = p^{2} c^{2} + m^{2} c^{4} ~.
\end{equation}
For a particle traveling at speed $c$, $E = pc$, and the rest mass is $m = 0$. To test general relativity with the propagation features of gravitational waves a dispersion relation can be assumed,
\begin{equation}
E^{2} = p^{2} c^{2} + A_{\alpha} p^{\alpha} c^{\alpha} ~.
\end{equation}
A massive graviton would be reflected with $\alpha = 0$ and $A_{0} = m_{g}^{2} c^{4}$, but other possible modifications to the dispersion relation would be testable for $\alpha \ne 0$.
LIGO and Virgo have made such an assumption in their examination of the gravitational wave signals that they have observed. The demonstrated small difference between $c$ and $v_g$, and the very large distance from the source allows for the gravitational wave amplitude (in the frequency domain) to be expressed simply 
\begin{equation}
\tilde{h}(f) = B(f) e^{i \Phi(f)} ~ ,
\end{equation}
and the modifcation to the dispersion relation introduces a slight change to the phase term, $\delta\Phi(f)$. This phase modification is introduced into the parameter estimation routines. Posterior probability distributions are then generated for the $A_{\alpha}$ parameters; this is done for one particular value of $\alpha$ at a time for values between 0 and 4, including non-integer values~\cite{LIGOScientific:2019fpa}.

LIGO and Virgo have used the 10 gravitational wave signals from binary black holes detected in O1 and O2 to constrain $A_{\alpha}$ for a number of $\alpha$ values. The limit on the graviton mass is consequently constrained to $m_{g} \leq 4.7 \times 10^{-23}$ eV/$c^{2}$~\cite{LIGOScientific:2019fpa}. 
A similar analysis using the gravitational wave signal data from the binary neutron star merger GW170817 gives a limit on the graviton mass of $m_{g} \leq 9.51 \times 10^{-22}$ eV/$c^{2}$~\cite{Abbott:2018lct}. The phase modification increases $\delta\Phi(f)$ with the distance to the source, and since the binary black hole mergers are at a larger distance, they provide a better limit on $m_{g}$. 
A total of 31 events from O1, O2 and O3a have been analyzed in this fashion, producing a graviton mass limit of 
$m_{g} \leq 1.76 \times 10^{-23}$ eV/$c^{2}$  with 90\% credibility.
For comparison, the limit set on the graviton mass from studies of the solar system ephemerides is $m_{g} \leq 3.16 \times 10^{-23}$ eV/$c^{2}$~\cite{PhysRevLett.123.161103,Bernus:2020szc}.

\section{Rates and Populations}
\label{RandP}
Based on  all observed gravitational wave signals, one can use statistical methods to estimate the rates of binary black hole or binary neutron star mergers, and describe the population of these systems in the universe. With 50 announced gravitational wave detections to date it is possible to use statistical methods 
to make statements about how often compact binary mergers happen, and what  the probable formation scenarios for these binaries might be. In \citet{Abbott:2020gyp} LIGO and Virgo have recently completed such studies from the O1, O2 and O3a detections described in catalog GWTC-2 ~\citep{Abbott:2020niy}; these studies are based on the 47 compact binary mergers where the gravitational waves were detected with a false-alarm-rate of less than one per year. A similar rates and populations study was presented in \citet{LIGOScientific:2018jsj} based on the 11 compact binary produced gravitational events for O1 and O2 presented in the catalog GWTC-1~\citep{LIGOScientific:2018mvr}. We summarize here the latest methods used by LIGO and Virgo to make statistical statements about the rate of compact binary mergers and their possible formation scenarios.

\subsection{Binary Black Holes}
For the most recent LIGO-Virgo rates and populations study of binary black hole mergers presented in \citet{Abbott:2020gyp} it is demanded that both primary objects have masses exceeding $3 M_{\odot}$ at 90\% credibility, and detected with a false-alarm-rate less than one per year. This produces 44 events, although it excludes GW190814~\citep{Abbott:2020khf}, with its masses of $m_{1} = 23.2^{+1.1}_{-1.0} M_{\odot}$ and $m_{2} = 2.59^{+0.08}_{-0.09} M_{\odot}$; the small secondary mass is considered to be an outlier and difficult to explain via binary formation; see Sec.~\ref{subsubsec:DetGW190814}.

These studies have a number of important goals. One pertains to the mass and spin distributions of the black holes in these merging binary systems. Part of the complexity for this study concerns the possible mass limiting effects from (pulsational) pair-instability supernova processes~\citep{Spera:2017fyx,Farmer_2019}. As discussed with the detection of GW190521~\citep{Abbott:2020tfl,Abbott:2020mjq}, Sec.~\ref{subsubsec:DetGW190521}, it is difficult to explain the formation by stellar processes of black holes in the mass range $\sim 64 - 135 M_{\odot}$. 
For component masses up to $\sim 50 M_{\odot}$ one can imagine formation via common envelope evolution, which also produces nearly aligned spins; see \citet{Kalogera:1999tq}, \citet{2002ApJ...572..407B}, \citet{Dominik:2014yma} and \citet{2017PASA...34...58E}.
Black holes could also form via dynamical processes in dense environments; they would not be affected by (pulsational) pair-instability supernova processes if they were formed by previous mergers~\citep{1993Natur.364..421K,1993Natur.364..423S,PortegiesZwart:2002iks}. Such dynamical formation could create binary black holes where the distribution of spins for the component masses is isotroptic~\citep{Rodriguez:2016vmx}. Primordial black holes are another possible formation channel~\citep{10.1093/mnras/168.2.399,PhysRevD.94.083504}; in this case the black holes would not have significant spin, but what they do have would be isotropic in direction~\citep{Fernandez:2019kyb}.

The observation of gravitational wave events by LIGO and Virgo can also provide evidence as to the minimum mass of black holes formed by astrophysical processes, as opposed to primordial formation where any mass could, in principle, be possible. As already noted, a component mass for GW190814 is $m_{2} = 2.59^{+0.08}_{-0.09} M_{\odot}$; the question is whether this is a black hole or a very heavy neutron star. But even excluding GW190814, strong limits can be placed on the minimum black hole mass. Important research questions concern how gravitational wave observations can be used to distinguish between neutron star and low-mass black hole mass distributions, and to explain the formation of these compact objects \citep{Fishbach:2020ryj}.

Another important question pertaining to the masses of observed binary black hole systems pertains to the distribution of the mass ratio, namely $q = m_{2}/m_{1}$. The observations by LIGO and Virgo seem to indicate that  roughly equal mass binary systems are preferred, but there are important exceptions, as seen with GW190412 and GW190814; see Sec.~\ref{subsubsec:DetGW190814}.

The observed gravitational wave signals from binary black holes also provide important information on the spins of the initial component masses. Both the magnitudes of the spins and their orientation with respect to the obital plane give evidence as to the possible formation of the binary system.

Finally, it is not only important to estimate the rate of binary black hole mergers, but also whether this rate may change with redshift. For example, is there a similarity between the black hole merger rate and the star formation rate~\citep{Madau:2014bja}?

LIGO and Virgo are aleady in a position to give informative statements about all of these questions. Here we present a review of how these statistical studies are conducted, and summarize the latest results from the GWTC-2 catalog from \citet{Abbott:2020gyp}.

\subsubsection{Statistical Methods}
The recent LIGO-Virgo results in \citet{Abbott:2020gyp} were based on the modeling approaches of \citet{Thrane_2019,Mandel:2018mve,Vitale:2020aaz}. 
Bayesian hierarchical methods are used to estimate the parameters of the binary black hole population given $N_{\mathrm{det}}$  individual events  detected by  LIGO and Virgo; the data for these events are represented by $\d_{i}$, $i=1,\ldots,N_{\mathrm{det}}$.
This is accomplished by parameterizing the prior distribution of individual black hole merger parameters such as their masses and spins, putting a {\em hyperprior} distribution on these population parameters or 
hyperparameters. Then
 both individual parameters and hyperparameters are estimated and  the posterior distribution of the population parameters is extracted by marginalizing over the parameters of individual events. Now it also needs to be taken into account that  the number of detected black hole merger events $N_{det}$ is a random variable with a Poisson distribution with expectation given by
$N \xi(\Lambda)$ where $N$ denotes the total number of events expected during the observation period and $\xi(\Lambda)$ is the fraction of detectable binaries for a population with hyperparameter $\Lambda$. This detection fraction is estimated using injections as described in \citet{Abbott:2020gyp}.
 Let $\d_i|\btheta_i$ denote the observation of the $i$'th black hole merger given parameter vector $\btheta_i$  (with $p$ components, e.g.\ mass, spin, redshift)  and ${\cal L}(\d_i|\btheta_i)$ denote the likelihood for events $i=1,\ldots, N_{\mathrm{det}}$. The $\btheta_i$ are assumed to be conditionally  independent with joint population distribution depending on a hyperparameter vector $\Lambda$.
This model can be represented in a hierarchy of priors, where the first level of the hierarchical prior is comprised of the prior distributions of individual parameters $\btheta_i$ and $N_{\mathrm{det}}$ and the second level consists of the  hyperprior distribution of the  prior parameters $\Lambda$ and $N$:
\begin{eqnarray*}
\d_i|\btheta_i &\sim& {\cal L}(\d_i|\btheta_i) \quad \mbox{independent for } i=1,\ldots,N_{\mathrm{det}},\\
\btheta_i|\Lambda &\sim& \pi(\btheta_i|\Lambda)\quad \mbox{ independent for } i=1,\ldots,N_{\mathrm{det}},\\
N_{\mathrm{det}} &\sim& \mathrm{Poisson}(N\xi(\Lambda)),\\
\Lambda &\sim& \pi(\Lambda),\\
N &\sim& \pi(N)
\end{eqnarray*} 
Using Bayes' theorem and conditional independence and denoting $\btheta=(\btheta_1,\ldots,\btheta_{N_{\mathrm{det}}})$ and $\d=(\d_1,\ldots,\d_{N_{\mathrm{det}}})$,  the joint posterior distribution is given by:
\begin{eqnarray*}
\pi(\Lambda,\btheta,N_{\mathrm{det}}|\d) &\propto& (N\xi(\Lambda))^{N_{\mathrm{det}}} \e^{-N\xi(\Lambda)} \times\\
& &\prod_{i=1}^{N_{\mathrm{det}}} {\cal L}(\d_i|\btheta_i) \pi(\btheta_i|\Lambda)  \pi(\Lambda)\pi(N)
\end{eqnarray*}
and marginalizing over all $\btheta_i$ and $N$ (with $\pi(N)\propto 1/N$, a log-uniform prior allows for marginalization over $N$~\cite{Fishbach:2018edt,Mandel:2018mve}) gives the marginal posterior distribution of the population parameters:
\begin{equation}\label{eq:pop}
\pi(\Lambda|\d) \propto  \prod_{i=1}^{N_{\mathrm{det}}}  \frac{1}{\xi(\Lambda)} \left(  \int  {\cal L}(\d_i|\btheta_i) \pi(\btheta_i|\Lambda) d\btheta_i\right) \pi(\Lambda)
\end{equation}
The integrals in Eq.~\ref{eq:pop}  are the  marginal likelihoods for each detected event and can be estimated by importance sampling as described in Eq.~\ref{eq:importancesampling} with samples obtained from  importance density $q$ equal to the individual likelihood ${\cal L}(\d_i|\btheta_i)$ and a default prior $\pi_{\emptyset}$. This enables reusing posterior samples from each event that were obtained under a different prior rather than re-running the MCMC simulations again with the hierarchical prior.

The likelihoods are implemented in {\tt GWPopulation} \cite{PhysRevD.100.043030} and {\tt PopModels} available on GitLab~\footnote{\url{https://git.ligo.org/daniel.wysocki/bayesian-parametric-population-models}}.
Similarly, one can obtain the marginal posterior distribution of a parameter $\theta_j$ ($j=1,\ldots,p$, e.g. mass, spin, redshift) by marginalizing over $\Lambda$ and all other
parameters $\theta_{i-(j)}$:
\begin{equation}
\begin{split}
& \pi(\theta_j|\d) \propto \\
& \int \left[\prod_{i=1}^{N_{\mathrm{det}}}  \left( \int \frac{{\cal L}(\d_i|\btheta_i) \pi(\btheta_{i-(j)}|\Lambda) d\btheta_{i-(j)}} {\xi(\Lambda)}\right)  \right]\pi(\Lambda)d\Lambda
\end{split}
\end{equation}

\subsubsection{Binary Black Hole Models}
LIGO and Virgo have used models of various complexities to attempt to describe the mass distribution of the black holes in merging systems. We give a brief summary of the models used in~\citet{Abbott:2020gyp}. 

The simplest is the {\it Truncated} model, where there are two hard cut-offs between a minimum mass and a maximum mass, with a power law form for the primary (most massive) mass, and the mass ratio. The high mass cut-off is assumed to correspond to where the pair instability mass gap begins~\citep{Spera:2017fyx,Farmer_2019}. 
Note that the Truncated model need not finish where the pair-instability gap is. 
If there are merger products forming new 
binaries, the maximum mass would be higher. To find the mass gap in 
the presence of these mergers would require a model allowing for second 
generation black holes, such as in the study of \citet{Kimball:2020opk}.
This model depends on 4 parameters.

The {\it Broken Power Law} model makes slight modifications to the Truncated model. The hard lower mass cut-off is replaced with a smoothing function. There is also a break, at some mass, in the power law between the two cut-offs, thereby changing the slope of the mass distribution. In this way, the formation of black holes by means not prevented by pair-instability supernovae can be avoided. This model has 7 parameters.

The {\it Power Law + Peak} model is also a slight modification to the Truncated model. Again the hard cut-off at the low mass limit is replaced by a smoothing function. For large masses there is the addition of a Gaussian peak. This peak would try to address an excess of events that are limited from being more massive by pulsational pair instability supernovae~\citep{Talbot:2018cva}. This model has 8 parameters.

\begin{figure*}
\centering
\includegraphics[width=0.9\textwidth]{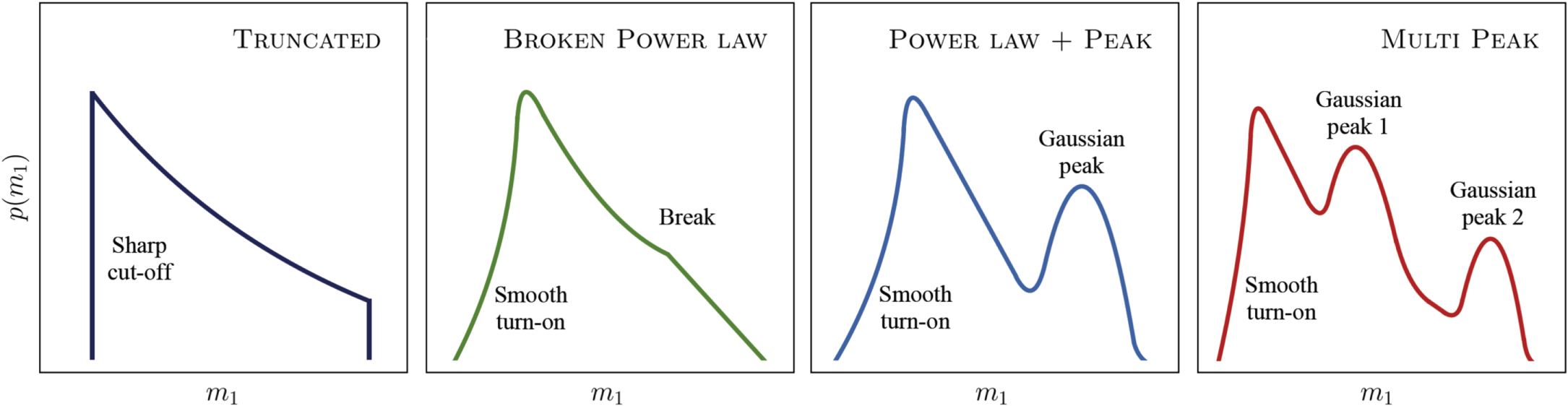}
\caption{A representation of different mass distribution models used in the study of \citet{Abbott:2020gyp} to describe the rates and mergers of binary black holes in the universe.     
See \citet{Abbott:2020gyp}, the source of this figure, for more details.}
\label{fig:R_and_P_models}
\end{figure*}

Finally, the {\it Multi Peak} model is an extension of the Power Law + Peak model. However, for the Multi Peak model two peaks are assumed. The assumption is that with this model hierarchical binary black hole mergers can be addressed, namely the possible population of second generation black holes. This model has 11 parameters. 
See Fig.~\ref{fig:R_and_P_models} for a graphical representation of these mass distribution models that have been used in \citet{Abbott:2020gyp}.

The next important question addressed by LIGO-Virgo is to whether the binary black hole merger rate depends on redshift~\cite{Fishbach:2018edt}. The simplest model assumes that there is no dependence on redshift; this is what is referred to as the {\it Non-Evolving} model, which has no parameters. A redshift dependance is addressed in the {\it Power-Law Evolution} model, where the binary black hole merger rate density is assumed to vary as a power-law with $(1+z)$. This model has one parameter.

Finally, the binary black hole mergers should describe the distribution of spins for the initial component masses. These distributions must describe the magnitude and direction of the spins. For the spin direction the reference is typically the orbital angular momentum vector, and one tries to describe the distribution of the {\it tilts} with respect to this. The models for the spin distributions try to encompass different binary black hole formation scenarios. The simplest model, called the {\it Default}, assumes the same spin distribution for each initial black hole component. A beta distribution, depending on mean and variance parameters, describes the spin magntitude~\citep{Wysocki:2018mpo}. The distribution for the tilt tries to describe two formation channels. For black hole binaries formed by dynamical processes the component for the tilt distribution of the progenitor black holes is isotropic. However for binaries formed in the isolated field of stellar progenitors  the spins are more likely to be aligned with the orbital angular momentum, and hence a Gaussian distribution (depending on some variance) is assumed about the tilt of zero. A mixing parameter describes the relative number of binary black holes systems formed between these field and dynamic scenarios. This spin distribution has been introduced in \citet{PhysRevD.96.023012}. The Default model depends on 4 parameters.

The {\it Gaussian} spin model works by describing spins in terms of the effective spin parameter $\chi_{\rm eff}$ and the effective precession spin parameter $\chi_{\rm p}$. A bivariate Gaussian distribution is used to describe the mean and standard deviation for $\chi_{\rm eff}$ and $\chi_{\rm p}$, plus the correlation between them. Hence this model has 5 paramters. 

The {\it Multi Spin} model simultaneously addresses both mass and spin distributions. The goal is to see if there is a variation in the binary black hole spin distribution as a function of mass. There is a hard cut-off for the masses, like the Truncated model, but there are also Gaussian components for the two initial masses, in addition to the power law. This is somewhat similar to the Power Law + Peak model. Each possible mass distribution can have its own independent spin distribution from the Default model. The Multi Spin model depends on 12 spin parameters and 10 mass parameters.

Comprehensive descriptions of all of these models can be found in \citet{Abbott:2020gyp}.

\subsubsection{Binary Black Hole Population Results from GWTC-2}
The LIGO-Virgo detections presented in \citet{Abbott:2020niy}
have led to a number of important conclusions that are reported in \citet{Abbott:2020gyp}. Gravitational wave observations are describing the rates of compact binary coalescences, and we can expect that future detections will continue to improve our knowledge about black holes and neutron stars in the universe.

The merger rate for binary black hole systems is reported to be $23.9^{+14.3}_{-8.6}$ Gpc$^{-3}$ yr$^{-1}$. This estimation has used the Power Law + Peak distribution for mass, and the Non-Evolving model for redshift distribution. The distribution for the most massive black holes is not well fit by a simple power law. Instead, there appears to be a break in the distribution (a change in the power-law, or the presence of a peak) in the neighborhood of $40 M_{\odot}$. This may be the influence of (pulsational) pair-instability supernova processes. The power-law distribution for the higher masses reaches up to a maximum of $86^{+12}_{-13} M_{\odot}$. The estimation of the low mass limit for black holes from the data from binary mergers is highly dependent on whether GW190814~\citep{Abbott:2020khf} is included in the study or not. If ignored, the studies give a preference for a lower black hole mass around $7 M_{\odot}$. However, with the lower mass in GW190814 estimated to be $2.59^{+0.08}_{-0.09} M_{\odot}$, and if that object is assumed to be a black hole, then the estimation for the lower mass limit for the black hole mass distribution falls to 2.6 to 3 $M_{\odot}$. 

The binary black hole merger rate as a function of redshift does not appear to be uniform. The analysis reveals a preference for models where the binary black hole merger rate increases with redshift, but not as fast as the increase in star formation rate with redshift~\cite{Fishbach:2021mhp}. Specifically, the Power-Law Evolution model was used to describe the redshift, while two mass models were used, namely Power Law + Peak and Broken Power Law. The merger rate is assumed to depend on redshift, $z$, as $(1 + z)^{\kappa}$. In such a case the binary black hole merger rate at $z = 0$ is estimated to be $19.3^{+15.1}_{-9.0}$ Gpc$^{-3}$ yr$^{-1}$. For the Broken Power Law mass model the redshift dependence for the merger rate goes as $\kappa = 1.8^{+2.1}_{-2.2}$, and $\kappa = 1.3^{+2.1}_{-2.1}$ for the Power Law + Peak model. It is interesting to compare these numbers with the star formation rate estimate of $\kappa = 2.7$ from \citet{Madau:2014bja}. Also, the probability that the binary black hole merger rate dependence $\kappa > 0$ is about 90\% for both mass distribution models.

The ensemble of the observed binary black hole merger signals seems to display an interesting distribution for the component spins. The measured distribution for the effective precession spin parameter peaks at $\chi_{\rm p} \approx 0.2$. This peak in the distribution is consistently present with the different spin models. Systems with a non-zero $\chi_{\rm p}$ will experience orbital precession. Consequently, this effect should not be expected to be rare in stellar mass binary black hole systems.

Distributions are also generated for the effective spin parameter $\chi_{\rm eff}$, measuring the component spins aligned with the orbital angular momentum. The mean of the $\chi_{\rm eff}$ distribution is $\mu_{\rm eff} = 0.06^{+0.05}_{-0.05}$, but with a standard deviation of $\sigma_{\rm eff} = 0.12^{+0.06}_{-0.04}$, implying that many systems have $\chi_{\rm eff} < 0$, or at least one component mass with its spin being misaligned by more than 90$^{\circ}$  from the orbital angular momentum. The estimate is that 12\% to 44\% of the binary black hole systems have a mass component with a spin misaligned by more than 90$^{\circ}$ from the orbital angular momentum. This suggests that these binaries would be formed from dynamic processes, whereas aligned spin systems could be from the field formation of isolated stellar progenitors. Note that analysis of \citet{Roulet:2020wyq} finds less support for negative $\chi_{\rm eff}$, possibly due to tidally torqued stellar progenitors.

Finally, the observed spin distributions do not seem to depend on the masses of the progenitors. A black hole formed by the merger of two roughly equal mass progenitors would have an effective spin of approximately $\chi \approx 0.7$. This would lead to the assumption that heavier black holes would have more spin. The preference for higher spin for the heavier black holes is not statistically significant. 

The distributions for the mass, redshift and spins for binary black holes will become better understood with future observations. The full presentation of the population properties of binary black holes is given in \citet{Abbott:2020niy}. 

\subsection{Binary Neutron Stars}
LIGO and Virgo have observed gravitational wave from two neutron star mergers, GW170817~\citep{TheLIGOScientific:2017qsa} and GW190425~\citep{Abbott:2020uma} as reported in GWTC-2~\citep{Abbott:2020niy}. With the assumption that the initial neutron stars are non-spinning, and that there is a uniform distribution for their masses between $1 M_{\odot}$ and $2.5 M_{\odot}$, the rate of binary neutron star mergers is then estimated to be $320^{+490}_{-240}$ Gpc$^{-3}$ yr$^{-1}$. With the additional estimate that there is 1 Milky Way equivalent galaxy (MWEG) per 1 Mpc$^3$~\citep{Kopparapu:2007ib},  this rate then becomes $32^{+49}_{-24}$ MWEG$^{-1}$ Myr$^{-1}$~\citep{Abbott:2020niy}.

\subsection{Neutron Star -- Black Hole Binaries}
The gravitational wave observations of two binary systems containing a neutron star and a black hole have recently been reported by LIGO and Virgo~\cite{LIGOScientific:2021qlt}; see Sec. \ref{subsubsec:DetGW200105_GW200115}. From these two events, and assuming that they represent neutron star - black hole binary systems, a merger rate density of $45^{+75}_{-33}$ Gpc$^{-3}$ yr$^{-1}$ is calculated. Another analysis takes into consideration other events observed by LIGO and Virgo where the primary mass is in the range of 2.5 to 40 $M_{\odot}$ and the secondary mass is in the 1 to 3 $M_{\odot}$ range. Under these conditions the neutron star - black hole binary merger rate density is estimated to be
$130^{+112}_{-69}$ Gpc$^{-3}$ yr$^{-1}$. See \citet{LIGOScientific:2021qlt} for more details on these derivations.

\section{Other Signal Searches for LIGO and Virgo}
\label{Other-Sources}
There are many other sources of gravitational waves targeted by LIGO and Virgo. Here we summarize the signal searches and associated parameter estimation for short and long duration transients (bursts), continuous waves (from pulsars), and a stochastic gravitational wave background. These types of signals have yet to be observed, but sophisticated methods are in place for attempts at detection and then associated parameter estimation. In the absence of a detection,  limits on various parameters have been set. 

\subsection{Stochastic Gravitational Wave Background} 
\label{subsec:stochastic}
The superposition of many independent gravitational wave sources will produce a stochastic gravitational wave background~\cite{ChristensenNelson2019Sgwb}. Just as there is a cosmic microwave background, it is likely that  gravitational waves were created in the early universe, creating a stochastic background that could be observable today. These cosmological sources would be quantum fluctuations in space-time during inflation, phase transitions, or cosmic strings. Astrophysical processes over the history of the universe could also create a stochastic background. These sources would include binary black hole and binary neutron star mergers, supernovae, pulsars, magnetars, or other processes. The observation of gravitational wave events from binary black hole and binary neutron star mergers~\cite{LIGOScientific:2018mvr} implies that there is a stochastic background that may be measureable by LIGO-Virgo in the coming years~\cite{LIGOScientific:2019vic}.

Another assumption is that the stochastic gravitational wave background is isotropic, namely that the characteristics of the gravitational waves are independent of their direction. As such, the LIGO-Virgo correlation analysis tries to determine the energy density of gravitational waves in the universe. 
The normalized energy density of the stochastic background is expressed as
\begin{equation}
\Omega_{\rm GW}(f) = \frac{f}{\rho_c} \frac{{\rm d} \rho_{\rm GW}}{{\rm d} f} ~ ,
\end{equation}
where $\rho_{\rm GW}$ is the energy density of gravitational waves, the closure density of the universe is $\rho_c=3 H_0^2 c^2/(8\pi G)$, $c$ is the speed of light, $G$ is Newton's constant, and we use $H_0=67.9\ {\rm km\ s^{-1}\ Mpc^{-1}}$ \cite{Ade:2015xua}.
One can assume a power law form for the energy density, 
    \begin{equation}
    \label{eq:Omega_spec}
   \Omega_{\rm GW}(f) = \Omega_{\alpha} \left(\frac{f}{f_{\rm ref}}\right)^\alpha ~ .
    \end{equation}
The reference frequency $f_{\rm ref}$ provides location for which the limit on $\Omega_{\alpha}$ is set.

The level of the stochastic gravitational wave signal will be far below the detector noise in the LIGO and Virgo interferometers. The assumption, however, is that there is a common signal present in the different detectors, whereas the noise is independent between the detectors. A correlation analysis using the data from pairs of detectors will allow for the extraction of the common signal~\cite{PhysRevD.46.5250,Romano:2016dpx}. Consider first a simple model where the signal in detector $k$
\begin{equation}
\label{eq:sig_and_noise}
s^{(k)}(t)= h(t) + n^{(k)}(t), \quad k=1,2
\end{equation}
 is the sum of a gravitational wave signal, which is the same in all detectors, and noise $n^{(k)}(t)$, which is unique to each detector. It is assumed that $h \ll n^{(k)}$. A simple correlation between the data from two detectors would then give~\citep{PhysRevD.46.5250}
\begin{equation}
\label{eq:simp_cov}
\begin{split}
\mbox{Cov}(s^{(1)}(t), s^{(2)}(t)) & = \mbox{Cov}( h(t) + n^{(1)}(t), h(t) + n^{(2)}(t)) \\
& = \mbox{Cov}(h(t) ,h(t)).
\end{split}
\end{equation}
This again assumes that the noise in the two detectors is independent and not correlated (exceptions are explained below). In reality, the detectors are located at different locations on the Earth. Hence there is a reduction of the correlation due to the non-alignment of the detectors, and their physical separation. This effect is encompassed in what is known as the overlap reduction function
\begin{equation}
\gamma(f) = \frac{8}{5 \pi} \int_{0}^{2 \pi} d\phi \int_{0}^{\pi} \sin{\theta} \big[F_{1+}F_{2+} + F_{1+}F_{2+} \big] \cos{\mathbf{k} \cdot \mathbf{x}}
\end{equation}
where the detector response functions for two detectors, 
see Eq.~\ref{eq:Fplus} and Eq.~\ref{eq:Fcross},
depend not only on the direction of the gravitational wave source, $(\phi,\theta)$ in an Earth centered coordinated system, but also on their positions
on the Earth and relative orientations. For an isotropic stochastic gravitational wave background the waves of frequency $f$ come from all directions, namely
\begin{equation}
\label{eq:GW-kvector}
\mathbf{k} = \frac{2 \pi f}{c} (\sin{\phi} \sin{\theta}, - \cos{\phi} \sin{\theta}, \cos{\theta})
\end{equation}
and $\mathbf{x}$ is the vector from the vertex of detector 1 to the vertex of detector 2.
 
The LIGO-Virgo stochastic background search produces a cross-correlation statistic, $\hat C^{IJ}(f_a)$, defined as
\begin{equation}
\hat C^{IJ}(f_a) = \frac{2}{T} \frac{{\rm Re}[ \tilde{s}^{*}_{I}(f_a) \tilde{s}_{J}(f_a)]}{\gamma_{IJ}(f_a) S_{0}(f_a)}
\end{equation}
where $I$ and $J$ refer to particular detectors in the network, $f_a$ is the frequency in bin $a$, $T$ is the length of time used to calculate the Fourier transform, $\gamma_{IJ}(f_a)$ is the normalized overlap reduction function between detectors $I$ and $J$~\citep{PhysRevD.46.5250}, 
$\tilde{s}_{I}(f_a)$
is the Fourier transform of the strain time series in detector $I$ (see Eq.~\ref{eq:FFT_signal}), and the spectral shape for a flat background is given by
\begin{equation}
S_{0}(f_a) = \frac{3 H_{0}^{2}}{10 \pi^{2} f_{a}^{3}} 
\end{equation}
\citep{LIGOScientific:2019vic}.
The expected value of the cross-correlation statistic is such that in each frequency bin
\begin{equation}
\big< \hat C^{IJ}(f_a) \big> = \Omega_{\rm GW}(f_a) ~ .
\end{equation}
The variance of the cross-correlation statistic, assuming that the gravitational wave signal is much smaller than the detector noise, is given by
\begin{equation}
\sigma_{IJ}^2(f_a) = \frac{2}{T \Delta f_a} \frac{P_{I}(f_a) P_{J}(f_a)}{\gamma_{IJ}^{2}(f_a) S_{0}^{2}} ~ ,
\end{equation}
where
$\Delta f_a$ is the frequency resolution between the bins $f_a$ and $f_{a+1}$, $P_{I}(f_a)$ is the one-sided noise PSD for detector $I$.

The next step is to create an optimal estimator of the energy density of the stochastic background, namely~\citep{LIGOScientific:2019vic}
\begin{equation}
\label{eq:opt-est}
\hat{\Omega}^{IJ}_{\rm ref} = \frac{\sum_{a} w(f_a)^{-1} \hat C^{IJ}(f_a) \sigma_{IJ}^{-2}(f_a)}{\sum_{a} w(f_a)^{-2} \sigma_{IJ}^{-2}(f_a)}
\end{equation}
with
\begin{equation}
\label{eq:sigmaIJ}
\sigma_{IJ}^{-2} = \sum_{a} w(f_a)^{2} \sigma_{IJ}^{-2}(f_a) ~ 
\end{equation}
and the optimal weighting factors for the spectral shape $\Omega_{\rm GW}(f)$ are
\begin{equation}
\label{eq:weight}
w(f) = \frac{\Omega_{\rm GW}(f_{\rm ref})}{\Omega_{\rm GW}(f)}
\end{equation}
for the fixed reference frequency $f_{\rm ref}$, as for example the power law shape defined in Eq.~\ref{eq:Omega_spec}.

With multiple detector baselines (3, for example for LIGO-Virgo), the final estimator is
\begin{equation}
\label{eq:final-estimator}
\hat{\Omega}_{\rm ref} = \frac{\sum_{IJ} \hat{\Omega}^{IJ}_{\rm ref} \sigma_{IJ}^{-2}}{\sum_{IJ} \sigma_{IJ}^{-2}} ~ ,
\end{equation}
\begin{equation}
\label{eq:final-sigma}
\sigma^{-2} = \sum_{IJ} \sigma_{IJ}^{-2}
\end{equation}
where the sum is over the different independent baselines $IJ$.

One can define a log-likelihood function that compares the stochastic background from model $M$ with the cross-correlation from the data; it is assumed that the detector noise is Gaussian:
\begin{equation}
\label{eq:log-like-SGWB}
  \ln \mathcal{L}(\hat C^{IJ}_a|\Theta) = - \frac{1}{2} \sum_{IJ,a}\frac{(\hat C^{IJ}_a-\Omega_{\rm GW}^{(M)}(f_a|\Theta))^2}{\sigma_{IJ}^2(f_a)},
\end{equation}
where the parameters for the model are represented by $\Theta$. A cosmologically produced stochastic background in the LIGO-Virgo frequency band is predicted to be approximately flat, $\alpha = 0$, while
a background created by binary black hole and binary neutron star mergers throughout the history of the universe would have $\alpha = 2/3$ in
the LIGO-Virgo observational band.
Prior probabilities are then defined and the parameters can be estimated. While this statistical approach used by LIGO and Virgo is a combinations of frequentist and Bayesian methods, it has been shown that the generated results are the same as what can be derived from a fully Bayesian analysis~\citep{Matas:2020roi}.
This LIGO and Virgo parameter estimation method for the stochastic background search is based on the presentation of~\citet{PhysRevLett.109.171102}; this approach was recently expanded to
use nested sampling to create the posterior distributions~\cite{Callister:2017ocg}.

No stochastic gravitational wave background has been detected by LIGO-Virgo,  so upper limits have been set that depend on $\alpha$ and are based on the data from observing runs O1, O2 and O3. This includes marginalization
over the calibration uncertainties.
 The data in the 20 Hz to 1726 Hz band was used.
For $\alpha = 0$ a 95\% credible level upper limit was set at $f_{\rm ref} = $ 25 Hz to be $\Omega_{0} < 5.8 \times 10^{-9}$, and for $\alpha = 2/3$ the upper limit for $f_{\rm ref} = 25$ Hz is $\Omega_{2/3} < 3.4 \times 10^{-9}$~\citep{LIGOScientific:2019vic,Abbott:2021xxi}. Just as there are anisotropies in the cosmic microwave background, there could be such anisotropies in the stochastic gravitational wave background. LIGO and Virgo also search for an anisotropic gravitational wave background. No such background has been observed~\cite{LIGOScientific:2019gaw,Abbott:2021jel}. It is likely that an astrophysically produced stochastic background would be anisotropic~\cite{PhysRevD.98.063509}.

General relativity predicts two polarizations for gravitational waves. This is similar to electromagnetic radiation. Gravitational waves have a quadrupole form, and one says that they have a tensor polarization. Alternate theories of gravity predict scalar and vector polarizations as well. A Bayesian parameter estimation method using nested sampling, including model comparison, has been developed in order to search for a stochastic gravitational wave background containing scalar, vector polarizations~\cite{Callister:2017ocg}. This has been applied to the Advanced LIGO O1~\cite{Abbott:2018utx} and O2~\cite{LIGOScientific:2019vic} data.
A Bayesian odds was computed for having a stochastic background of any polarization present in the data versus Gaussian noise. Another computed Bayesian odds compared models having vector and scalar polarizations in comparison to the general relativisitic prediction of just having the tensor polarizations.
No background was detected for any polarization and upper limits have been set on their energy densities.
Using the O3 LIGO-Virgo data and doing a marginalization over the polarization spectral index $\alpha$ the upper limit on a scalar polarization stochastic gravitational wave background is 
$\Omega^{(S)}_{GW}(25 Hz) \leq 2.1 \times 10^{-8}$ 
at the 95\% credible level,  while for the vector polarization the limit is $\Omega^{(V)}_{GW}(25 Hz) \leq 7.9 \times 10^{-9}$~\cite{Abbott:2021xxi}.

The search for a stochastic gravitational wave background depends on correlations between two or more detectors since the signal level is far below the noise in an individual detector. An assumption in the analysis is that the noise in the different detectors is independent, and not correlated. Two co-located detectors would have the best relative orientation for detection sensitivity. But being co-located would likely lead to correlated noise. This was the case for the initial LIGO S5 stochastic search involving the two co-located H1 (4 km) and H2 (2 km) interferometers~\cite{Aasi:2014jkh}. It was impossible to eliminate correlated noise below 460 Hz. Consequently the upper limit on the energy density of the stochastic gravitational wave background was set at $\Omega_{\rm GW}(f) < 7.7 \times 10^{-4} (f/ 900 Hz)^3$. With the LIGO and Virgo (and KAGRA) sites separated by thousands of kilometers one might naively expect that there would be no correlated noise. However this is not the case. The Schumann resonances are formed by electromagnetic standing waves in the spherical cavity between the surface of the Earth and the ionosphere~\cite{1952ZNatA...7..149S,1954NW.....41..183S}. The cavity is excited by lightning strikes around the world~\cite{PRICE20041179}. Magnetometers at the LIGO and Virgo sites have confirmed the presence of correlated magnetic fields~\cite{Thrane:2013npa}; correlations have also been observed when including the KAGRA site in the network~\cite{Coughlin:2018str}. 
Coincident magnetic {\it bursts}, short duration transients (10s to 100s of milli-seconds in duration), have also been observed in the network~\cite{Kowalska-Leszczynska:2016low}. Different mitigation schemes have been proposed~\cite{PhysRevD.90.023013,Coughlin:2016vor}. The LIGO-Virgo searches for an isotropic stochastic background now ensure that correlated magnetic noise is not corrupting the search. 

A Bayesian parameter estimation approach for addressing correlated magnetic noise in the LIGO-Virgo stochastic background searches has been introduced in~\citep{Meyers:2020qrb}. The models used for the parameter estimation consider the sprectral shape and amplitude of a stochastic background, namely Eq.~\ref{eq:Omega_spec}, and also the magnetic noise contamination from the structure of the Schumann resonances and the magnetic noise transfer functions in LIGO and Virgo. An artificial signal injection study showed that it was possible to estimate the parameters describing a stochastic gravitational wave background and the magnetic field transfer function in the three LIGO-Virgo gravitational wave interferometers. This method has been used to address possible coherent magnetic field coupling in the LIGO-Virgo O3 stochastic background search; the Bayes factor for the presence of magnetic contamination was sufficiently low, showing that this possible correlated noise did not affect the study~\cite{Abbott:2021xxi}. 

The method used in~\citet{Meyers:2020qrb} to use parameter estimation to distinguish a stochastic gravitational wave background from magnetic contamination can also be used to distinguish different stochastic backgrounds with different slopes, or a stochastic background described by a broken power law, as could be produced by phase transitions in the early universe. This method is described in~\citet{PhysRevD.103.043023}.  The study shows that spectral separation for multiple backgrounds could be difficult for the Advanced LIGO - Advanced Virgo network, but could be possible with third generation detectors such as the Einstein Telescope~\cite{ET_Punturo} or Cosmic Explorer~\cite{Abbott_2017_3G,Reitze:2019dyk}. The third generation detectors will be able to observe almost all binary black hole mergers in the observable universe, plus many binary neutron star mergers~\cite{PhysRevLett.118.151105}; these events can then be subtracted from the data and allow for observing a cosmologically produced stochastic background at the level of $\Omega_{GW} \sim 3 \times 10^{-12}$ at 15 Hz~\cite{Sachdev:2020bkk}. However using parameter estimation techniques for spectral separation the sensitivity limit can be reduced to $\Omega_{GW} \sim 4.5 \times 10^{-13}$ for cosmic strings and $\Omega_{GW} \sim 2.2 \times 10^{-13}$ for broken power law models from phase transitions at 25 Hz~\cite{PhysRevD.103.043023}.

The ability to use parameter estimation for spectral separation of the stochastic gravitational wave background has been addressed in many studies. We summarize those here that are applicable to LIGO-Virgo, or third generation ground based detectors.
A frequentist method to conduct component separation method was presented in~\cite{2016JCAP...04..024P}; this uses Fisher Information matrices and maximum likelihood  estimation, and is proposed to replace MCMC methods.
Filters for broken power laws have been proposed to distinguish various stochastic backgrounds~\cite{Ungarelli_2004}.
Yet another approach~\cite{Biscoveanu:2020gds} attempts to directly observe the stochastic background produced by binary black hole mergers over the history of the universe. In order to do this one must cut the data into time segments and assign a probability for the presence of a binary black hole produced gravitational wave signal. Bayesian parameter estimation methods are used to generate the probability for the presence of such a signal, especially concentrating on low amplitude signals that would not be directly detected by standard signal search methods~\cite{Smith:2017vfk}.

\subsection{Continuous Wave Signals}
\label{subsec:CW}
A rotating neutron star that is highly magnetized will often emit regular pulses of electromagnetic radiation. These are known as pulsars.
Such a rotating neutron star with some sort of non-axisymmetric shape will emit gravitational waves~\cite{PhysRevD.20.351,10.1046/j.1365-8711.2002.05180.x}. 
These signals would be quasi-periodic, long duration, having amplitudes and frequencies that are essentially constant. The gravitational wave signal would be sinusoidal at signal frequency $f_{s}$ related to the neutron star rotation frequency $f_{r}$ by $f_s=2f_r$.
There are other mechanisms that could produce gravitational waves at other frequencies, harmonics and the rotation frequency itself~\cite{10.1111/j.1365-2966.2009.16059.x,Glampedakis:2017nqy}. For this review we will concentrate on signals obeying the relationship $f_s=2f_r$.

There are different strategies for searching for a continuous gravitational wave signal from a rotating neutron star. A targeted search attempts to find a signal from a known pulsar. In this case the location of the source on the sky is known, as well as the rotational information (such as the rotational frequency and the phase) from electromagnetic observations. A directed search concentrates on interesting locations on the sky, such as a supernova remnant; such a location may contain a rotating neutron star, but there is no information about the parameters associated with the rotation. An all-sky search attempts to find continuous gravitation wave signals at every location on the sky, and then over a large range of parameters pertaining to rotation. LIGO and Virgo have conducted numerous searches for continuous wave signals, although no detection has been made. So far in the first three observational runs of the advanced detector era these include~\citet{AbbottB.P.2017Fsfg,AbbottB.P.2019Nsfg,Authors:2019ztc,PhysRevD.100.122002,Pisarski:2019vxw,Abbott:2020lqk}.

The observed gravitational wave signal from a rotating neutron star would have the form,
\begin{multline}
\label{eq:CW_h}
h(t)= F_+(t,\psi)h_0\frac{1}{2}(1+\cos^2\iota)\,\cos\phi(t) \\
+ F_{\times}(t,\psi)h_0 \cos\iota \, \sin\phi(t)~ ,
\end{multline}
where the magnitude of the gravitational wave is $h_0$,
$\psi$ is the  polarization angle, $\iota$ the angle between the line of sight and the pulsar's spin axis, and $\phi(t)$ describes the 
phase evolution. $F_+(t,\psi)$ and $F_{\times}(t,\psi)$ represent the response of an interferometric detector to the gravitational wave of the two polarizations, and the polarization angle is $\psi$; these response functions also account for the sky position and the orientation of the detector~\cite{PhysRevD.58.063001,Abbott:2003yq}. 
The detected signal will have a phase evolution that can be expressed as a Taylor series
\begin{multline}
\label{eq:CW_phase}
\phi(t) = \phi_{0} + \\
2\pi \left[f_{\rm s}(T - T_{0}) + \frac{1}{2}\dot{f_{\rm s}} (T - T_{0})^{2} 
+ \frac{1}{6}\ddot{f_{\rm s}}(T - T_{0})^{3} + \ldots\right]
\end{multline}
where the signal arrival time at the solar system barycenter is
$T=  t + \delta t= t + \frac{\vec{r} \cdot \vec{n}}{c}  +
\Delta{T}$, the signal phase for the fiducial
time $T_{0}$ is $\phi_{0}$, the detector position with
respect to the barycenter of the solar system is $\vec{r}$, the unit vector
pointing toward the pulsar is $\vec{n}$, the speed of light is $c$, and relativistic corrections to the arrival
time are within
$\Delta{T}$~\cite{RevModPhys.66.711,ChristensenNelson2004Mafe,Abbott:2003yq}. 
The time derivative $\dot{f_{\rm s}}$ is very small for the majority of pulsars, and timing noise is typically much larger than $\ddot{f_{\rm s}}$.
When radio observations give information on $f_{\rm s}$ and $\dot{f_{\rm s}}$, heterodyning by multiplying the data by $\exp[-i\phi(t)]$, followed by low-pass filtering and resampling will then
produce a model with four unknown parameters: $h_0, \psi, \iota$, and $\phi_0$. The uncertainties of the frequency and frequency derivative give
two other parameters~\cite{ChristensenNelson2004Mafe,Umstatter2004,PhysRevD.72.102002}.

MCMC methods were first applied to such gravitational wave signals using simulated data in \citet{ChristensenNelson2004Mafe}. A Metropolis-Hastings algorithm~\cite{Gilks1996} was first applied to a signal described by the four parameters $h_0, \psi, \iota$, and $\phi_0$. In such a case the sky position and signal frequency are assumed to be known from electromagnetic observations. The MCMC method was successfully demonstrated. A fifth parameter was then added, the  uncertainty  in the frequency  of  the  source, $\Delta f$. This would be a situation where a potential source has known location, but with unknown rotation frequency. A presumed rotating neutron star at the locations of SN1987A~\cite{Zhang:2018dez,Cigan_2019} or Scorpius X-1~\cite{1967ApJ...148L...1S} would be examples of this application. A subsequent study then expanded the number of parameters to six by also including the uncertainty of the frequency derivative of the rotating neutron star, $\Delta \dot{f}$~\cite{Umstatter2004}.    
The study presented methods to sample a posterior distribution with delayed rejection, reparametrizaton, and simulated annealing. This is an improved method for searching for a signal at a known sky location, but lacking information on the neutron star rotation parameters.
While one MCMC routine would run per computing core, an extension of the technique would be to have many MCMCs running on different computer processors, changing the rotation parameters prior probabilities and giving a search over a larger range of frequencies.
The Bayesian methods were further extended in~\citet{PhysRevD.72.102002} where they developed a more complete structure for a signal search and parameter estimation with gravitational wave data, and applied to known radio pulsars. They also applied the method for data from multiple detectors, looking for a coherent signal.

A new version of parameter estimation for continuous wave signals from known pulsars was developed and presented in~\citet{Pitkin:2012yg}; this used nested sampling~\cite{skilling2006}. The possibilities for model selection were also discussed. This method was realized with LALInference~\cite{Veitch:2014wba}. Further developments were presented in~\citet{Pitkin:2017qfy}.

The Advanced LIGO - Advanced Virgo signal searches have used these Bayesian methods, along with subsequent technical developments. For the search of the LIGO O1 data, the Bayesian nested sampling method was applied in the search for gravitational waves from 200 known pulsars~\cite{AbbottB.P.2017Fsfg}.

The original raw L1 and H1 data has a sampling rate of 16,384 Hz, but was then heterodyned according to the phase evolution known from electromagnetic observations. Low frequency noise was removed with a low pass filter, and then downsampled to one data point per minute. The bandwidth of the search was then $1/60$ Hz. The uncertainty in the phase evolution for a pulsar were addressed with the applied nested sampling routine. 

\begin{figure}
\includegraphics[width=0.5\textwidth]{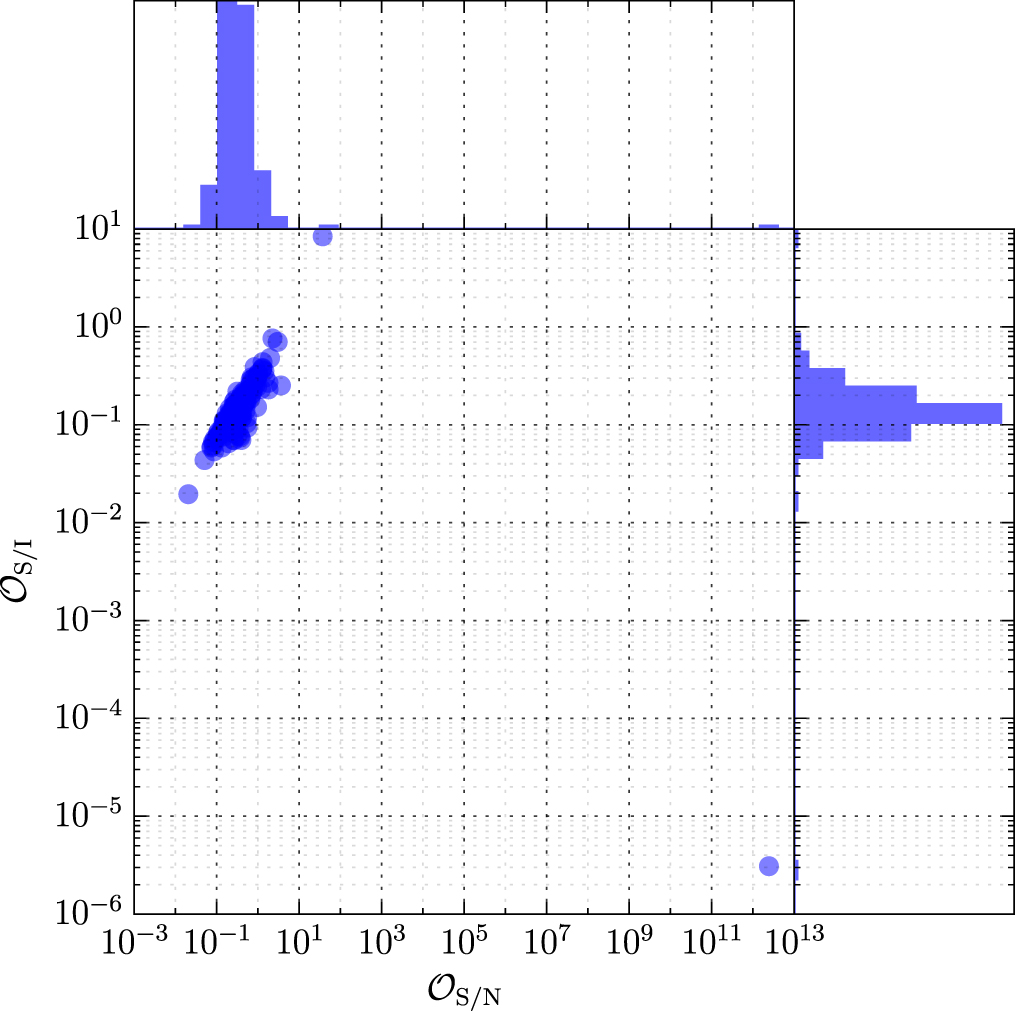}
\caption{The probability ratios $\mathcal{O}_{\mathrm{S/I}}$ and $\mathcal{O}_{\mathrm{S/N}}$ for the search for continuous gravitational wave signals from 200 known pulsars using Advanced LIGO O1 data.     
See \citet{AbbottB.P.2017Fsfg}, the source of this figure, for more details.}
\label{fig:O1_CW_prob}
\end{figure}

Using the data from two detectors also provided the ability to calculate various Bayes factors for the comparison of different models such as the presence of a coherent signal in both detectors, incoherent and different signals in the two detectors, or just independent Gaussian noise in each detector. $\mathcal{O}_{\mathrm{S/N}}$ represents the ratio of the probability that two detectors have a coherent continuous-wave signal to the probability that both detectors just contain independent Gaussian noise. $\mathcal{O}_{\mathrm{S/I}}$ represents the ratio of the probability that two detectors have a coherent continuous-wave signal to the probability that each detector has independent signals or noise with respect to the other detector.
For the 200 pulsars in the study of \citet{AbbottB.P.2017Fsfg} the distribution of $\mathcal{O}_{\mathrm{S/N}}$ and $\mathcal{O}_{\mathrm{S/I}}$ values are displayed in Fig.~\ref{fig:O1_CW_prob}. The pulsar PSR J1932+17
had a value of $\mathcal{O}_{\mathrm{S/I}} \approx 8$, however it was not claimed that this was significant for a detection, especially considering Jeffreys scaling~\citep{Jeffreys61}.
The pulsar PSR J1833--0827 had $\mathcal{O}_{\mathrm{S/N}} \approx 2.5 \times 10^{12}$, but an insignificant $\mathcal{O}_{\mathrm{S/I}} \approx 3 \times 10^{-6}$, which was claimed to be from interference in the data. 
No other signals had significant probabilities for containing a continuous wave signal.
As such, no gravitational wave signal was claimed to be observed from these 200 pulsars~\cite{AbbottB.P.2017Fsfg}. 

This nested sampling method was then extended to a search for not just a signal at twice the rotation frequency (from the from both the $l=m=2$ mass quadrupole mode), but also the rotation frequency (from the $l=2, m=1$ mode) itself. A narrowband time series is made for both frequencies, and this Bayesian analysis then searches for both signals together coherently; the pulsar inclination angle parameter and the polarization angle are assumed to be the same for both frequencies. If a pulsar had a rotation {\it glitch}, as observed from electromagnetic observations, within the time period of the gravitational wave observations, then an additional parameter is added to the analysis, namely an unknown phase offset after the glitch.
Data was used from Advanced LIGO observing runs O1 and O2, and the search targeting 222 known pulsars with rotation frequencies of 10 Hz or larger.
No gravitational wave signal was found at either frequency for any of the pulsars, and various limits were placed for physical parameters associated with the pulsars~\cite{Authors:2019ztc}.

Using LIGO-Virgo data from O3, combined with the data from O1 and O2, this same Bayesian search has been conducted for gravitational wave signals from five pulsars. While again no signals were observed, important limits have been placed on the pulsars' equitorial ellipticities, limiting them to be less that $10^{-8}$~\cite{Abbott:2020lqk}.
A search has also been done with O3 data for continuous gravitational waves from young supernova remnants~\cite{LIGOScientific:2021mwx}, as well as an all sky search for isolated neutron stars~\cite{LIGOScientific:2021tsm}; again no signals have been detected.

A method to search for non-tensorial polarizations in continuous gravitational wave signals has been developed. For alternative theories of gravity, this typically involves the addition of vector and scalar polarization modes.
First, there exists a means to detect gravitational wave signals of any mixture of polarizations, and measure the polarization content~\cite{PhysRevD.91.082002}.
Using nested sampling one can then implement model selection~\cite{Pitkin:2017qfy}; this can be a comparison for the presence of a gravitational wave signal versus just noise, or if a signal is present, the polarization content~\cite{PhysRevD.96.042001}. The method was used to analyze the Advanced LIGO O1 data for possible continuous gravitational wave signals from 200 known pulsars. The search targeted possible tensor, vector and scalar polarizations. No signals of any of the polarizations were detected, so upper limits were placed for the amplitudes of signals for the different polarizations~\cite{Abbott:2017tlp}.

Convolutional neural networks were presented in \citet{PhysRevD.100.044009} as a means to conduct a search for continuous gravitational waves. 
The method appears to be promising in terms of the speed of the analysis. However, more work is needed in order to improve the probability of detection.

Using an MCMC to optimize the F-Statistic, a technique has been developed for a hierarchical follow-up of potential continuous gravitational wave events produced in semi-coherent searches over a large parameter space~\cite{Ashton:2018ure}. This method uses an affine-invariant ensemble sampler~\cite{Foreman_Mackey_2013}.

\subsection{Core-Collapse Supernovae}
One of the most important possible signal source for LIGO and Virgo would be from a core-collapse supernovae (CCSN)~\cite{Gossan_2016,Radice_2019,abdikamalov2020gravitational}. 
Gravitational wave observations from CCSN can aide in our ability to discern the explosion mechanisms of stars. The extremely complicated nature of CCSN creates a challenge for estimating the physical parameters.
The last few years have seen much progress in the development of numerical codes for simulating the physics of CCSN, including gravitational wave emission~\cite{BMueller:2020}.

LIGO and Virgo have not yet observed gravitational waves from CCSN~\cite{PhysRevD.101.084002}. CCSN offer an unique opportunity to conduct multimessenger astronomy, with the chance to observe gravitational waves, electromagnetic radiation, and neutrinos from the CCSN.
In fact, the importance of multimessenger astronomy was displayed with the electromagnetic and neutrino observations of SN1987A~\cite{PhysRevLett.58.1490,PhysRevLett.58.1494}.
The timescale for a CCSN produced
gravitational wave signal is short, about 1 s or less.
CCSN are an important target for LIGO and Virgo gravitational wave observations~\cite{Abbott:2019prv,PhysRevD.101.084002}.

The emission of electromagnetic radiation from a CCSN can be delayed, from seconds to days, due to the high densities of charged particles present. The photons are forced to do a random walk through the material in order to exit~\cite{2011ApJ...728...63R,Waxman2017}.
On the other hand, gravitational waves and neutrinos can exit instantly from the core of the star, and they will carry important information about the physical processes, such as the core collapse and the revival of the shock-wave~\cite{Kuroda_2017}.
It is impossible to derive analytical expressions for gravitational waves from CCSN that capture the complexity involved with all the physical processes: high energy particle physics, nuclear physics, general relativity, thermodynamics. The predicted gravitational wave signals are a product of the numerical simulations, but these can take months to run for a single waveform capturing all of the complicated physics in the three dimensions~\cite{Bruenn:2018wpz}. Parameter estimation methods will need to work with these conditions, so imaginative methods are required.

The work of \citet{Summerscales_2008} used a maximum-a-posteriori approach to attempt to separate the gravitational wave signal produced by a CCSN from the detector noise. 
They justify the Gaussian assumption for the likelihood using the principle of maximum entropy~\cite{PhysRev.106.620,PhysRev.108.171} as the Gaussian distribution  maximizes the entropy amongst all distributions with the same mean and variance and can thus be interpreted as the most conservative in this class of distributions.
Instead of sampling from the posterior distribution, they find the parameter values that maximize the posterior distribution.
The gravitational waveforms derived via the inference methods would be compared to a catalog of simulated waveforms. The catalog would presumably contain a large number of predicted signals generated by covering a large volume of the physical parameter space.
To do this, they would compare their estimated waveforms to the CCSN gravitational wave signal waveforms from the catalog of \citet{ott:2004}. 
The assumption is that the waveform derived from inference would be most similar to the simulated signal in the catalog which has physical characteristics most in agreement with the real CCSN.
To quantify the success of the routine, the authors used a simple cross-correlation between the gravitational waveforms from inference and the catalog of simulated CCSN waveforms. 

A method using principal component analysis (PCA) was introduced to decompose the signals in a CCSN gravitational wave catalog, and to create an orthonormal basis vector set~\cite{Heng_2009}. With the CCSN catalog used~\cite{PhysRevLett.98.251101} (using a large range of rotation rates) 12 principal component vectors were sufficient to reconstruct the catalog waveforms with a \emph{match} exceeding 0.9.

The use of PCA was extended in \citet{Rover2009} where they used a 
Metropolis-within-Gibbs sampler~\cite{GelmanAndrew2014} to
reconstruct CCSN gravitational waveforms using principal component
regression (PCR). The CCSN catalog that was used had various progenitor masses, initial spins, initial differential spins, and nuclear equations of state~\cite{dimmelmeier:2008}. The PCA eigenvectors are first calculated from the waveform catalog.
A signal from the catalog (but not used in creating the PCA eigenvectors) was injected into simulated LIGO detector noise.
With Bayesian inference, information about the signal was derived. This was accomplished by generating the posterior probability distribution functions of the PCA eigenvector amplitudes and the pulse arrival time. Attempts were made to use the reconstructed signal and the amplitudes of the PCA eigenvectors to give information about the CCSN physical parameters. 
The match between the reconstructed signal and the waveforms from the catalog was quantified with a $\chi^2$ value.
This study had limited success in making a clear association to the physical parameters.

The reconstruction of CCSN signals using Bayesian PCR was further developed in \citet{EdwardsThesis} with a birth-death RJMCMC~\cite{Green1995711}.
In such an analysis the number of principal components was not fixed in advance but treated as an additional parameter to be estimated. Model selection was addressed via model averaging.

It was subsequently shown in \citet{Edwards2014} that one can derive important astrophysical information from the PCA coefficients derived with the PCR methods described in \citet{Rover2009}.
By sampling from the posterior predictive distribution one can derive credible intervals for some physical parameters, including 
the ratio of the rotational kinetic energy to gravitational energy of the inner core at the bounce.
Two supervised machine learning methods were also applied so as to classify the
precollapse differential rotation profile, and it was shown that the techniques are effective at discriminating between different rapidly rotating progenitors to the CCSN.
The study displayed a constrained optimization approach for model selection which provided a value for the appropriate number of principal components for the Bayesian PCR models.

\begin{figure}
\includegraphics[width=0.5\textwidth]{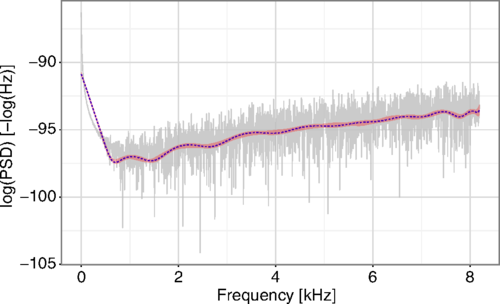}
\caption{An example of estimating the noise PSD while also reconstructing a CCSN signal. Assuming an Advanced LIGO noise curve, this figure displays the estimation of the log PSD for this noise. In shaded pink (small band) is the 90\% credible region, while the posterior median is in dashed blue (line). The solid gray (large band) is the log periodogram.    
See \citet{PhysRevD.92.064011}, the source of this figure, for more details.}
\label{fig:ccsn_noise}
\end{figure}

\begin{figure}
\includegraphics[width=0.5\textwidth]{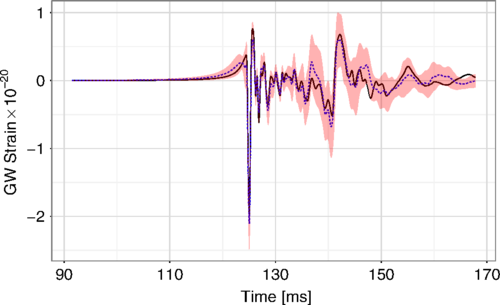}
\caption{Displayed here is the reconstruction of a gravitational wave signal from rotating core CCSN, in particular, the signal A1O10.25 from \citet{abdikamalov:2014}, that was embedded in simulated Advanced LIGO noise with a SNR $= 50$. The shaded pink band is the 90\% credible interval, the dashed blue line is the posterior mean, while the solid black line is the A1O10.25 signal.  
See \citet{PhysRevD.92.064011}, the source of this figure, for more details.}
\label{fig:ccsn_signal}
\end{figure}

As decribed in Sec.~\ref{PSD}, the estimation of the noise PSD is critical for gravitational wave parameter estimation.
For a short duration transient gravitational wave signal, the most accurate parameter estimation of the signal qualities would occur if the noise PSD is estimated simultaneously with the signal parameters. The study of \citet{PhysRevD.92.064011} displays parameter estimation methods for CCSN while simultaneously estimating the noise PSD.
Often in gravitational wave signal searches it is assumed that the noise present has a Gaussion distribution, is stationary, and that the PSD has been determined from observations at a time different from that when a signal is present. However the observed LIGO and Virgo data can violate these assumptions~\cite{LIGOScientific:2019hgc}. Hence, an incorrect estimation of the noise PSD would in general affect parameter estimation results.
This was addressed in \citet{PhysRevD.92.064011} where  a Bayesian semiparametric method employed a nonparametric Bernstein polynomial prior for the noise PSD with weights given from a Dirichlet process distribution and then the Whittle likelihood provided an update. A Metropolis-within-Gibbs sampler~\cite{GelmanAndrew2014} provided the posterior samples. In addition to the noise estimation, a rotating CCSN gravitational wave signal was injected into simulated Advanced LIGO noise, and the reconstruction parameters were estimated using the PCR method described in \citet{Rover2009,Edwards2014}. An example is displayed in Fig.~\ref{fig:ccsn_noise} and Fig.~\ref{fig:ccsn_signal}. A CCSN signal (A1O10.25 from the catalog \citet{abdikamalov:2014}) was injected into simulated Advanced LIGO noise with SNR $= 50$. The CCSN signal was reconstructed, Fig.~\ref{fig:ccsn_signal}, while simultaneously estimating the noise PSD, Fig.~\ref{fig:ccsn_noise}.

The physics behind a CCSN is complicated and complex. The way in which the explosion happens is not yet totally explained. Different models exist, and lead to different parameter estimation results. For example, one mechanism that has been proposed is a neutrino driven explosion; this would apply to slowly rotating progenitors. Another possibility is a magnetorotational driven explosion, which applies for progenitors that are
rapidly rotating. A comprehensive review of CCSN explosion mechanisms is presented in \citet{janka:2012}.
The resulting gravitational wave signals from these mechanisms are different, and could be distinguished by parameter estimation. The study of \citet{PhysRevD.86.044023} showed that one could differentiate between different magnetorotational explosion mechanisms. Assuming the Advanced LIGO sensitivity, they claim that this differentiation could be done for CCSN in the Milky (distances less than 10 kpc). For neutrino driven explosions their method could differentiate between different models for CCSN source distances up to 2 kpc.
The parameter estimation method applies PCR and nested
sampling \citep{skilling2006}; a calculated Bayesian evidence supports
the most probable CCSN mechanism for the explosion.
\citet{powell:2016} continued this line of research with the use of nested sampling for providing evidence for CCSN explosion mechanisms. This was tested by embeding simulated signals in real LIGO detector data in order to understand the effects of nonstationary and non-Gaussian noise. 
They claim with their method the Advanced LIGO - Advanced Virgo network at design sensitivity could establish if the mechanism for the explosion is neutrino driven for CCSN in the galaxy, and rapidly-rotating core collapse
to the Large Magellanic Cloud.
Then the subsequent study presented in \citet{powell:2017} used three-dimensional CCSN simulations and a means to reject noise transients.
The importance for applying these methods to the data from third generation gravitational wave detectors was presented in \citet{Roma:2019kcd}.

The study of \citet{abdikamalov:2014} presented two methods to use CCSN produced gravitational wave data to 
estimate parameters. One was the total angular momentum of the inner core at the bounce. Another was the ratio of the inner core's rotational kinetic energy to gravitational energy. Their model for the CCSN included the effects of rotation, gravitation, electron capture during collapse, and the nuclear equation of state. They also assume noise at the level of Advanced LIGO at design sensitivity.
Their first method uses a template bank from their simulations (124 models) and then they use matched filtering to estimate the total angular momentum to 20\% for rapidly rotating cores, and 25-35\% for slowly spinning cores. A Bayesian nested sampling model selection technique was applied to estimate the differential rotation. The Bayesian approach used PCA and expanded on the work of \citet{Rover2009} and \citet{PhysRevD.86.044023}, and was able to successfully describe the pre-collapse differential rotation profile.

A frequentist approach to the problem of parameter estimation for gravitational waves observed from CCSN was presented in \citet{Engels:2014nua}. PCA is applied to CCSN waveform catalogs.
A least squares solution relates the PCA eigen-solutions and the physical parameters. The method was successfully demonstrated for CCSN signals injected into detector noise, and then identifying important CCSN parameters that are responsible for the form of the gravitational wave signal.

The study of \citet{Bizouard:2020sws} used frequentist parameter estimation methods to address the information that can be extracted from proto-neutron stars formed after a CCSN. Specifically, the information comes from the observed g-mode signal (gravity modes)~\citep{Kokkotas:1999bd}. From the time-evolution of the g-mode, as described in \citet{PhysRevLett.123.051102} and potentially observable in the gravitational wave signal, it was shown that it is possible to estimate a relationship between the evolving proto-neutron star (PNS) mass $M_{PNS}$, and radius $R_{PNS}$. 
They show that one can observe how the ratio $M_{PNS}/R^{2}_{PNS}$ evolves with time.
The model with the best fit to the data (time-frequency evolution) is chosen by the Akaike information criterion~\citep{10.2307/4356217,https://doi.org/10.1111/1467-9868.00353}. Assuming the design sensitivity noise levels for the Advanced LIGO -- Advanced Virgo network, it is stated that the mass-radius evolution for a PNS could be observable for sources within the Milky Way. For the third generation detectors, such as Cosmic Explorer and Einstein Telescope, the observable range could extend to around 100 kpc. 

In the study of \citet{Edwards:2020hmd} deep convolutional neural networks were tested on simulated gravitational wave signals from CCSN with rotating progenitors in order to determine the nuclear equation of state.
The study used the 1834 gravitational waveforms from \citet{PhysRevD.95.063019}, which were generated from simulations using axisymmetric general-relativistic hydrodynamic scenarios. These simulations consisted of 98 rotation profiles and 18 equations of state.
With the convolutional neural network framework of \citet{Edwards:2020hmd}, 
which examined the temporal and visual patterns in the gravitation waves from rotating CCSN, correct  classifications at the level of 64\% to 72\% were achieved. If the signal set is then reduced to the five equations of state with the largest probability estimates for a given test signal, the identifaction success rises to 91\% to 97\%. The effects of the inclusion of detector noise will be the goal of a future study using these convolutional neural network methods.  

An information-theoretic approach to the detection of unmodeled short-duration transient gravitational wave signas is presented in \citet{Lynch:2015yin}. After a detection by a short duration transient signal search, parameter estimation is done with MCMC and nested sampling.
Ad-hoc models are used, such as sine-Gaussians, Gaussians and damped sinusoids; the parameters associated with these models are then estimated. It is claimed that the generated Bayes Factors can improve the detection efficiency.

The BayesWave analysis pipeline~\cite{Cornish:2014kda,Cornish:2020dwh}, described in Sec.~\ref{subsec:BayesWave},
can be used to reconstruct unmodeled transient gravitational wave signals. It can also be used to differentiate between gravitational wave transient signals (coherent across the detector network) and transient noise (incoherent across the detector network). An accurate reconstruction of the gravitational wave from a CCSN will certainly be an asset in studies pertaining to understanding the signal source.

\subsection{Long Duration Transients}
An important potential gravitational wave signal is a long duration unmodeled transient. Signal search pipelines have been developed to look for these signals in LIGO-Virgo data~\cite{PhysRevD.83.083004,PhysRevD.91.104021}. These searches target signals lasting tens of seconds, up to potentially weeks in duration. Pattern recognition methods are used to search for structure in time-frenquecy maps produced from the data. Long-duration gravitational wave signals have been a target for LIGO-Virgo~\cite{Abbott:2017muc,Abbott:2019heg}.
\citet{CoughlinM2014Mfeo} presented a method for estimating model parameters from the observed time-frequency map of the data using nested sampling.

This method was further developed to look for gravitational wave signals after the merger of a binary neutron star coalescence~\cite{Banagiri:2019lon}. Assuming that there was a massive neutron star remnant that survived for some time after the merger, there could be oscillations in the remnant, thereby producing gravitational waves. The production of gravitational waves in such a scenario is difficult to prediect, especially with respect to the phase of any oscillation. 
The study presents a means to probe for the presence of long duration post-merger gravitational wave signals, and to put limits on or measure various properties of the post-merger remnant.
A phase-agnostic likelihood is produced which only uses the spectral content of the signal. Simulated data with the O2 Advanced LIGO noise sensitivity is used. The physical parameters which are attempted to be constrained are the gravitational wave amplitude, the start time of the signal, the spin-down timescale, a model dependent spin-down parameter that can change the overall spin-down rate, and the initial gravitational wave frequency. It is assumed that the sky position and distance to the source are known, similar to GW170817. Nested sampling was used, specifically the PyMultiNest package of~\citet{Buchner:2014nha}. Depending on the models for the post-merger remnant, the study demonstrates the potential to constrain various paramereters, such as the ellipticity of the remnant, or its braking index.

\subsection{Cosmic Strings}
Gravitational waves offer an unique way to display new physics, and one example would be cosmic strings. These are remnants of a false vacuum, and manifest themselves as one-dimensional topological defects. If they exist they would have been made after a spontaneous  symmetry  phase  transition~\citep{Kibble_1976,Vilenkin:2000jqa}.
Many different field theories could be responsible for producing cosmic strings; Grand Unified Theories in the early universe could produce them~\citep{Kibble_1976,Vilenkin:2000jqa}. They could be produced when inflation is terminating~\citep{Sakellariadou:2009ev}.

Cosmic string loops have periodic oscillations, and as they do so, they produce gravitational waves. The gravitational waves are created by cusps, kinks and
kink-kink collisions. The gravitational waveforms are calculable~\citep{Damour:2000wa,Damour:2001bk,Damour:2004kw}. The gravitational wave signals and emitted power depend on the string tension; this is typically represented as $G \mu$, where $G$ is Newton's constant and $\mu$ is the linear mass density of the string ($c = 1$ assumed).

Because the waveforms for cosmic string produced gravitational waves can be calculated, LIGO and Virgo search for their signals with a template based search. No such gravitational waves have been detected by LIGO and Virgo~\citep{Abbott:2017mem,Abbott:2019prv,Abbott:2021ksc}. The absence of the detection of a stochastic gravitational wave background also constrains cosmic string parameters~\citep{Abbott:2017mem,LIGOScientific:2019vic,Abbott:2021ksc}. If a short duration gravitational wave signal from a cosmic string were detected, parameter estimation methods could certainly be applied using the know form of the waveforms~\citep{Kuroyanagi:2012wm}.

Bayesian parameter estimation methods can be used to constrain cosmic string parameters with the results from a stochastic gravitational wave background search. Given a model for cosmic string formation and the string tension, a stochastic background can be predicted. 
Various different cosmic string models $M$ exist \citep{Blanco_Pillado_2014,Lorenz_2010,Auclair_2019} which will produce a stochastic gravitational wave background and an associated energy density, $\Omega_{\rm GW}^{(M)}(f_a;G\mu,N_k)$, that depends on the string tension $G \mu$, the frequency $f_a$, and $N_k$ is the number of kinks per cosmic string loop oscillation. These would the be inserted into Eq.~\ref{eq:log-like-SGWB}, using for the parameters $G \mu$ and $N_k$.
\begin{equation}
\label{eq:likelihood}
  \ln \mathcal{L}(\hat C^{IJ}_a|G\mu,N_k) = - \frac{1}{2} \sum_{IJ,a}\frac{(\hat C^{IJ}_a-\Omega_{\rm GW}^{(M)}(f_a;G\mu,N_k))^2}{\sigma_{IJ}^2(f_a)}.
\end{equation}
For priors, in the LIGO-Virgo obsering run O3 cosmic string analysis the string tension prior is log uniform for $10^{-18} < G \mu < 10^{-6}$. Because no cosmic strings signals were detected, let alone any stochastic background, various ranges of the $G \mu - N_{k}$ parameter space are excluded for different models~\cite{Abbott:2021ksc}. Note that a cosmic string origin for GW190521 was investigated, but a $log_{10}$ Bayes factor of $\sim 30$ strongly favors a binary black hole origin for the signal over a cosmic string origin~\citep{Abbott:2020mjq}.

\section{Parameter Estimation Packages For Gravitational Waves}
\label{Packages}

Here we give a brief description of major software packages that have been made available for parameter estimation of gravitational waves. This is not an exhaustive list.
Further references to smaller packages or code for specific tasks such as spectral density or population parameter estimation can be found in the respective sections of this review.

\subsection{LALInference}
\label{sec:LALInference}
LALInference \citep{Veitch:2014wba} is the original and primary tool that is currently used by LIGO and Virgo for parameter estimation of gravitational wave signals.
It provides a flexible and  open-source toolkit available at \url{https://lscsoft.docs.ligo.org/lalsuite/}. It consists of a C library with postprocessing functions implemented in {\sc Python}. It can make use of all waveform approximants implemented in the LSC Algorithm Libary (LAL) and provides implementations of two independent samplers: a parallel tempering MCMC scheme \cite{Gilks1996} and nested sampling \cite{SkillingJohn2012Bcib}.

\subsection{PyCBC}
\label{sec:PyCBC}
In contrast to LALInference which is written in C, 
PyCBC \cite{Biwer_2019}  is a Python-based suite of functions for parameter estimation of compact binary coalescence signals.  It is an open-source toolkit available on GitHub.
Several waveform models are available either directly implemented in PyCBC or via calls to LAL and the Whittle  likelihood function Eq.~\ref{eq:Whittle} is used. The user can choose between three ensemble MCMC samplers: {\tt emcee} \cite{Foreman_Mackey_2013}, its parallel-tempered version {\tt ptemcee} \cite{VousdenW.D.2015Dtsf} and {\tt kombine}, a kernel-density-based, embarassingly parallel ensemble sampler  \url{https://github.com/bfarr/kombine}. PyCBC is also using a dynamical nested sampling algorithm~\cite{Nitz:2021uxj}, \textit{dynesty}~\cite{Speagle_2020}.

\subsection{Bilby} \label{sec:Bilby}
The package {\sc Bilby}  is a collection of {\sc Python}-based routines for Bayesian parameter estimation of gravitational waveform parameters.
It aims to provide a more user-friendly suite of computational tools than LALInference through modularisation. Its core library is not specific to the analysis of gravitational waves but can be used for general parameter estimation problems.  The core library passes user-defined likelihood and prior to various samplers and returns the results in an {\tt hdf5} file. The Python software package {\tt PESummary} \cite{hoy2020pesummary} can be used for processing and visualising the results. The user has the choice of the MCMC samplers {\tt emcee} \cite{Foreman_Mackey_2013}, {\tt ptemcee} \cite{VousdenW.D.2015Dtsf}, PyMC3 \cite{SalvatierJohn2016PpiP},  and various versions of nested samplers 
such as dynesty (the default)~\cite{Speagle_2020}, Nestle~\cite{nestle}, CPNest~\cite{john_veitch_2021_4470001}, PyMultiNest~\cite{Buchner:2014nha}, PyPolyChord~\cite{Handley:2015fda}, UltraNest~\cite{BuchnerJohannes2016Astf,Buchner_2019}, DNest4~\cite{Brewer:2016scw}, and Nessai~\cite{williams2021nested,nessai}.
Its gravitational wave specific library allows the user to define their own waveform models but includes standard waveform approximants for  transient signals via the LALSimulation package
 \cite{lalsimulation}. The standard likelihood is Eq.~\ref{eq:Whittle} and implementations of the current gravitational wave detectors, their location, orientation and PSDs are provided. A further functionality is provided by enabling hierarchical Bayesian inference on populations. A parallelized version of  nested sampling implemented in {\tt pBilby} enables the use of hundreds or thousands of CPUs of a high-performance computing cluster and  yields a reduction in computation time that scales almost linearly with the number of parallel processes \citep{10.1093/mnras/staa2483} and provides an efficient implementation for population inference \cite{PhysRevD.100.043030}.
Bilby also has a MCMC sampler, Bilby-MCMC~\cite{Ashton:2021anp}.
A detailed description of this software package which is available from the git repository \url{https://git.ligo.org/lscsoft/bilby/}, can be found in \citet{Ashton:2018jfp} and \citet{Romero-Shaw:2020owr}.

\subsection{BAJES}
\label{subsec:BAJES}
The Python-based  package {\tt bajes} \citep{Breschi2021} implements a Bayesian inference pipeline for compact binary coalescence transients that has the flexibility to combine different datasets and physical models. Similar to {\sc Bilby}, it provides a user-friendly modular software with minimal dependencies on external libraries but specific functionalities for multimessenger astrophysics. Algorithms for sampling from the posterior distribution are based on  {\tt emcee} \cite{Foreman_Mackey_2013}, parallel tempering MCMC with a variety of proposal distributions,
and nested sampling \citep{skilling2006}.

\subsection{RIFT}
\label{subsec:RIFT}
RIFT stands for {\sc Rapid parameter inference on gravitational wave sources via Iterative FiTting} and provides fast methods to infer parameters of 
coalescing, precessing compact binary systems. The algorithm is based on original work by \citet{PhysRevD.92.023002} and described in detail in \citet{Lange:2018pyp} and \citet{O_Shaughnessy_2017}. RIFT achieves a considerable saving in computation time by a combination of various strategies: considering  candidate signals on a regular grid of the parameter space and interpolating the likelihood values over the grid;  marginalizing the likelihood over the extrinsic parameters using an adaptive Monte Carlo integration scheme; given training data,  interpolating the log marginal likelihood using Gaussian processes; using an adaptive Monte Carlo method to sample from the marginal posterior distribution of the intrinsic parameters; and finally iterating the fitting and sampling procedures  over revised training data.
Using graphical processing units for some of the elementary operations of the RIFT algorithm, a further substantial improvement of computation time can be obtained \citep{Wysocki_2019}. The source code is available at \url{https://github.com/oshaughn/research-projects-RIT}.

\subsection{BayesWave}
\label{subsec:BayesWave}
Bayeswave is an open-source suite of C++ functions scripted in  Python. It has been designed to robustly estimate gravitational wave signals, noise and instrumental glitches without relying on any prior assumptions of waveform morphology. Its main importance lies in the morphology-independent waveform reconstruction. It is available on GitLab \url{https://git.ligo.org/lscsoft/bayeswave}.
BayesWave models instrumental transients and burst signals using a Morlet-Gabor continuous wavelet frame where the number and placement of the wavelets is variable and estimated by a transdimensional RJMCMC algorithm. It can simultaneously estimate the noise power spectral density using the BayesLine algorithm  \cite{Littenberg:2014oda}.
Detailed descriptions of the methodology are given in \citet{Cornish:2014kda} and \citet{Cornish:2020dwh}.

\subsection{BAYESTAR}
\label{sec:BAYESTAR}
The rapid determination of the source location (sky position and distance) is of huge  importance for multi-messenger astronomy. BAYESTAR \cite{Singer_2016,PhysRevD.93.024013} provides such a rapid sky-localization code that yields Bayesian estimates of the 3-dimensional information within minutes. This is accomplished by fixing the values of the intrinsic parameters to the values from the detection pipeline, computing the posterior distribution of the extrinsic parameters and  approximating the marginal posterior distribution of the sky location via numerical integration.
Bayestar exploits the fact that almost all of the information needed for producing a sky position estimate can be extracted from the matched fiter trigger, plus the detectors' signal arrival times, amplitudes, and phases. Hence Bayestar need not use the intrinsic parameters when estimating just the sky localization.

\section{Conclusions}
\label{Conclusions}
Beyond the detection of gravitational waves, Bayesian statistical methods have played  a pivotal part in estimating the physical parameters of gravitational waveform models.
We have given a comprehensive review of Bayesian methodology used for parameter estimation with an in-depth focus on Bayesian computational techniques for characterizing the posterior distribution of waveform parameters. Simulation-based posterior computation  methods in their various forms such as Gibbs sampling, Metropolis-Hastings algorithms, Hamiltonian MCMC, adaptive MCMC, RJMCMC, nested sampling, parallel tempering and combinations thereof have been the pre-dominant techniques used for estimating the waveform parameters of signals observed by ground-based detectors.  General methods for accelerating the convergence of MCMC algorithms \citep{robert2018accelerating}  as well as reduced order models and surrogate waveform models \citep{CanizaresPriscilla2015AGWP,Setyawati_2020} will prove essential for future applications, in particular for third generation observatories \citep{smith2021bayesian}. Moreover, the newly emergent machine-learning  methodology based on neural networks and deep learning offers promise for gravitational wave applications. For short burst signals like the black hole mergers that have been observed by LIGO/Virgo, it has so far been adequate to estimate the instrumental noise characteristics in terms of its power spectrum separately from the waveform parameters and treating the power spectrum as fixed for the purpose of parameter estimation. State-of-the-art MCMC methods now also quantify the remaining uncertainty in the spectral density estimates and estimate both noise power spectrum and waveform parameters simultaneously using a semiparametric approach. For longer duration signals, it will be important in the future to take the time-varying nature of instrumental noise into account and develop robust  methods for estimating evolutionary spectra. Bayesian methods for checking model assumptions such as stationarity \citep{Cornish:2014kda,LIGOScientific:2019hgc,EdwardsMatthew2020IaAN}  and for assessing  model fit such as posterior predictive checks \citep{GelmanAndrew2014}, some of which are already included in \cite{Abbott:2020gyp}, will need close attention and further development as tools  to scrutinize the  validity  of results.

This review includes a summary of results and conclusions for the various detections during the first three observations runs  of LIGO/Virgo  that could be drawn from parameter estimation.  An exhaustive overview of Bayesian model comparison methods is given and  based on these, conclusions from tests of general relativity are detailed. The hierarchical Bayesian modeling approach and its application to estimating the rates of compact binary mergers as well as to inference on possible formation scenarios is explained. With  ever increasing sensitivity of the ground-based detectors, the number of detections in future observing runs will be surging and hierarchical Bayesian methods for estimating population parameters such as their merger rates, mass spectra and spin distributions,  will become more and more important.

\section{Acknowledgments}
We thank Christopher Berry, Sylvia Biscoveanu, Gregorio Carullo, Nathan Johnson-McDaniel, Geraint Pratten, Richard O'Shaughnessy and Daniel Williams for helpful comments on the manuscript.
R.M.\ gratefully acknowledges support by the James Cook Fellowship from Government funding, administered by the Royal Society Te  Ap\={a}rangi and  DFG Grant KI 1443/3-2. The  work of N.C.\ is supported by NSF Grant No. PHY-1806990. This paper has been given LIGO Document No. P2100108.

\bibliography{main}

\end{document}